# Maintaining Consistency of Data on the Web

## Dissertation

Conducted for the purpose of receiving the academic title
"Doktor der Sozial- und Wirtschaftswissenschaften"

### Advisors

**Gerti Kappel**
Vienna University of Technology
Institute of Software Technology and Interactive Systems (E188)
Business Informatics Group

**Michael Schrefl**
Johannes Kepler University at Linz
Department of Business Informatics
Data & Knowledge Engineering

Submitted at the Vienna University of Technology
Institute of Software Technology and Interactive Systems
Business Informatics Group
by

**Martin Bernauer**

Vienna, December 2004

# Eidesstattliche Erklärung

Ich erkläre hiermit an Eides Statt, dass ich die vorliegende Dissertation selbständig verfasst, andere als die angegebenen Quellen und Hilfsmittel nicht benutzt und die aus anderen Quellen entnommenen Stellen als solche gekennzeichnet habe. Diese Dissertation habe ich bisher weder im Inland noch im Ausland in irgendeiner Form als Prüfungsarbeit vorgelegt.

Wien, im Advent 2004

# Abstract

Increasingly more data is becoming available on the Web, estimates speaking of 1 billion documents in 2002. Most of the documents are Web pages whose data is considered to be in XML format, expecting it to eventually replace HTML, the current lingua franca of the Web, e.g., by XHTML.

A common *problem* in designing and maintaining a Web site is that data on a Web page often replicates or derives from other data, the so-called base data, that is usually not contained in the deriving or replicating page. Two properties of Web sites account for this situation. First, the hypertext structure of a Web site not necessarily coincides with the structure of its underlying conceptual domain model, thus it may be necessary to present a single data item on several pages. Second, the content of pre-generated Web pages is often drawn from legacy systems, usually relational databases. In this case Web pages replicate data items from databases.

Consequently, replicas and derivations become inconsistent upon modifying base data in a Web page or a relational database. For *example*, after modifying a product's price in the database, already pre-generated Web pages offer the product at an out-dated price. Or, after assigning a thesis to a student and modifying the Web page that describes it in detail, the thesis is still incorrectly contained in the list of offered thesis, missing in the list of ongoing thesis, and missing in the advisor's teaching record.

The thesis presents a *solution* by proposing a combined approach that provides for maintaining consistency of data in Web pages that (i) replicate data in relational databases, or (ii) replicate or derive from data in Web pages. Upon modifying base data, the modification is immediately pushed to affected Web pages. There, maintenance is performed incrementally by only modifying the affected part of the page instead of re-generating the whole page from scratch.

The proposed approach provides for consistent, up-to-date Web pages any time. It is efficient by providing incremental page maintenance techniques, generic by maintaining consistency of XML data in general, flexible by reacting to modifications in Web pages of other businesses, transparent by maintaining a business' autonomy in managing its data, open by allowing future extensions to be built on top of it, and extensible by enabling the integration of arbitrary legacy systems.

# Kurzfassung

Im Web sind zunehmend mehr Daten verfügbar, wobei Schätzungen eine Milliarde Dokumente im Jahr 2002 nennen. Es wird davon ausgegangen, dass die meisten dieser Dokumente im XML Format vorliegen, da zu erwarten ist, dass HTML von XML als Lingua franca des Webs abgelöst wird, z.B. in der Form von XHTML.

Ein verbreitetes *Problem* im Entwurf und der Wartung von Websites liegt darin, dass Webseiten sehr oft andere Daten replizieren oder aus anderen Daten abgeleitet werden. Diese anderen Daten, die auch als Basisdaten bezeichnet werden, sind für gewöhnlich nicht Bestandteil der replizierenden oder ableitenden Seite. Zwei Eigenschaften von Websites zeichnen für dieses Problem verantwortlich. Erstens muss sich die Hypertext-Struktur einer Website nicht notwendigerweise mit dem darunterliegenden konzeptuellen Domänenmodell decken. Daher kann es notwendig sein, ein und dasselbe Datenelement auf mehreren Seiten darzustellen. Zweitens wird der Inhalt vorgenerierter Seiten oft aus Legacy Systemen bezogen, wie relationalen Datenbanken. Webseiten replizieren in diesem Fall Datenelemente von Datenbanken.

Folglich werden Replikas und Ableitungen inkonsistent, wenn sich Basisdaten ändern, die in Webseiten oder relationalen Datenbanken gespeichert sind. Zum *Beispiel* zeigt eine vorgenerierte, ein Produkt anbietende Seite nach der Änderung des Preises in der Datenbank dieses zu einem falschen Preis an.

Die Dissertation bietet eine *Lösung* an, indem ein kombinierter Ansatz zum Erhalt der Konsistenz jener Webseiten vorgeschlagen wird, die (i) Daten aus relationalen Datenbanken replizieren, oder (ii) Daten aus Webseiten replizieren bzw. sich aus ihnen ableiten. Werden Basisdaten geändert, wird die Änderung sofort zu betroffenen Webseiten weitergeleitet. Dort wird deren Wartung inkrementell vollzogen, indem nur die von der Änderung betroffenen Teile modifiziert werden, anstatt die Seite komplett von neuem zu generieren.

Der vorgeschlagene Ansatz bietet so konsistente und aktuelle Seiten zu jeder Zeit. Er ist u.a. effizient durch die Anwendung inkrementeller Seiten-Wartungstechniken, generisch durch den Erhalt der Konsistenz in XML Daten im Allgemeinen und flexibel durch das Reagieren auf Änderungen in Webseiten anderer Unternehmen.

# Contents

# Chapter 1

# Introduction

## Contents

Increasingly more data is becoming available on the Web. Statistics report 4.8 million Web sites providing 500 million documents as of 2000 [124] or 9 million Web sites[1] providing almost 1 billion[2] documents as of 2002. Most of the data is contained in Web pages and is often change-dependent on data in other Web pages or relational databases. Thus, changes to a Web page or database should entail modifications in change-dependent Web pages to *maintain consistency*. While maintaining consistency is a topic that has been extensively addressed in the database literature, such research is still rare in the area of the Web. The aim of this thesis is to explore consistency maintenance in the realm of the Web.

---

[1]Reported by Online Computer Library Corporation (OCLC), available at `http://www.oclc.org/research/projects/archive/wcp/stats/size.htm`.

[2]The number of documents is estimated using the site/document ratio as of 2000.

The thesis[3] comprises three parts, each dealing with one of the three settings in which consistency of data on the Web should be maintained. As will be discussed in detail in this chapter in course of the definitions and the problem statement, these settings address consistency between Web pages and relational databases, within Web pages, and in document flows. The latter refers to the exchange of XML documents in business transactions and personal ad-hoc data exchange.

Part I introduces an approach that maintains consistency between Web pages and relational databases (Section 2), and subsequently presents the approach's realization with off-the-shelf relational database technology (Section 3). Part II proposes an approach that maintains consistency within Web pages (Section 4), with a focus on composing events that may cause inconsistencies (Section 5), and realizing the metaschema with XML (Section 6). Part III presents a two-layered approach for maintaining consistency in document flows, which consists of a generic layer to trace document flows (Section 7), and a layer that adapts the approach from Part II to maintain consistency in document flows (Section 8).

The rest of this chapter is devoted to underlying definitions (Section 1.1), the problem statement (Section 1.2), the proposed contribution (Section 1.3), and related work (Section 1.4).

## 1.1  Definitions

Before problems and solutions in maintaining consistency of data on the Web can be discussed, we need a clear understanding of the notions of "data on the Web" and of "maintaining consistency" of it. This section aims at providing that understanding in the following subsections. For the reader who is already familiar with the area, the following paragraphs, which summarize the section, may suffice as definitions.

The notion of "data on the Web" refers to static data stored in Web pages, where the data is XML data that appears in all three design dimensions of Web applications, i.e., content, hypertext, and presentation. Data in the content and hypertext dimension may originate from relational databases. In the realm of this thesis, only those Web pages are considered that are a view over other data. This notion applies to Part I and II of the thesis, while Part III uses an extended notion of data on the Web, namely one that refers to XML documents that may be arbitrarily moved around on the Web instead of being allocated statically at a Web server, thus forming document flows on the Web. For more details see Subsection 1.1.1.

---

[3]The thesis is based on a number of papers, namely [19, 20, 21, 22, 23, 132, 133].

The notion of "consistency of data on the Web" refers to view consistency, meaning data being consistent with data it is derived from. With the above notion of data on the Web, consistency is threefold: (1) consistency between relational databases and the Web, (2) consistency within data on the Web, and (3) consistency within document flows on the Web. Figure 1.1 on page 8 depicts the first two aspects graphically. For more details see Subsection 1.1.2.

### 1.1.1 Data on the Web

Data on the Web has been *semi-structured* from its beginning [7, 35]. And although the Web was intended to be able to handle arbitrary data formats, which is reflected in the content negotiation feature of the Hypertext Transfer Protocol (HTTP [60]), the main data format used turned out to be the Hypertext Markup Language (HTML [148]). According to its inventor, the design of HTML was a bit of a hack [24], which, however, did not hinder its success. It has been tidied up later on by defining it as a document type definition specified in the Standard Generalized Markup Language (SGML [87]). Semistructured data is often referred to as "self-describing", allowing it to come along without a schema, and usually mixes structured data with unstructured data, i.e., text, making it an ideal format for data on the Web. The structure of semi-structured data usually varies among data instances due to lacking schemas or appropriately designed schemas.

Throughout the thesis, the notion of "data on the Web" refers to *static data* on the Web. Static means that the data requested by a user has been stored in the same form before the request is made. Data on the Web is contained in Web pages being the units that users request and servers deliver. *Static Web pages* are Web pages that contain only static data and are stored on Web servers and are delivered as-is upon requesting them. On the contrary, dynamic data or dynamic Web pages are generated by an application upon a user's request. They are also said to be generated on-the-fly. During generation, various sources may be taken into account such as a user's input, her profile, and data residing in relational legacy databases.

Static Web pages are predominant in two settings. First, when a legacy database is not employed to store the data contained in Web pages but the data is stored directly in the pages themselves. This is common for small to medium sized Web sites. Second, when a legacy database is employed and Web pages are not generated on-the-fly but are pre-generated before users' requests, mainly to increase performance and reduce response latencies. Such techniques are employed for large Web sites only. Generally, delivering Web pages with minimal response latencies upon a user's request is considered to be the most important design requirement for Web pages [117]. Thus, and because databases are widely used to store a Web site's content, efficiently publishing database content on the Web is a key success factor for Web sites.

On the contrary, with dynamic Web pages response latencies are likely to occur, because dynamically generating pages is expensive [49, 137, 170]. Most Web sites experience problems of response latencies, because they use the naive approach of generating Web pages on-the-fly. To improve the situation, several techniques have been proposed in theory and practice to speed up the construction of Web pages, such as pooling database connections, using prepared statements, or employing load balancing. However, these techniques have two major shortcomings. First, when they focus on efficient database access, they ignore that constructing an HTML or XML page is often more expensive than retrieving its content from a database [170]. Second, when they focus on efficient page construction, they ignore that dynamically generating a page is more expensive than reading a pre-generated page from disk.

In literature, pre-generation of static web pages is a topic of *data-intensive Web sites* with pre-generation being considered a design principle [42]. When pre-generating pages, one has them ready built upon users' requests whereby the shortcomings of generating pages on-the-fly are overcome. Furthermore, approaches for pre-generating Web pages often feature a schema-based approach for defining Web pages (e.g., see [137, 170]). Thus they provide for a more declarative way of specifying Web pages, drastically easing definition and maintenance of pages compared to programmatically generating them on-the-fly. Approaches for data-intensive Web sites can be divided into two categories, ones that support data integration from various sources such as relational databases and the Web, and others that do not. The first category needs an internal data model, to which integrated sources are mapped. Giving two early examples of such approaches, Araneus [10, 110] and Strudel [58] use an internal semistructured data model. Aside of concentrating on how to integrate and model Web sites, both provide for pre-generation of Web pages. The second category comprises approaches dealing with pre-generation of semistructured data, i.e., Web pages directly from relational databases without any intermediate representation, such as [23, 99, 125, 133, 137, 170]. These approaches discuss alternatives in realizing the pre-generation process and their implications. For more information on data-intensive web sites the interested reader is referred to [64].

Nowadays, data on the Web seen by users can be considered to be mostly *XML data*, i.e., data in the format of the Extensible Markup Language (XML [147]). The data is either already in XML format or likely to be transformed to XML in the future to eventually replace HTML (see [114]). Since 1998, when XML was first standardized by the W3C, research has been active on XML issues. As such, for example, Araneus switched to XML (see [109]) and the pre-generation of Web pages directly from XML databases has been addressed (e.g., in [5]).

Looking at the structure of a Web site as a set of Web pages in more detail, it can be characterized by three *design dimensions*, namely content, hypertext, and presentation (e.g., see [11, 40, 129]). The *content* dimension describes the data and relationships between pieces of it provided on Web pages, i.e., the Web site's content, while the *hypertext* dimension describes the composition of data to Web pages and navigation paths between them. Finally, the *presentation* dimension describes how Web pages are presented to the user. These dimensions are reflected in hypermedia design methods that deal with them in separate phases in their process model, e.g., OOHDM [135], RMM [86], and the Araneus Design Methodology [11]. Consequently, languages for modelling Web applications distinguish between these dimensions as well, e.g., WebML [41].

Data in the three design dimensions may appear in different formats. Under above premise that data seen by users is XML data, data in the hypertext and presentation dimension is XML data. Data in the content dimension can be XML data, however, since XML is a rather new format and data is often stored in legacy systems such as databases, it can be relational or other data as well. Due to commonplace ad-hoc constructions of Web pages, data in the hypertext dimension does not necessarily have a counterpart in the content dimension, being referred to as "*Web-only content*", e.g., data describing a lecture may solely be stored in the hypertext dimension. Moreover, it is not necessary that a dedicated document stores a Web page's data in each dimension. For example, a single XML document, such as an XHTML document and an XML document that includes its own stylesheet as a part of it, comprises data from a page's hypertext and presentation dimension.

Various vocabularies can be used to express XML data in the respective design dimension. Data in the content dimension can be expressed in a vocabulary addressing the Web application's domain or a domain-independent general purpose vocabulary. This also applies for data in the hypertext dimension, where additionally navigation can be expressed using standard linking vocabularies like XLink [155] and XPointer [156]. Finally, data in the presentation dimension can be expressed using standard vocabularies like XHTML [159], CSS [146], SVG [162], or XSL [152] (also known as XSL-FO). Data in the presentation dimension confers not only to data that is directly visible to the user, but also to data used in constructing the Web site's presentation.

Data in the hypertext dimension often originates from *relational legacy databases* as indicated earlier. This has led to several approaches for providing XML views over relational databases in the database literature. In the context of the Web this has led to the aforementioned approaches for data-intensive Web sites that are capable of pre-generating Web pages from relational database content. Both kind of approaches aim at publishing relational data as XML, however, with a different focus. For details see Subsection 1.4.2.

Finally, only Web pages are relevant that are defined as a *view* over other data. Together with the characteristics of data on the Web discussed so far, this implies that data in a Web page's hypertext or presentation dimension must be a view, i.e., dependent on data possibly contained in other Web pages or in relational databases. Note that Web pages that are not at least partially a view are not of interest, while the amount of data that is a view may vary from very little to the complete Web page. Most likely, data in a Web page's hypertext dimension will be a view (a) on data in other Web pages' hypertext dimension or (b) on data in the content dimension. One may safely assume that data in the presentation dimension or data defining navigation will be dependent on other data less frequently.

The last part of the thesis (Part III, see also Subsection 1.2.3) uses an *extended notion* of data on the Web. It relaxes the characteristics of data discussed so far in that it abandons the necessity of data being contained in Web pages and rather deals with XML documents in general. Thus talking of data appearing in any design dimension is not appropriate anymore when the refined notion of data is used. Moreover, it extends the notion of data in that it deals with documents that are stored on varying locations in a network, thereby forming *document flows*. This resembles the flow of data in the course of business transactions or ad-hoc data exchange.

Summarizing, if not stated otherwise, the notion of "data on the Web" subsumes the following characteristics: (i) static data stored in Web pages, where (ii) the data is XML data, which (iii) may appear in all three design dimensions of Web applications, i.e., content, hypertext, and presentation, where (iv) data in the content and hypertext dimension may originate from relational databases, and finally, where (v) only Web pages are relevant that are a view over other data.

### 1.1.2  Consistency of Data on the Web

This subsection briefly surveys different notions of consistency, especially those common in the database area, and subsequently identifies the notion of consistency that is addressed by the thesis.

The notion of *consistency*, generally meaning an "agreement or harmony of parts or features to one another or a whole"[4], has different meanings in different research areas. For example, in the area of interorganizational workflows, a consistent workflow execution refers to participating businesses having the same knowledge of whether an interaction as part of the workflow has failed or succeeded. Differently, in the area of Web caching, consistency refers to cached Web pages having the same content as their original counter

---

[4]Merriam-Webster Online Dictionary, `http://www.m-w.com`

Web page, regardless of whether the latter is a static or dynamic Web page.

Looking at the notion of consistency in the database area, it has the following overloaded meanings.

First, *type consistency* means that data is consistent with types and additional constraints. Thus, the data's structure adheres to the type's structure (e.g., a tuple has a value for each of its schema's attributes), the data's values adhere to the type's definitions (e.g., `124` is a valid `ssnr number(10)`), and the data's values adhere to additional constraints (e.g., `123` satisfies `CHECK ssnr<999`). For details on type consistency see, e.g., [36].

Second, *view consistency* means that data is consistent with data it is derived from, i.e., with data it is a view of. This means that view data $d_v$, which depends on base data $d_b$ due to replication or derivation, must be updated upon modification of the base data. Defining $d_v$ using view function $v$ as $d_v := v(d_b)$ this means that upon modifying $d_b$ to $d_b'$, the view data is only consistent with the base data if its value is $d_v' = v(d_b')$. For details on view consistency see, e.g., [75].

Third, *consistency under concurrent access* means that data is consistent after concurrent access to it. This means that if two or more applications access the same data concurrently (not simultaneously but interleaved) the value of the data after the applications end is the same as it would have been if the applications had been executed sequentially. Database theory deals with this kind of consistency in serializability theory [73, 168].

Type consistency and consistency under concurrent access are already widely dealt with in the XML (database) literature. Moreover, these two notions of consistency are closer linked to the area of databases than to the Web. Briefly sketching existing solutions, several schema languages allow to express types in XML such as DTD [147], XML Schema [157, 158], and Relax NG [118]. In addition, constraints may be specified by these languages as well, or special purpose languages such as Schematron [89] and SchemaPath [108]. Protocols for serializable concurrent access to XML data are under development, such as the ones proposed in [76, 77], which leverage ideas from relational databases.

Remembering that data on the Web refers to Web pages that are views, one easily recognizes that the notion of view consistency neatly provides the needed characteristics for describing *consistency of data on the Web*. As such, it provides for the notion of consistency the thesis addresses. As will be pointed out later in Section 1.4, this is not an area without any research results, however, we have approached the issue differently from others.

Figure 1.1 summarizes the notion of maintaining consistency of data on the Web by depicting its first two aspects: (1) consistency between relational databases and the Web, and (2) consistency within data on the Web. The third aspect, (3) consistency within document flows on the Web, is left out for conciseness. The figure depicts consistency maintenance by arrows pointing from base data to derived data. As one can see, consistency is to be

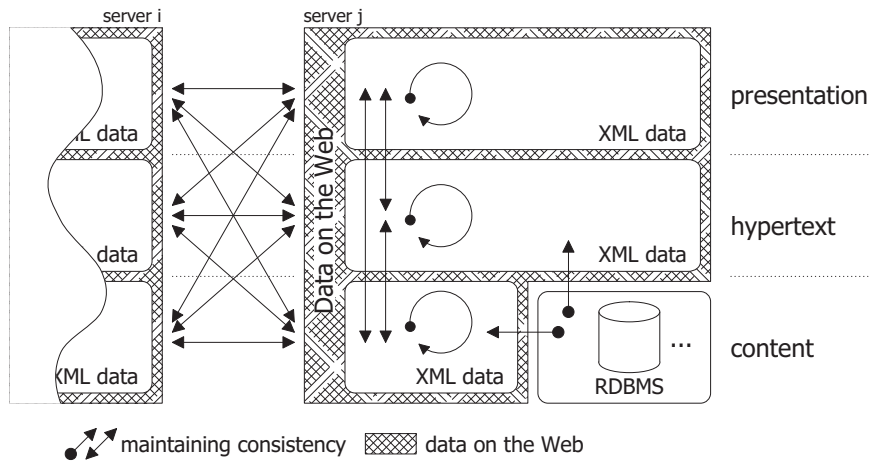

Figure 1.1: Maintaining Consistency of Data on the Web

maintained across and within the design dimensions of Web pages and across
and within different Web servers. Note that maintaining view consistency
within legacy systems such as relational databases is not addressed, since
this has been dealt with extensively in database literature and elsewhere.

## 1.2    Problem Statement

Now having the definition of "consistency of data on the Web" at hand,
we can break down the problem of maintaining consistency of data on the
Web into the following three parts, each one being dealt with in a separate
subsection.

### 1.2.1   Consistency between Relational Databases and the Web

When pre-generating Web pages from relational databases, it has to be de-
cided on how to model Web pages, how to model the database, and how
to map between the two, having far-reaching consequences. First, the ex-
pressivity of the models and the mapping determines the kind of Web pages
that can be modelled. Moreover, the models determine the kind of mappings
that can be designed. Second, the kind of mapping determines the main-
tainability of a Web site in terms of the amount of manual action that needs
to be performed upon a modification in the database to adjust the map-
ping. Third, the models and the mapping determine possible consistency
techniques, varying in their update granularity and kind of consistency they
can provide. All of these aspects in pre-generating Web pages and their
consequences are discussed in the following.

The *database* can be modelled either *instance-based* or *class-based*. This addresses the possibility to modularly design the chunks of data that represent the content of Web pages. A simple approach will allow to define such chunks individually only, i.e., being instance-based. An approach should employ a class-based technique, generating the corresponding instances automatically. This will reduce the complexity of underlying database queries and simplify mapping knowledge and its maintenance in terms of easily determining "dirty" pages, i.e., pages that contain out-dated content.

Also *Web pages* can be modelled either *instance-based* or *class-based*. Instance-based refers to modelling every single page by its own, while class-based refers to modelling only classes of Web pages. A class of Web pages (short a "page class") comprises a set of pages that share the same structural, navigational, and presentational features, just like the extent of a class in an object-oriented database collects objects of the same kind. An approach should support class-based models, because they again simplify mapping knowledge and its maintenance. In addition, an approach may also support instance-based models, however, this is less important.

Possibilities to define the *mapping knowledge* between the database and Web pages can be distinguished along two dimensions. First, the mapping may be *implicit*, e.g., defined within database queries for determining the content of a page, or *explicit*, e.g., in the form of metadata. Explicit mapping knowledge facilitates the maintenance of and reasoning about it. Second, following the models for databases and Web pages, mappings may be *instance-based*, mapping database instances to Web pages, or *class-based*, mapping database classes to page classes. Class-based mappings result in the least amount of mapping knowledge, minimized design and maintenance costs, and lowest coupling between the database and Web pages.

The maintained *consistency* between Web pages and database content can be *weak* or *strong*. Under weak consistency Web pages may contain out-dated content, while strong consistency ensures up-to-date content on Web pages. Weak consistency is usually solved by periodically polling for database updates and carrying out re-generation of affected pages in case of detected updates. If an implicit mapping knowledge is used it is likely that only weak consistency can be maintained (e.g., as in [170]). Maintenance of weak consistency is also said to be "pull-based". Strong consistency mechanisms ensure immediate updates of Web pages upon a database modification. The mechanism usually exploits explicit mapping knowledge to determine the pages that need to be modified. Strong consistency is usually "push-based", pushing database modifications to dirty pages. If only one kind of consistency is supported by an approach, strong consistency is obviously more important. With either consistency, special attention needs to be drawn to update all affected pages upon a single database modification and that those changes become visible to users at once, so that the Web pages are consistent among themselves. This is also referred to as *multiple* *view consistency* in the literature [174].

Web pages can be modified to be up-to-date with different *update granularities*. First, a dirty page can be re-generated from scratch by "pulling" the page's content from the database and generating a new page replacing the dirty one. This involves execution of many unnecessary queries and maintenance by *completely* re-generating Web pages that contain for their largest part unchanged content, resulting in high maintenance costs. Second, a dirty Web page can be modified *incrementally* by only updating the "dirty" part of it. This is likely to reduce maintenance cost, because no unnecessary data needs to be read from the database.

Summarizing, the **problem** in maintaining consistency between relational databases and the Web lies in (i) designing the languages for modelling relational databases and Web pages, so that (ii) a mapping that is easy to maintain can be designed, i.e., an explicit and class-based mapping, which (iii) allows to design a mechanism that at least provides for strong consistency with incremental update granularity.

### 1.2.2   Consistency within Data on the Web

Data on the Web can be *change-dependent* on other data in different organizational settings. The base data, i.e., the data depended on may reside within the same Web page, or on other Web pages within the same business or on Web pages of another business. If base data does not reside within the same Web page, it is referred to as "remote data". If other businesses are involved, they may be cooperating, i.e., agreeing and supporting replication or derivation of data, or not cooperating, being neutral or hostile to others replicating or deriving data from theirs. Intuitively, the organizational setting that is to be supported by an approach influences its technical realization. The thesis' focus is on base data within the same business or in cooperating businesses, while also allowing the integration of data from hostile businesses.

Various *examples* for change-dependencies in different organizational settings come to mind. Within the same business, an employee's data, e.g., his/her phone number or job title is often replicated on several Web pages. Across businesses, a travel agency selling trips that use a certain accommodation may want to raise the price of the trips if the price for an overnight stay in that accommodation rises. Or, an author or reader of a book may be interested when a review is added at Amazon, a university department may want to publish job offers at its Web site provided by a job agency that could be interesting to its students, or a citation index like CiteSeer[5] may want to derive its citations directly from publication announcements made at universities' Web sites. An example for a possibly hostile derivation is Geizhals[6], a service that collects and compares offers of different businesses for the same product.

---

[5]`http://citeseer.nj.nec.com`

*Reasons* for change dependencies are manifold. Technically, the intent is to adapt data according to remote data, to have data replicated or derived from remote data available in a new context, or to combine and/or compare remote data locally. More concrete, it can be used for example to add value by presenting remote data in a new context, like publishing relevant job offers at a department's Web site; to provide a preview of linked content or an annotated link; to provide a navigation context in order not to get lost in hyperspace; to promote remote data by presenting it on several pages; and finally to add value by combining and/or comparing remote data, e.g., comparing products.

To achieve consistency, storing data in a relational database and using approaches for maintaining consistency between the database and Web pages is often not a viable option. This would only be viable in the organizational setting of a single business, and even there it may not be easy to negotiate and run a shared database. Moreover, there exists content for which it is not reasonable to store it in a database. For such Web-only content, i.e., content in Web pages that is not drawn from underlying databases, a light-weight Web-based solution to Web content management would be a better alternative. This should not replace but complement database functionality.

Additional kinds of change dependencies exist to the ones on data in other Web pages. First, data may be *change-dependent on time.* An approach should support for adapting Web pages to points and periods in time, e.g., for automatic adding of up-to-date content or removing out-dated content. For example, an announced job offer on a department's Web site is to be removed after the application deadline is over. Second, data may also be *change-dependent on data in legacy systems* such as relational databases or workflow systems. This resembles the functionality of data-intensive Web sites as far as data in relational databases is concerned. Nevertheless, a uniform interface to maintain consistency to arbitrary legacy systems is desirable. And third, data may be *change-dependent on a Web page's history*, i.e., on previous states of a page. For example, a department publishes jobs announced by a job agency, which may announce a job multiple times, only if the department has not published the job before.

Based on the above considerations and the properties of the Web, four desired characteristics can be derived that an approach for maintaining consistency of data on the Web should feature. First, change dependencies between Web pages are usually characterized by autonomous control over participating pages, since they often belong to autonomous businesses. Thus, businesses being involved in replication or derivation of data must not be constrained in their *autonomy* in how they store and manipulate pages in

---

[6]`http://www.geizhals.at`

any way. Second, ideally there exists *low coupling* between the systems that
maintain consistency, so that one business is not (vitally) affected by actions
another business undertakes, such as adding/removing a Web page or going
offline for some period. Third, the possibility to *integrate legacy systems*
in a uniform way should be supported to adapt Web pages according to,
e.g., data residing in relational databases or workflow systems. Fourth, the
approach should be *general* so that it can be applied to manage data in any
of the three design dimensions of content, hypertext, and presentation.

Summarizing, the **problem** in maintaining consistency within data on
the Web lies in designing an approach supporting change dependencies on
data in Web pages, on data in legacy systems, on time, and on Web pages'
histories. It should (i) preserve businesses' autonomy in managing their Web
site, (ii) result in a low coupling between systems holding dependent data,
(iii) provide a uniform interface to legacy systems, and (iv) maintain con-
sistency in all design dimensions of Web pages, namely content, hypertext,
and presentation. As with consistency between relational databases and the
Web, such an approach should provide at least for strong consistency with
incremental update granularity.

### 1.2.3   Consistency in Document Flows on the Web

This subsection uses an extended notion of data on the Web, as described
at the end of Subsection 1.1.1. Thus, when talking of an approach herein
that shall maintain consistency of data on the Web, it is referred to data in
the refined sense.

Web pages often change their location, i.e., their URL over time, a prop-
erty of Web pages not considered in this thesis yet. An approach providing
for consistency of data on the Web, however, should consider this property,
which is supported by the following observations. First, every person who
is using the Web regularly will admit that she is often confronted with "404
Not Found" or similar due to links to pages whose locations have been mod-
ified. Second, there are prominent calls to refrain from modifying URLs
of Web pages or at least to handle them properly [25, 160]. Third, URI
schemes have been proposed to hide changing locations of pages by provid-
ing for location-independent identifiers that can be translated to location-
dependent identifiers (for an overview see [153, 169]). Examples are URNs
[79] and DOIs [85].

Looking at XML documents in general instead of Web pages, it can be
observed that in certain situations XML documents also change their loca-
tion over time. Thereby they form *physical* document flows between nodes
in the network, being externally observable. This is different from many
systems that seem to support such flows, which, however, in fact only pro-
vide for *virtual* document flows. They allocate documents permanently at
a central repository and provide access from varying nodes, e.g., a work-flow engine usually stores workflow data in a central repository, causing the workflow data to flow only virtually between nodes, i.e., clerks' desks. Contrarily, on the Web there is a physical document flow where documents are moved physically from node to node. Note that documents participating in virtual document flows do not change their location and that consistency management is thus already covered by dealing with the problems in the previous two subsections.

Today, physical document flows take place via the Web in business transactions and personal ad-hoc data exchange. First, rather recently, in the area of Web services, approaches such as BPEL [17] and ebXML [2] provide for the specification of business transactions between businesses. In the realm of a business transaction XML documents such as a purchase order and an invoice are exchanged. Second, since the advent of the Web, ad-hoc data exchange between individuals has become a daily routine. The data exchange is carried out by handing over documents in situations where people want to exchange documents but cannot rely on the same information infrastructure, e.g., because they are working for different businesses. Instead, they use emails with attachments, instant messaging, or filesharing applications such as FTP.

Physical document flows have the following *advantages* over virtual document flows: First, they provide for *lower coupling* because participants do not rely on a central repository which is a possible central point of failure and which is usually difficult to provide when documents flow across businesses' boundaries. Second, they provide for *higher mobility* in manipulating documents going hand in hand with emerging mobile computing devices, assuming that the functionality otherwise provided by the central repository, such as active behavior and versioning, is available locally at each involved node of the document flow. Third, they provide for *higher autonomy* by allowing to implement security aspects such as authentification and authorization locally where documents are stored. Fourth, lower coupling and higher autonomy result in *higher flexibility* in whether and how a document flow takes place, thus providing among others for spontaneous ad-hoc document flows.

A peer-to-peer (P2P) infrastructure is a pre-requisite for supporting physical document flows. Thus an approach for physical document flows has the *disadvantages* connected to P2P infrastructures, which are the following. Above all, global properties are hard to guarantee in P2P systems, such as global availability of shared data and globally constrained access to data according to security policies. It is also difficult to maintain a globally consistent state in P2P systems, a topic distributed consistency protocols are aiming at. Therefore, a P2P system is usually more complex to design and implement and has to deal with several challenges connected with P2P systems, such as maintaining network control, guaranteeing network security, and avoiding freeloaders (see [131]). Thus, if the extended approach is not the approach of choice in a given setting, one may switch to virtual document flows instead. However, if physical document flows are to be used, a P2P infrastructure cannot be circumvented.

An approach that maintains consistency of data on the Web should be capable of maintaining consistency of data participating in document flows. As such it should provide consistency *across and agnostic to document flows*, where "across" addresses consistency between documents of which at least one participates in a document flow, and "agnostic" means that it should be possible to access a document transparently of whether it participates in a document flow or not, referred to as *document flow transparency*. Examples are consistency between Web pages irrespective of changes to the page's URLs and consistency between XML documents irrespective of their participation in document flows.

Summarizing, the **problem** in maintaining consistency in document flows on the Web lies in designing a P2P infrastructure that preserves the properties of the original approach for providing autonomy, low coupling, and strong consistency with incremental update granularity in spite of the extension. Additionally, the extended approach must provide access to documents transparently of whether they participate in document flows or not.

## 1.3   Contribution

The thesis provides approaches that solve the problems described in Section 1.2. Its contributions are providing for the following:

- Maintaining consistency between relational databases and the Web by realizing an already proposed approach called "Self-maintaining Web Pages" (SMWP), first described in [134]. The conceptual model presented therein and detailed in [133] is mapped to a (relational) realization model and implemented using off-the-shelf relational database technology. Moreover, a performance evaluation compares SMWP to related approaches. For details see Subsection 1.3.1.

- Maintaining consistency within data on the Web by proposing an approach called "Active XML Schema" (AXS). It provides for enriching XML schemas[7] with specifications to maintain consistency upon occurring events reflecting modifications in XML instances. By dealing with XML documents in general, the approach can be used for any design dimension of Web pages. For details see Subsection 1.3.2.

- Maintaining consistency in document flows by proposing a two-layered approach, extending AXS to deal with XML documents that are physically moved across a network. The first layer provides for "Traceable

---

[7]Throughout the thesis, "XML Schema" refers to the schema language proposed by the W3C in [157, 158], and "XML schema" refers to a schema expressed using XML Schema.

Document Flows" (TDF), which allows to trace the flow of a document across a network, while the second layer employs AXS on top of it. For details see Subsection 1.3.3.

All approaches use event-condition-action (ECA) rules as a technique to maintain consistency. An ECA rule is triggered upon an event occurrence that matches its event description. If its condition applies, its action is executed. ECA rules are more commonly referred to as "triggers" and are well known in relational databases where they have proven useful (see Subsection 1.4.1 for an application).

### 1.3.1 Maintaining Consistency between Relational Databases and the Web

The SMWP approach solves the problem of maintaining consistency between relational databases and the Web as described in Subsection 1.2.1.

For class-based modelling of relational databases, the SMWP approach exploits and extends well established concepts from distributed database systems (DDBS). The analogy between a database-backed Web site and a distributed database, which has one central site and several remote sites, is the following: While the database containing content to be published on Web pages is seen as the database at the central site, Web pages are seen as databases at remote sites at which content from the central site is replicated. In particular, the concepts of fragmentation and allocation [43, 122] are used to build subsets of relations and allocate them on Web pages.

Web pages in SMWP are defined using class-based models by exploiting the observation that most often they display content according to some parameters. For example pages enlisting products in categories with one page per category, or publications in departments with one page per department. A page class is used to describe the structure of its instances at the schema level and is parameterized, while a page is an instance of the class with bound parameter values. For example, a publications page for the department "Business Informatics Group" is an instance of its page class that has "department" as its parameter.

The SMWP approach has the following advantages, compared to closely related approaches (for more information on closely related approaches see Subsection 1.4.3):

- *Class-based database design:* in contrast to other approaches, which do not explicitly model the underlying database and see every tuple as a single instance (thus being instance based), SMWP is class based in that it introduces an explicit data fragmentation design defining classes. A class comprises defined chunks of content (i.e., sets of tuples) each forming an instance and representing parts of Web pages. Class-based modelling has the advantages of providing re-usable database components, simplifying queries to retrieve a page's content, and simplifying the mapping knowledge, thus easing definition and maintenance of pages.

- *Class-based Web page design:* by modelling pages at the class level, definition and maintenance of pages is simplified. Related approaches partly model pages at the instance level.

- *Class-based mapping knowledge:* the amount of the database to Web mapping knowledge is significantly reduced by using a class-based mapping. This is possible by class-based modelling of both the database and the Web site. Thus design and maintenance costs are minimized. Consequently, by defining the database to Web mapping between classes (not instances), the SMWP approach features the lowest coupling between the database and Web pages.

- *Incremental update granularity:* while other approaches update a page by completely re-generating it from scratch, causing unnecessary server-load, the SMWP approach incrementally updates a page by only modifying XML elements that are to be updated due to a database modification.

- *Strong consistency:* by pushing database updates immediately to affected Web pages, they contain up-to-date content any time. Moreover, this takes off server-load originating in periodically polling the database for updates. Updates to multiple pages caused by a single database update become visible to the user at once.

The thesis shows how the SMWP approach can be realized using off-the-shelf relational database technology. After a quick tour of the approach (see Section 2.1), it presents a declarative language for data fragmentation and Web page design and the maintenance of both (see Section 2.2). It shows how statements, which are issued using this language, are translated into SQL statements on the employed database (see Section 3.1 and 3.2). It presents predicate based parameters, which are a conceptual extension to the SMWP approach increasing its modelling power, and their realization (see Section 3.3). Finally, a performance evaluation is presented, comparing the SMWP approach to related approaches (see Section 3.4).

The work presented in the thesis accompanies previous theory papers on SMWP (see [133, 134]). The conceptual model presented therein, which comprises among others two kinds of fragment classes, fragments, page classes, pages, and algorithms for incremental maintenance thereof, is mapped to a realization model that employs an off-the-shelf relational database system, thus comprising relations (see Section 3.1) and triggers (see Section 3.2). This work marginally overlaps the theory papers in that it deals with algorithms that have the same intent, i.e., with algorithms that provide for incremental maintenance of data. The trigger algorithms presented herein, however, substantially differ from algorithms presented in the theory papers because they operate within a different, i.e., the realization model. Moreover, additional algorithms necessary for realization are presented in this thesis (see Subsections 3.2.1 and 3.2.2). Furthermore, the schema evolution problem is addressed. It is explained how the set of maintenance triggers has to be modified with each schema modification statement. We show this explicitly for the creation of new fragments. Dropping a fragment requires to undo the changes made when creating the fragment.

Several approaches have been proposed in literature for publishing *relational data as XML*. For more details on these *distantly related* approaches see Subsection 1.4.2. They can be considered distantly related because none of them provides for self-maintainable, incremental maintenance of materialized XML views. Even more important, none of these approaches deals with modelling the database or modelling Web pages. Using [145], the closest of the distantly related approaches, to solve the problem of maintaining consistency between relational databases and the Web instead of developing SMWP from scratch was not an option for one simple reason: SMWP was proposed two years before [145]. Still, after an extension of [145] to support class-based modelling of databases, Web pages, and the mapping, it would not provide for a self-maintainable view maintenance, i.e., one that uses the update as only information to modify the XML view. A property from which one can expect increased efficiency compared to a mechanism that additionally uses base tables and auxiliary data structures for incremental maintenance.

SMWP as an application of relational database triggers is a so called *generated extender*, according to [39]. For details regarding triggers in relational view maintenance see Subsection 1.4.1. Extenders provide enhanced database functionality, e.g., for maintaining replicated data. In case of SMWP, triggers are generated, i.e., derived from declarative specifications given by means of SMWP's schema definition language. While [44] derives triggers for incrementally maintaining views that are defined by SQL statements, SMWP can be seen as its equivalent for replicated fragments, i.e., deriving triggers for maintaining (horizontal and derived) fragments.

### 1.3.2 Maintaining Consistency within Data on the Web

The Active XML Schema (AXS) approach solves the problem of maintaining consistency within data on the Web as described in Subsection 1.2.2.

First, AXS is *general* in that it can be used to maintain consistency in any of a Web page's design dimensions, i.e., in the content, hypertext, and presentation dimension. Since data on the Web is XML data in all design dimensions, this is simply achieved by designing AXS to maintain consistency of XML documents in general.

Second, AXS provides for *strong consistency* with *incremental update granularity* by providing for *ECA rules* or *triggers*. By defining a trigger one can manually specify how to update a physical data item (by the action) upon the modification of another physical data item (represented by an event). That is why triggers are said to provide re-active behavior (or short active behavior). We have chosen to use triggers instead of other view definition mechanisms such as queries because usually more complex views can be incrementally maintained using triggers than when defined by queries. Moreover, triggers for incremental maintenance can be derived from arbitrary declarative definitions of views, not just queries, e.g., from view correspondence assertions proposed in [144], being a good starting point for future extensions. By bringing triggers to XML also other application areas of triggers are brought to XML such as implementing business rules. Old techniques for providing active behavior in XML are replaced, like immediate checking a document after an event occurred, or periodically polling a document for changes. Other approaches for providing triggers in XML data have been proposed (see Subsection 1.4.5), all being referred to as approaches for "active XML".

AXS per se does not provide for class-based modelling of data as SMWP does. Instead, AXS only provides for instance-based modelling of XML documents because we were not able to identify general replication and derivation patterns in XML documents that are so prominent as the ones in SMWP. Nevertheless, providing for instance-based models is a good starting point for later extensions by domain-dependent class-based models.

Active behavior is provided by extending XML schemas such that active behavior comes with an XML schema and is used for all document instances of it. This is line with the current trend in Web-based information systems to learn from years of research in conceptual modelling and to apply schemas for modelling and implementing Web content. Such an approach provides for the reuse and interoperability of not only structure, but also of active behavior. Moreover, active behavior defined at the schema level is easier to design, to implement, and to maintain than it would be if it was defined separately. Related approaches for active XML do not store active behavior with a schema.

Third, by employing an *event based model*, AXS supports all kinds of change dependencies. Regarding the events themselves, the approach builds on experience from conceptual modelling of business rules by the means of situation/activation diagrams [101]. These diagrams are characterized by modelling events as first class objects that have an event type and are collected into event classes. Beyond supporting different kinds of events (e.g., mutation events, calendar events, and abstract events), they allow to schedule future events. Thereby *change dependencies on data in Web pages* and *change dependencies on time* are supported by AXS, being extensible by specializing abstract events.

AXS provides a rich set of event kinds to be able to define rules on a most extensive set of events. AXS provides for (a) *calendar events* which occur at points in time; (b) *method events* which occur upon performing operations defined with AXS schemas (see Subsection 4.1.1); (c) *primitive mutation events* which occur upon a data modification, e.g., upon insertion of an element; (d) *composite mutation events* which are composed from primitive mutation events by a special event composition language, as it is the case with active database systems [123]; (e) *logical events* which are defined declaratively by querying the extension of other event classes; (f) *abstract events* which can be specialized to represent arbitrary events not addressed by (a)-(e). Opposite to [30], which derives rules to be triggered from the code that performs a data modification (i.e., an XQuery query), AXS determines the rules directly from occurred mutation events representing data modifications due to the code's execution, being a more natural approach. The latter applies for all kinds of events, not just mutation events. The set of event kinds provided by AXS is by far more extensive than in any related approach for active XML, which only provide for primitive mutation events.

By taking a *document centered* approach, *change dependencies on a Web page's history* are supported. This means that events occurring at a document are stored in the document, forming its event history. This mimics traditional paper form processing, where all events related to a business case are recorded with its form and can be queried when decisions on the case have to be made.

Fourth, *Autonomy* of participating businesses is maintained by utilizing the publish/subscribe protocol from event based systems for *event communication*. Thereby events that occur at publishing documents are delivered to subscribing documents, allowing them to react accordingly. Moreover, by communicating events asynchronously, the loosely coupled nature of distributed Web pages is taken into account. Pessimistic transaction protocols from distributed databases (for details see Subsection 1.4.4), which guarantee consistency, are not suited for an adaption to AXS since they restrict autonomy. On the contrary, because optimistic protocols do not limit autonomy, adapting them for consistency management on the Web seems a viable option. After optimistic transaction executions, however, a roll-back by means of compensating transactions may still occur due to the behavior of other participants. Since this is considered a harmful interference, optimistic protocols are not adapted either. Nevertheless, extending the proposed AXS approach with support for optimistic protocols for use in restricted environments may be desirable.

Fifth, systems storing dependent documents have a *low coupling* by using an event-based infrastructure and constraining relationships between documents. The latter refers to defining all components of a rule (i.e., event, condition, action) *local* to the document they are specified for. Local means that a rule may only react to a local event, a condition may only test the local document including its event history, and an action may only modify the local document. Aside of providing for low coupling this maximizes design autonomy. Sharing events and actions is only possible via asynchronously communicating events between documents.

The two related approaches that provide materialized XML views over XML data (see Subsection 1.4.5) are not well suited for maintaining consistency of data on the Web. They are intended rather for database-internal use. [55] uses auxiliary data structures to maintain relationships between tuples in the algebra tree and the base XML data, assuming internal identifiers on base XML data. This would result in an undesired high coupling between Web pages. [144] does not provide for self-maintainable views, resulting in a higher coupling due to necessary access to base data in case of an update.

Seventh, the event based infrastructure also provides for support of *change dependencies on data in legacy systems* using a *uniform interface to legacy systems*. A *wrapper* for a legacy system, e.g., a relational database, makes modifications or other events within the system available as AXS events. Depending on the functionality provided by the wrapped legacy system, the wrapper may need to periodically poll the legacy system for changes. Rules in AXS which react to events need not be aware of the source of the triggering event (i.e., the wrapped system), thus providing *event source transparency*.

Compared to related approaches for active XML (again, for details see Subsection 1.4.5), AXS has several unique features. First, active behavior is specified with XML schemas, while related approaches leave this question open. Second, the event-based infrastructure and asynchronous event delivery provides for reacting to events that have occurred remotely. Third, AXS's event set is the richest one, also due to provision of user-definable composite mutation events. On the negative side, AXS has not yet dealt with rule analysis as Bailey et al. did, e.g., in [12].

### 1.3.3 Maintaining Consistency in Document Flows on the Web

The approach for traceable document flows that is enriched with active behavior (ATDF) solves the problem of maintaining consistency in document flows on the Web as described in Subsection 1.2.3. Since document flows use an extended notion of data on the Web (see the end of Subsection 1.1.1), it is only natural that the ATDF approach extends the Active XML Schema (AXS) approach which maintains consistency of data in the not extended sense. The extension is two-layered because the first layer is useful on its own, i.e., without the active extension of the second layer.

The *first layer* provides for *traceable document flows* (TDFs). Trace-ability is important for various aspects. In commercial settings traceability provides for legally relevant properties, such as non-repudiation. From an application's point of view it may be useful to have traces of a document flow available, not only to influence an application's behavior, e.g., by a document's location, but, assuming that documents are persistent, to have a document's history available. Finally, from a user's point of view, she simply may want to discover where a document has been moved to after accessing it last time. The layer was designed to be applicable for any document not just XML documents, because traceable documents are useful on their own, as is exemplified in Chapter 7.

TDFs proposes an infrastructure model, defining how document flows can take place, in parallel with an ontological framework. The *infrastructure model* for document flows is a P2P model so that a shared, central information infrastructure is not necessary for document exchange. Basically, every document has a globally unique identifier which is preserved when sending it across the network. Because a document may be edited by its current owner and re-distributed afterwards, the model keeps track of various versions of the document spread across the network. When re-distributing a document, metadata describing its flow so far and metadata describing the document itself are distributed along with it. Moreover, an *ontological framework* is presented for describing document flows and documents, revealing the flows and descriptions of the documents (referred to as annotations) by using an open data format for their representation. With it, users can query metadata describing document flows, can query annotations, and can determine the past and current content of documents. By employing Semantic Web technology, i.e., OWL [165], the metadata is described non-proprietarily and can be used by any application.

The *second layer* extends traceable document flows with active behavior by employing AXS on top of the first layer (ATDF). Since AXS communicates events from publishing to subscribing documents according to the publish/subscribe protocol, and documents as part of TDFs are stored at varying locations over time, events are published in documents at varying locations and have to be delivered to documents at varying locations. Therefor an event routing algorithm is proposed that ensures event delivery despite of varying locations, using a hybrid P2P model, i.e., a P2P model with some form of centralization. Event routings can be manually adjusted. Moreover, two optimized event routing algorithms are presented, which make event delivery more robust to offline peers and minimize network traffic.

ATDF ensures peers' *autonomy* by providing access to documents irrespective of other document flows and of the behavior of other peers. *Low coupling* is maintained by using appropriate data structures to store document flows. By providing event routing algorithms that deliver events from publishing to subscribing documents irrespective of document flows they participate in, *strong consistency* with *incremental update granularity* (already provided by AXS) is maintained *across document flows*. ATDF provides for *document flow transparency* by providing access to a document and delivering events to/from it irrespective of its participation in document flows.

## 1.4   Related Work

This section surveys related work. After giving an introduction by providing a general overview of views in relational databases in Subsection 1.4.1, related work for the problem of maintaining consistency between relational databases and data on the Web is described in Subsection 1.4.2 and 1.4.3. Subsequently, related work for the problem of maintaining consistency within data on the Web is described in Subsection 1.4.4 and 1.4.5. Additional references to related work are given along the description of the proposed approaches in Parts I, II, and III where applicable.

### 1.4.1   Views in Relational Databases

Dealing with consistency of data it is natural to look at relational databases in general to see which techniques they have developed. Databases provide for (i) view consistency aside of providing for (ii) type consistency by relational schemas and constraints (among others defined by SQL assertions) and (iii) consistency under concurrent access by transactions and their ACID properties, which refer to atomicity, consistency, integrity, and durability. The consistency for which "C" in ACID stands for refers to a database state that is consistent in terms of all three notions of consistency (where view consistency is only guaranteed as far as the database provides for it).

Views are widely used in relational databases. Their intent is usually to separate the way users and applications see the data from the way the data is represented, as dealt with by the ANSI/SPARC three-level schema architecture [142]. Thus, the users' view, also referred to as external schema, may remain unchanged in case of modifications to the logical schema (logical data independence) and physical schema (physical data independence). Among others, views can be used to gain data independence, to control data access, to integrate data, and to increase performance.

In database systems a view is defined by an SQL statement over base tables. Since a view itself has a relational schema, they can be reused by other views and queried using SQL. When a query is issued against a view, the database's query processor rewrites the query and issues the rewritten query against the base tables if the view is a *virtual view*. On the opposite, the view can be a *materialized view*, i.e., the tuples contained in the view are physically stored on disk like tuples contained in tables. Then the query can be issued directly against the materialized view. Performance is increased if the cost of maintaining the materialized view and query execution is less than the cost of rewriting queries and their execution on base tables, where the cost of queries on the materialized view is usually less than the cost of their rewritten equivalent on base tables.

Several ready-made techniques for maintaining materialized views in databases have been proposed. Most easily, a view can be re-generated from scratch upon a modification in a base table. Alternatively, only parts of the view can be updated that are affected by the base table modification. The latter is also referred to as *incremental view maintenance*. Several approaches for incremental view maintenance of relational data have been proposed. For an overview the interested reader is referred to [75]. The approaches differ in various aspects, for example in the amount of information used to determine an incremental update: the modification may be used in combination with base relations and/or with the view, and possibly with auxiliary data structures. Views that are maintained using only the modification, the view, and key constraints are called *self-maintainable*. All approaches impose restrictions on the definitions of views, e.g., restrict to SPJ queries[8] to ensure incremental maintainability.

Maintenance of complex views, however, often has to be manually implemented using *database triggers*. This applies when more expressive power is needed in their definition than SQL provides, if non-relational data structures are to be held consistent with base data, or if applications maintaining external views are to be informed of modifications of the database to maintain a consistent state. An overview of trigger applications is given in [39], classifying triggers in one dimension along the categories (i) triggers embedded into the database kernel, (ii) triggers providing extended database functionality, and (iii) triggers supporting external applications. Examples are maintaining materialized views for the first category, maintaining replicated data and external data for the second, and keeping Web applications consistent for the third. The second dimension classifies whether triggers are (a) hand crafted or (b) generated. The latter refers to the common situation where procedural triggers are derived from declarative specifications to exhibit some behavior, as for example in [44], where triggers are derived from SQL statements to maintain materialized views. A brief introduction to materialized view maintenance using triggers can be found in [56].

### 1.4.2 XML Views over Relational Databases

Integration of XML data with relational data is approached in literature from two directions. First, starting from XML data, it is *stored* in a relational database to reuse the relational database's functionality in building a transactional persistent XML database. Here, it has to be decided on how the XML data is stored and indexed and how XML queries on the loaded (virtual) XML data are translated to queries on the relational data (e.g., see [74]). Obviously, the storage model influences query translation and execution. Second, starting from relational data, it is *published* as XML data to make it available on the Web or to facilitate data integration with other sources. Here, the focus lies on designing the mapping from relational to XML data and on how to efficiently generate XML from arbitrary relational schemas. This is different from the first area where the mapping is often fixed, however, the two areas are not distinct in the issues they address. For an overview of integrating XML with relational data, e.g., see [91].

---

[8]Select-project-join queries, an algebra without operators such as $\cup$, $-$, and $\rho$.

Clearly, the area of publishing relational data as XML is related to the thesis' aim of maintaining consistency between relational data and Web pages. Several approaches have been proposed in literature in this area, the most prominent ones being SilkRoute [59] and XTABLES [65], which was formerly known as XPERANTO [37]. SilkRoute and XPERANTO use a query language to define views, while recent approaches also use different techniques, i.e., [18] uses attribute translation grammars and [145] uses view correspondence assertions. As of today, commercial database products of major vendors IBM, Microsoft, and Oracle support defining restricted XML views over the database (see [65, 91] for an overview). While none of the products from practice take materialization of XML views into account, two approaches from literature do, namely SilkRoute, which supports materialization of XML views including heuristics to optimize database access, and [145], which derives triggers from view definitions for incremental view maintenance.

The approaches for XML views, however, lack necessary properties for maintaining consistency between relational databases and the Web as described in Section 1.2.1. First, the approaches do neither model the database nor Web pages, thus allowing for an instance-based mapping only. The approaches could be adapted to support class-based modelling and mapping. And second, most important materialization of views, i.e., pre-generation of Web pages, is missing in all approaches except two. Materialization could, however, be added naively by storing the evaluated view on disk and triggering a re-evaluation upon a database update. A more sophisticated extension could determine whether a database update makes a re-evaluation of the view necessary. Still, this would only provide for a complete, not an incremental update granularity, which is only provided by [145] (SilkRoute re-generates materialized views from scratch). [145], however, does not provide for self-maintainable view maintenance.

### 1.4.3 Data-intensive Web Sites

Several related approaches for data-intensive Web sites have been proposed that support pre-generation of Web pages [49, 99, 137, 170]. An in-depth comparison of these approaches with SMWP has already been published in [133], thus the interested reader is referred thereto for further information.

Table 1.1 summarizes the differences between the approaches. Most of the properties have already been discussed when Subsection 1.2.1 introduced the problem of maintaining consistency between relational databases and the Web. The three new properties are the following. First, *materialization policy* defines which data can be materialized, being data in the database (relational) or data on the Web as XML or HTML pages. Second, *materialization flexibility* refers to whether and who chooses among the different materialization policies. Third, *maintenance flexibility* refers to whether and who chooses to maintain a Web page either by complete regeneration or incrementally. The last column of the table denotes an "ideal" approach, which supports all desired property values. This does not show which of a property's possible values are the most desirable ones, e.g., that a class-based database design is more desirable than an instance-based one (important in case only one is supported). For such details the reader is referred to [133].

### 1.4.4 Distributed Databases

Naturally view consistency is also an issue in distributed databases (DDBS). Distributed databases can be divided into two categories. First, in a *top-down* DDBS a new distributed application is created by implementing and distributing the local applications and systematically distributing data along with them. The intent is to allow each local application to access the needed data at minimal cost. The cost model usually takes network traffic as most important, regarding remote access more expensive than local one, and remote writes more expensive than remote reads. An example for a top-down DDBS is an airline reservation system. Second, a *bottom-up* DDBS is created for data integration purposes, i.e., to make already distributed data available in terms of a global mediated schema. Examples are data warehousing in general or integrating Web data from different sites and integrating a business' legacy systems like databases, ERP, and CRM systems in particular.

Consistency issues, however, do not arise in every DDBS but depend on the kind of distributed data. *Partitioned data*, as the name indicates, refers to a set of disjunct data partitions which are allocated at different nodes. In particular, fragmentation [43, 122] (with disjoint fragments) has been proposed as a technique to partition data in DDBS. Since partitioning does not result in data being replicated, view consistency is not an issue. *Replicated data* constitutes the general case of distributed data, where copies of a logical data item are distributed. Then, mutual consistency among the copies has to be maintained. *Derived data* can be seen as special case of replicated data, where not a carbon copy of a data item is distributed but aggregated or derived data based on it.

Distributed databases designed bottom-up have special interest in *view*

Table 1.1: Approaches for Data-Intensive Web Sites with Pre-Generation of Web Pages

| | | Approaches | | | | | |
|---|---|---|---|---|---|---|---|
| | | [49] | [99] | [137] | [170] | [133] | ideal |
| Database Design | instance-based | ✓ | ✓ | ✓ | ✓ | ✓ | ✓ |
| | class-based | − | − | − | − | ✓ | ✓ |
| Web Design | instance-base | ✓ | ✓ | − | − | − | ✓ |
| | class-based | − | − | ✓ | ✓ | ✓ | ✓ |
| Mapping Knowledge | implicit | − | − | − | ✓ | − | − |
| | explicit | ✓ | ✓ | ✓ | − | ✓ | ✓ |
| Materialization Policy | relational | − | ✓ | − | ✓ | ✓ | ✓ |
| | XML | − | − | − | ✓ | ✓ | ✓ |
| | HTML | ✓ | ✓ | ✓ | ✓ | − | ✓ |
| Materialization Flexibility | user-defined | ✓ | − | − | ✓ | ✓ | ✓ |
| | system-determined | − | ✓ | − | ✓ | − | ✓ |
| Maintenance Policy | pull-based | − | − | − | ✓ | ✓ | ✓ |
| | push-based | ✓ | ✓ | ✓ | − | ✓ | ✓ |
| Maintenance Flexibility | user-defined | − | − | − | − | ✓ | ✓ |
| | system-determined | − | − | − | − | − | ✓ |
| Maintenance Mechanism | complete | ✓ | ✓ | ✓ | ✓ | ✓ | ✓ |
| | incremental | − | − | − | − | ✓ | ✓ |
| Page Freshness | with delays | ✓ | − | − | − | ✓ | ✓ |
| | w/o delays | − | − | − | − | ✓ | ✓ |

Legend: ✓... supported, −... not supported *materialization.* This stems from the fact that queries against the mediated schema, most likely the ones in a data warehouse, are often complex and time consuming, making view selection and subsequent view materialization attractive. Because literature on view maintenance, however, has dealt with relational views over relational data, it is not applicable for maintaining consistency of data on the Web.

Distributed databases have proposed several *distributed transaction protocols* to maintain mutual consistency in replicated data. For the execution of operations on logical data items it is necessary to translate them into operations on replicated copies. To maintain consistency, concurrent executions of operations on replicated data must be equivalent to a serial execution on a non-replicated data item, a property known as *one-copy serializability*. It guarantees for view consistency and consistency under concurrent access in replicated data. For an overview of protocols in non-partitioned networks see [26, 27], for protocols in partitioned networks [53, 54]. Partitioning occur due to communication failures which fragment the network into isolated subnetworks.

Each protocol has different properties regarding the *trade-off* between *correctness* and *availability*. As noted in [54]: "Since it is clearly impossible to satisfy both goals simultaneously, one or both [correctness or availability] must be relaxed to some extent, depending on the application's requirements." Davidson et al.'s notion of correctness subsumes mutual consistency of replicated data, and their notion of availability refers to the extent of being able to locally execute a transaction despite of a partitioned network. In the context of this thesis these notions are similar to (view) consistency and autonomy, thus there is a *trade-off* between *consistency* and *autonomy*.

Autonomy is an absolute requirement for maintaining consistency of data on the Web (see Subsection 1.2.2). As the trade-off indicates, this comes at the cost of consistency. In DDBS, *optimistic protocols* have been proposed that do not limit autonomy, allowing access to local data irrespective of other locations being off-line and network partitioning. However, compensating transactions may be necessary to restore global consistency. Contrarily, pessimistic protocols prevent inconsistencies by limiting autonomy, e.g., an update of a logical data item may only be performed if all or a majority of the locations storing a copy of it are available, excluding them from being adapted to maintain consistency of data on the Web.

### 1.4.5   XML Views over XML Data

Approaches for *incremental view maintenance* of XML views over XML data have been proposed lately. First, [55] uses their proprietary query algebra XAT to internally represent XQuery queries. After evaluating a query, every operator in the algebra tree stores its results from applying the operator on its input data. After an update to base data, the modification is propagated through the algebra tree, yielding in an incremental update of the view. The maintenance algorithms take as input the update and auxiliary data structures, which store among others dependencies between tuples in the algebra tree. Second, [144] uses view correspondence assertions, which express equivalence between portions of the base data's schema and the view's schema, for view definition. Therefrom triggers are derived to incrementally update the view upon a modification to base data. The view definition technique has limited expressiveness compared to XQuery queries, e.g., it does not provide for aggregation and natural joins and allows only limited restructuring. More distantly related, [8] proposes incremental maintenance of views over semi-structured data in the context of the Lore database and OEM data model. Views are defined using extended Lorel queries and the maintenance algorithm takes as input the update, auxiliary data, and the post-update database state for maintenance.

Alternatively, as in relational databases (see Subsection 1.4.1), triggers can be used for view maintenance. A range of approaches have been proposed to provide for triggers on XML documents. Bonifati at al. demonstrate the use of the active paradigm for the implementation of alerters, personalization of web pages, and for view maintenance in [31]. In [30] they propose XQuery triggers resembling semantics of SQL triggers. Algorithms are presented that expand XQuery queries that modify whole subtrees to determine which triggers will be triggered, even before actually executing the query. Bailey et al. [12, 13] propose triggers for XML as well, but focusses on trigger analysis and optimization. This work is related to trigger analysis in relational databases using triggering graphs and activation graphs [15]. Commonly these approaches use the notion of "active XML" to refer to XML extended by triggers. It should not be mixed up with Abiteboul's notion of active XML [6], which refers to XML documents extended by Web service invocations.

Finally, in the context of software engineering, [176] proposes an approach to define consistency rules between XML documents. A consistency rule identifies portions of a source document and portions of a destination document and defines a set of conditions on them that must apply. The approach does not provide for maintaining consistency (only for detecting), however, the declarative specification by consistency rules may be a good starting point for alternative view definitions in the context of data on the Web and for derivation of procedural triggers on the XML documents to maintain consistency.

# Part I

# Consistency between Relational Databases and the Web

# Chapter 2

# Self-Maintaining Web Pages (SMWP)

## Contents

The chapter gives a quick-tour of the Self-Maintaining Web Pages (SMWP) approach and its underlying concepts in Section 2.1. In particular, it briefly presents the concepts of fragmentation schemas for data fragmentation design and of page schemas for web page design. The interested reader is referred to [133, 134] for more detailed descriptions. Subsequently, a novel language is presented in Section 2.2 that allows to easily define and maintain fragmentation and page schemas.

## 2.1 A Quick Tour of the SMWP Approach

We illustrate newly introduced concepts throughout the section by applying them to an exemplary online wine shop, whose conceptual model is depicted on the left hand side of Figure 2.1 by an UML [130] model. The wine shop manages contents concerning wines on sale (class Wines), the wineries they are produced by (class Wineries), and regions the wineries are located in (class Regions). A transformation of the UML model to a relational model is depicted in the middle of Figure 2.1. It comprises relations Wines, Wineries, and Regions (primary keys are underlined, foreign keys in italics), which are referred to as *application relations* throughout the rest of the thesis' Part I.

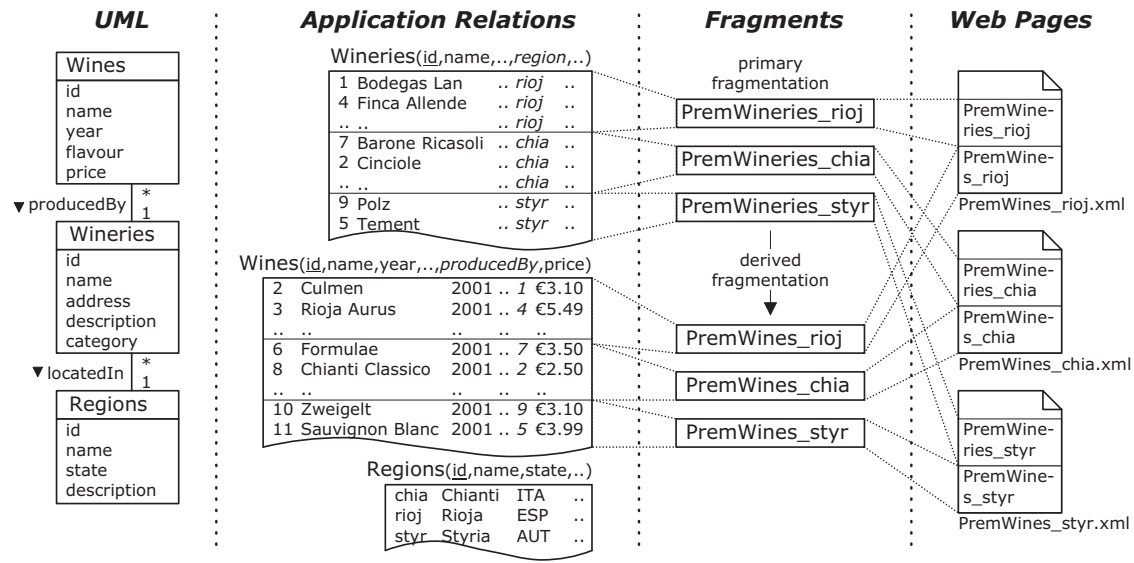

Figure 2.1: Simple Fragmentation

### 2.1.1   Data Fragmentation Design

As mentioned before, the SMWP approach utilizes concepts from DDBS. In particular, the concepts of fragmentation and allocation are exploited and extended.

#### Simple Fragmentation

To decompose relations into fragments, three different kinds of fragmentation, namely *horizontal fragmentation*, comprising *primary* and *derived* fragmentation, and *vertical fragmentation* as well as combinations thereof, called *mixed or hybrid fragmentation*, have been proposed [43, 122]. In the following, we will focus on horizontal fragmentation, since the other kinds of fragmentation can be integrated straightforwardly in the same way.

*Primary fragmentation* builds subsets of the tuples (rows) of a relation or fragment, called the *fragmentation base*, by applying a *selection predicate* which must evaluate to true for the tuple to be part of the primary fragment. *Derived fragmentation* takes into account that in some cases the fragmentation of a relation cannot be based on a property of its own attributes, but is derived from another relation, called the *derivation base*.

⊙ *Example 1.*   The wine shop wants to provide web pages, each enlisting wineries that are categorized as premium and their wines of a given region (cf. Figure 2.1). Therefore primary fragments of Wineries are defined for each region, e.g., primary fragment PremWineries_rioj builds a subset of application relation Wineries by applying selection predicate category="premium" and region="rioj", collecting all premium wineries that are located in Rioja. Fragmentation of Wines by regions cannot be based on a property of its own attributes, thus derived fragmentation is used to build subsets of Wines according to the primary fragments of Wineries, e.g., derived fragment PremWines_rioj builds a subset of Wines, collecting all wines that are produced in Rioja by a premium winery. Finally fragments are mapped to web pages, using a canonical mapping to serialize tuples as XML elements.

#### Fragmentation Schema

The fragmentation approach as introduced for distributed databases [43, 122] is rather inflexible and not fully adequate for its use in the web setting. If relations are split into logical fragments for presentation on web pages, the number of fragments will typically be large with each fragment being relatively small. Problems arise if the addition of a tuple gives rise to new fragments or if a relation should be partitioned alternatively according to some new criterion. For example, if wineries and wines of a new region are added to the product line of the wine shop, a new primary fragment of

Wineries, a new derived fragment of Wines, and a new web page as well as
the mapping knowledge for mapping these fragments to the web page must
be defined.

This inflexibility is addressed by introducing *parameterized fragment
classes*. Like similar objects are collected into classes in object-oriented
design, similar fragments are collected into fragment classes. They define
common characteristics of their fragments, such as selection predicate and
schema. To define the contents held by fragments at the class level, frag-
ment classes are parameterized. A fragment class comprises a fragment for
each parameter value from the parameter domain. A fragment is created by
instantiating a fragment class with a parameter value; it contains those tu-
ples of some fragment of the fragmentation base class for which the selection
predicate applies and that "match" the parameter's value.

⊙ *Example 2.*    Using fragment classes instead of fragments, the wine
shop can define data fragmentation more easily. Figure 2.2 depicts the wine
shop's fragment classes PremWineries<region> and PremWines<region>
whose instances are created and maintained automatically, resembling the
fragments depicted in Figure 2.1. Fragment class Regions<id> is newly
introduced, each of its fragments stores contents about a single region. Fig-
ure 2.2 uses UML and stereotypes to model fragment classes. Stereotype FC
marks fragment classes, **frag-base** marks associations to fragmentation base
classes, and **deriv-base** marks associations to derivation base classes.

In order to treat application relations and fragment classes uniformly,
we assume that an implicit fragment class with no parameters is defined for
each application relation, which contains this relation as a single fragment.
Because these fragment classes constitute the "roots" of a fragmentation
schema, each possibly serving as a fragmentation base class, they are called
*root fragment classes*.

⊙ *Example 3.*        As depicted in Figure 2.2, root fragment
classes Regions<>, Wineries<>, and Wines<> represent application re-
lations Regions, Wineries, and Wines. They serve as fragmentation
base classes for fragment classes Regions<id>, PremWineries<region>, and
PremWines<region>.

A modification of a relation has to be propagated to all fragment classes
that contain the modified tuple as a replica. The SMWP approach provides
algorithms for *propagating* these *modifications incrementally*. Thereby only
the replicated tuples are modified, there is no need to regenerate fragments
from scratch to reflect the modifications.

A brief introduction to primary and derived fragment classes is given in
Section 2.2 along with the presentation of a schema definition language.

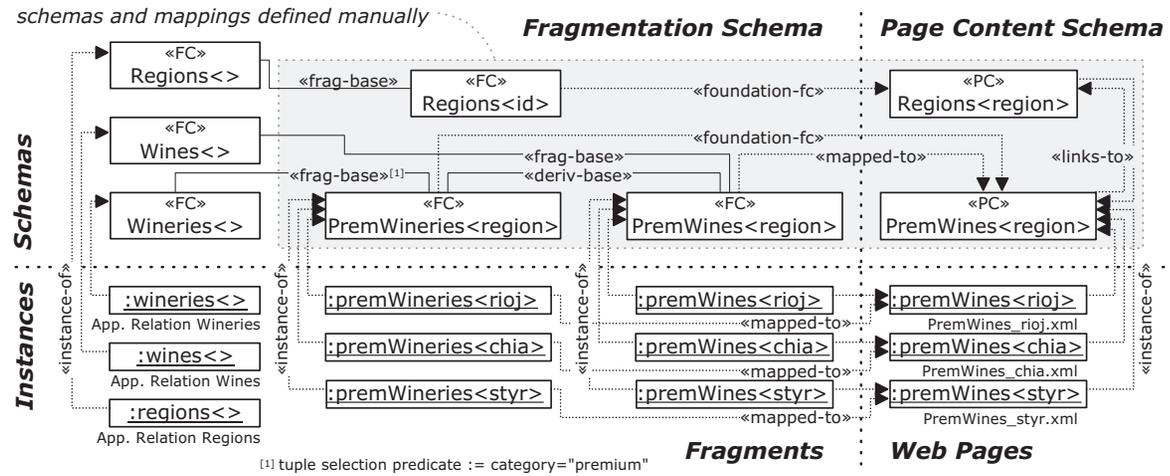

Figure 2.2: Parameterized Fragmentation

### 2.1.2   Web Page Design

Like similar fragments are collected into fragment classes, pages of the same kind are collected into page classes. A *page class* is specified by a *page schema*, comprising a *page content schema* and one or several *page presentation schemata*. Page classes comprising more than one page are defined with one or several parameters, where the parameter's values uniquely identify one page of the page class.

Bringing database contents to the web, fragments are allocated to web pages. Instead of defining this allocation for individual fragments and individual pages, fragment classes are mapped to one or several page classes. A page content schema defines among others, which fragment classes are mapped to this page class. A page class then comprises one page for each fragment from a distinguished mapped fragment class, called the *foundation fragment class*. Pages are pre-generated upon the definition of their page content schema. The contents of a page is determined by reading fragments of mapped fragment classes (including the foundation fragment class) and by transforming them into XML. We use a simple canonical mapping to serialize fragments and tuples therein as XML elements (cf. [134]).

⊙ *Example 4.*   The wine shop's page class PremWines<region> is depicted in the right upper half of Figure 2.2. As with fragment classes, UML stereotypes are used to model page classes in Figure 2.2. Stereotype PC marks page classes, foundation-fc marks associations to foundation fragment classes, mapped-to marks associations to mapped fragment classes, and link-to marks links between page classes. Page class PremWines<region> comprises a page for each fragment from its foundation fragment class PremWineries<region>, containing a fragment from the latter and an according fragment of fragment class PremWines<region>.

A modification of a fragment has to be propagated to all pages it is mapped to. Instead of re-generating affected pages completely, as previous approaches [49, 99, 137, 170] do, the SMWP approach incrementally modifies pages. Thereby only those XML elements of a page are modified that represent the modified tuple. As performance evaluations have shown [133], propagating modifications on fragments incrementally to pages generally outperforms re-generating pages from scratch.

For each page class one or several page presentation schemas define the presentation of its pages by the means of XSLT [150] or CSS [146]. The separation of content from presentation is common practice and enables one to define several presentations for the same content to take different user capabilities and preferences into account. Due to the focus of this chapter, we do not discuss page presentation schemas further but assume a default XSLT stylesheet that is used to format pages.

## 2.2   Schema Definition Language

Definition and maintenance of fragmentation and page content schemas is supported by a declarative schema definition language, providing statements for creating, modifying, and deleting fragment classes and page classes. To realize fragmentation with off-the-shelf database technology, issued statements are translated into SQL-DDL and PL/SQL statements. How this is done in detail is described in later sections. In the following we give a brief overview of the language and shortly review the concepts of fragment classes. A full definition of the language in EBNF is presented in the Appendix.

### 2.2.1   Data Fragmentation Design

A *primary fragment class* is defined upon a *fragmentation base class*. It comprises fragments and tuples therein that are taken over from the fragmentation base class. A primary fragment class "inherits" the parameters of its fragmentation base class and possibly adds new parameters. If a new parameter is added, fragments from the fragmentation base class are further partitioned according to the added parameter, where the number of sub-partitions is defined by the domain of the added parameter.

A selection predicate can be used in the definition of a primary fragment class to narrow the contents that is taken over from the fragmentation base class. First, a *fragment selection predicate* can be used to narrow the set of fragments that are taken over. It is specified by an SQL expression that may only refer to parameters of the fragment class. Second, a *tuple selection predicate* can be used to narrow the set of tuples contained in fragments that are taken over. Again, it is specified by an SQL expression, but this time it may only refer to tuple's attributes that are not used as parameters.

The statement to define a primary fragment class has the following syntax[1]: CREATE PRIMARY FRAGMENT CLASS $G{<}KL{>}$
        FRAGMENTATION BASE CLASS $F{<}L{>}$
        [ FRAGMENT SELECTION PREDICATE *SQLExpr* ]
        [ TUPLE SELECTION PREDICATE *SQLExpr* ].

⊙ *Example 5.*   To partition relation Wineries into fragments, where for each region a fragment exists that comprises all premium wineries of the respective region, primary fragment class PremWineries<region> is defined upon fragmentation base Wineries<> using tuple selection predicate category="premium", adding parameter region (as depicted in Figure 2.2 and shown below).

CREATE PRIMARY FRAGMENT CLASS PremWineries<region>

---

[1] In the following $F{<}{>}$, $F{<}L{>}$, $G{<}KL{>}$, $H{<}K{>}$, and $P{<}K{>}$ are variables. $P{<}K{>}$ conforms to non-terminal symbol *PCSignature*, while the others conform to non-terminal symbol *FCSignature*.

FRAGMENTATION BASE CLASS Wineries<>
TUPLE SELECTION PREDICATE {category="premium"};

⊙ *Example 6.*    To illustrate the use of fragment selection predicates, imagine that PremWineries<region> should only contain wineries from regions starting with the letter "A". To accomplish this, one would append the fragment selection predicate clause "FRAGMENT SELECTION PREDICATE {region LIKE 'A%'}" to the statement shown in Example 5.

A *parameter* must be defined before it can be used by a primary fragment class. Parameters become re-usable by defining them on root fragment classes. Once defined on a root fragment class, they can be used in the definition of any primary fragment class that is defined directly or indirectly via its fragmentation base classes upon the root fragment class. Furthermore, to achieve independence of a parameter's domain from the contents of the root fragment class upon which it is defined, the values of the primary key attribute of a *reference relation* keep the values of the parameter's domain. The statement to define a parameter has the following syntax:

CREATE VALUE BASED PARAMETER *Ident* ON *F*<>
   ((USE REFERENCE RELATION *Ident*(*Ident*)) |
   (CREATE REFERENCE RELATION)).

⊙ *Example 7.*    Parameter region must be defined upon root fragment class Wineries<>, before it can be used in the definition of primary fragment class PremWineries<region> (cf. Example 5). Because attribute region of Wineries<> references the primary key attribute id of application relation Regions, the latter is used as the reference relation to define the parameter's domain (as shown below).

CREATE VALUE BASED PARAMETER region ON Wineries<>
   USE REFERENCE RELATION Regions(id);

When defining a parameter two special cases may occur. First, the application relation of the root fragment class the parameter is defined upon may be used as the parameter's reference relation itself. This is possible if the parameter's domain is defined by the values of the primary key attribute of this application relation.

⊙ *Example 8.*    Parameter id must be defined before it can be used in the definition of fragment class Regions<id>. Because the values of primary key attribute id of application relation Regions define the parameter's domain, this application relation is used as the reference relation (as shown below).

CREATE VALUE BASED PARAMETER id ON Regions<>
    USE REFERENCE RELATION Regions(id);

The second special case occurs when for a given parameter no application relation can be used as a reference relation. This happens if all application relations have inappropriate primary key attributes. In this case a reference relation has to be created by using the **CREATE REFERENCE RELATION** clause with the statement to create the parameter (as shown below).

⊙ *Example 9.* To illustrate this situation, imagine that parameter **state** is defined upon root fragment class **Regions**<> to be used in the definition of primary fragment class **Regions**<state>. Because no application relation has an appropriate primary key attribute for **state**, a new reference relation has to be created as shown below.

CREATE VALUE BASED PARAMETER state ON Regions<>
    CREATE REFERENCE RELATION;

A *derived fragment class* is defined upon a *fragmentation base class* and a *derivation base class*. It comprises fragments and tuples therein that are taken over from the fragmentation base class, where each fragment from the fragmentation base is further partitioned according to the derivation base. The number of sub-partitions is determined by the domain of the derivation base's parameters. The fragment into which a tuple from a fragment of the fragmentation base class belongs is determined by joining it with tuples from fragments of the derivation base class. The derived fragment class "inherits" the parameters of both its fragmentation and derivation base class. The statement to define a derived fragment class has the following syntax, where "**AS Ident**" is used to define an alias (like in SQL) that can be used in *SQLExpr*:

CREATE DERIVED FRAGMENT CLASS $G<KL>$
    FRAGMENTATION BASE CLASS $F<L>$ [ AS *Ident* ]
    DERIVATION BASE CLASS $H<K>$ [ AS *Ident* ]
    JOIN BY *SQLExpr*.

⊙ *Example 10.* Derived fragment class **PremWines**<region> is defined upon fragmentation base class **Wines**<> and derivation base class **PremWineries**<region> (as depicted in Figure 2.2 and shown below). Thereby single fragment **wines**<> is partitioned according to the fragments of **PremWineries**<region>. The derived fragment class then comprises several fragments, each containing wines that were produced in the respective region by a premium winery (e.g., **premWines**<rioj> contains wines that were produced in **Rioja**).

```
CREATE DERIVED FRAGMENT CLASS PremWines<region>
   FRAGMENTATION BASE CLASS Wines<> AS wines
   DERIVATION BASE CLASS PremWineries<region> AS premWineries
   JOIN BY {wines.producedBy=premWineries.id};
```

## 2.2.2   Web Page Design

The page content schema defines page classes and the mapping of fragment classes to page classes. A *page class* is defined upon a foundation fragment class, which must possess the same parameters as the page class. For each fragment of the foundation fragment class, a page containing the fragment is created. Furthermore other fragment classes can be mapped to a page class, whereas the mapped fragment classes' parameters must be a subset or equal the page class' parameters. The statement to define a page class has the following, shortened syntax, where non-terminal symbol *PCFCMapping* details the mapping of a fragment class (see the Appendix and later explanations in this section):

```
CREATE PAGE CLASS P<K>
   FOUNDATION FRAGMENT CLASS PCFCMapping
   { FRAGMENT CLASS PCFCMapping }.
```

⊙ *Example 11.* Page class Regions<region> is defined upon foundation fragment class Regions<id> (as shown below). A parameter map (PARAMETER MAP clause as defined by *PCFCMapping*) defines the parameter id of the foundation fragment class to equal parameter region of the page class. The page class then comprises a page for each fragment of fragment class Regions<id>.

```
CREATE PAGE CLASS Regions<region>
   FOUNDATION FRAGMENT CLASS Regions<id>
      PARAMETER MAP id AS region;
```

When defining a page content schema it is possible to specify links between fragment classes (with symbol *PCFCMapping*), which are either mapped to the same page class (internal link) or different page classes (external link). Such meta data about links is used during the formatting process for proper nesting of fragments contained in the same page and determining links between pages.

⊙ *Example 12.* Page class PremWines<region> is defined upon foundation fragment class PremWineries<region> and fragment class PremWines<region> (as shown below). Each tuple of PremWineries<region> links internally to corresponding tuples of PremWines<region> (as specified by the INTERNAL LINK clause), and externally to a fragment of fragment class Regions<id> contained in a page of page class Regions<region> (as specified by the EXTERNAL LINK clause). Link targets, which are specified at the class level, are determined at the instance level as defined by the JOIN BY clause.

```
CREATE PAGE CLASS PremWines<region>
    FOUNDATION FRAGMENT CLASS PremWineries<region>
        AS premWinery
      INTERNAL LINK TO FRAGMENT CLASS PremWines<region>
          AS premWines
        JOIN BY {premWines.producedBy=premWinery.id}
        EXTERNAL LINK TO PAGE CLASS Regions<region> AS pc_region
          CONTAINING FRAGMENT CLASS Regions<id> AS fc_region
        JOIN BY {premWinery.region=pc_region.region
            AND premWinery.region=fc_region.id}
    FRAGMENT CLASS PremWines<region>;
```

# Chapter 3

# SMWP: From Theory to Practice

## Contents

The chapter shows how the SMWP approach can be realized using off-the-shelf relational database technology. It describes the realization of statical aspects by the relational representation of fragments and the storage of pages in Sections 3.1. Next, it shows the realization of dynamical aspects by triggers that maintain fragments and pages in Section 3.2. Subsequently, Section 3.3 presents how the SMWP approach known from [133, 134] is extended by the concept of predicate-based parameters and how they are realized. Finally, Section 3.4 describes the implemented prototype and presents results of a performance evaluation.

Several advantages are gained by employing the database system that stores contents published on Web pages to realize the SMWP approach. In particular, by using the same database system to also store meta data (e.g., about fragmentation) and Web pages, as well as to execute application code, the following advantages are gained. First, no proprietary software other than the database system itself has to be installed and maintained (opposed to previous approaches). Second, access to database contents is more efficient from inside the database than from outside, possibly from a remote computer. Third, by using the same storage system for relations, fragments, and pages, fragments and pages are modified in the same transaction as the triggering modification of a relation. Thereby modifications to several pages that were caused by a single modification to a relation become "visible" at once, resulting in consistent pages any time (contrarily, [99, 137, 170] do not deal with this aspect of consistency). Fourth, database tuning techniques (such as clustering) can be applied to meta data as well as database contents.

## 3.1 Relational Representation of Fragments and Pages

Off-the-shelf relational database technology is used to store fragments and pages as instances of fragment classes and page classes respectively. Because the number of fragments will be typically large with each fragment being relatively small, we do not realize each fragment but all fragments of one fragment class by a single relation. To be able to determine the fragment a tuple is contained in, fragment parameters are stored as part of the tuple. Thus a fragment is read by selecting tuples that have the fragment's parameter values.

It is crucial to determine well-suited techniques for realizing fragment classes. Database *views* would be a reasonable option, because primary and derived fragment classes build views over their fragmentation and derivation base classes. Unfortunately, triggers, which would have to be used to incrementally propagate updates on fragments to web pages, cannot be defined on views in major database products (such as Oracle or Microsoft SQL Server). However, some products allow to define triggers on a variation of views, known as *materialized views*. But they present another drawback: an incremental update mechanism (i.e., only the affected tuple is updated) for materialized views that join master tables is not provided but only a complete update mechanism (i.e., re-constructing the materialized view from scratch). Again, this prevents from incremental update propagation to web pages.

Therefore we have chosen to realize fragment classes by relations and associated update mechanisms between dependent fragment classes. Fragments $f<l>$ of fragment class $F<L>$ are stored in a single, so called *content* *relation*, named $FC\_F\_L$ after the fragment class it represents. Application relations are the starting point for newly defined fragment classes in that they serve as content relations of fragmentation base classes. In order to be able to use uniform prefixes (i.e., "FC_") for names of root fragment classes (which are application relations) and other fragment classes, synonym $FC\_F$ is defined for each application relation $F$ representing root fragment class $F<>$.

⊙ *Example 13.*    Figure 3.1 depicts the realization of the wine shop's fragmentation schema (cf. Figure 2.2). For the three application relations synonyms FC_Wines, FC_Wineries, and FC_Regions are defined to uniformly treat them like content relations FC_PremWineries_region, FC_PremWines_region, and FC_Regions_id, which store the respective fragment class' tuples.

To provide for independence of fragmentation from fragment class' current contents, meta data about fragmentation is stored separately from content relations. For each content relation $FC\_F\_L$ holding tuples of fragments, *fragmentation relation $FR\_F\_L$* stores the domain of parameters $L$. In that way fragments that currently do not comprise any tuples are supported as well. Such a fragment will be listed with its parameter values in $FR\_F\_L$ but will posses no tuples in $FC\_F\_L$.

⊙ *Example 14.*    Figure 3.1 depicts fragmentation relations FR_PremWineries_region, FR_PremWines_region, and FR_Regions_id, which store the respective fragment's parameters. To better motivate the use of fragmentation relations, imagine region saha in which no premium wineries are located. If no fragmentation relation had been used, fragments of fragment class PremWineries<region> would have been determined by content relation FC_PremWineries_region. Because no premium winery is located in region saha, the content relation would not contain tuples having saha as the value of attribute region. Thus fragment premWineries<saha> and page premWines<saha> would not have existed. On the other hand, by storing the parameter's domain in fragmentation relation FR_PremWineries_region, a tuple with the value saha exists therein independent of content relation FC_Prem_Wineries, and thus fragment premWineries<saha> and page premWines<saha> exist. This prevents a user who follows the link from page region<saha> to page class PremWines<region> to experience a HTTP 404 (Not Found) error or similar.

As described in Subsection 2.2.1, a parameter is re-usable by defining it on a root fragment class and independent of the fragment class' contents by using a reference relation to store the values of the parameter's domain. An application relation may serve as a reference relation (e.g., see Example 7 and Example 8), otherwise a reference relation is created (e.g., see Example 9). In either case, the reference relation for parameter $P$ defined upon root fragment class $F<>$ is named $RV\_F\_P$ (via a synonym if an application relation is used as a reference relation).

⊙ *Example 15.* Figure 3.1 depicts reference relations RV_Wineries_region and RV_Regions_id, which store the values of the domains of parameters region and id (both reference relations being a synonym for application relation Regions).

In the following subsections we describe how fragmentation schemas are realized by content, fragmentation, and reference relations in detail.

### 3.1.1   Parameters

Pre-requisite for the definition of a parameter $P$ upon root fragment class $F<>$ with content relation $FC\_F(X)$ is that $P{\in}X$. If $P$ is the primary key of $FC\_F$, the latter can be used as the parameter's reference relation, which stores the parameter's domain. Otherwise, to achieve independence of parameter $P$'s domain from contents held by $FC\_F$, another application relation can be used as the reference relation, where $P{\in}X$ must reference the primary key attribute of this relation.

⊙ *Example 16.* In the definition of parameter region upon root fragment class Wineries<>, application relation Regions is used as reference relation (cf. Example 7), because region is an attribute of FC_Wineries and attribute region references primary key attribute id of relation Regions. The parameter's domain is defined by primary key attribute values $\pi_{id}(\mathsf{FC\_Regions})$. In order to treat reference relations uniformly, synonym RV_Wineries_region is defined for relation Regions. If analogously to PremWineries<region> (cf. Example 5) fragment class StdWineries<region> was defined upon fragmentation base class Wineries<> using tuple selection predicate category="standard", the parameter definition of region would be re-used.

If no application relation can be used as a reference relation, a reference relation has to be created. Such a reference relation is named $RV\_F\_P$, has schema $P$, and is initialized with current values from the content relation of the root fragment class (i.e., with $\pi_P(FC\_F)$). Furthermore, a foreign key constraint is defined on attribute $P{\in}X$, referencing the newly created relation.

⊙ *Example 17.* Reference relation RV_Regions_state would be created for parameter state, which was introduced in Example 9 for illustrative purposes.

### 3.1.2   Primary Fragment Classes

When primary fragment class $G<KL>$ is defined upon fragmentation base class $F<L>$ a new content relation and a new fragmentation relation are

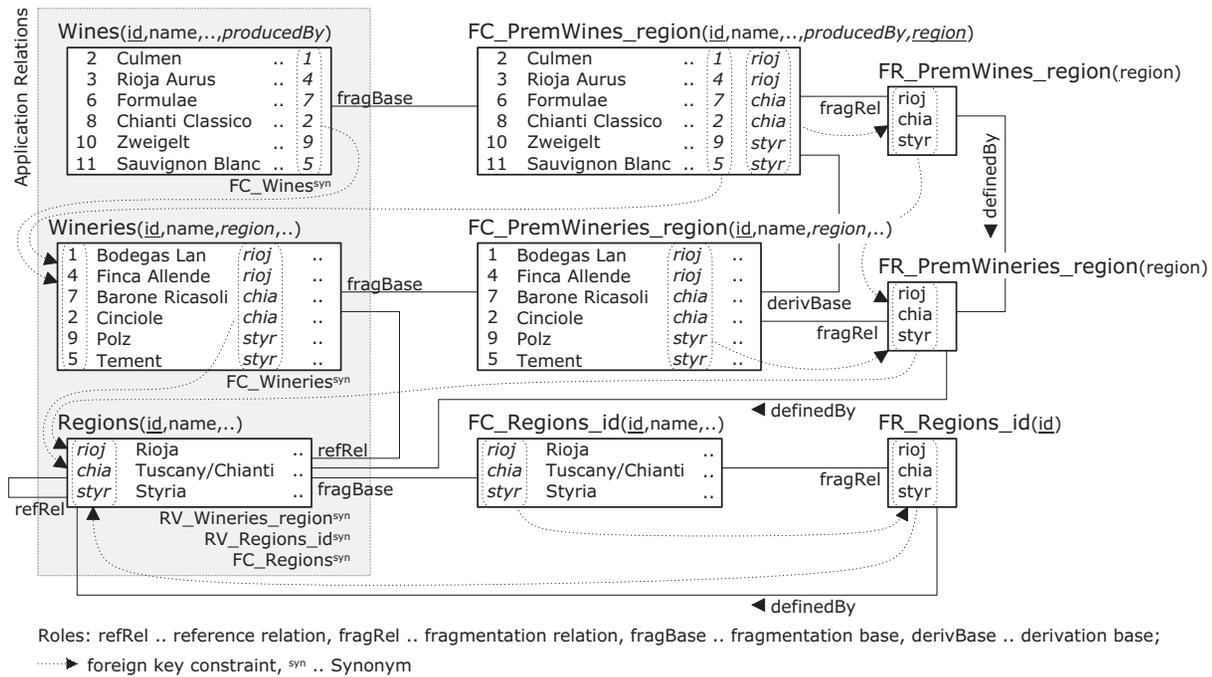

Roles: refRel .. reference relation, fragRel .. fragmentation relation, fragBase .. fragmentation base, derivBase .. derivation base;

········▶ foreign key constraint, syn .. Synonym created. They are initialized with tuples from the content and fragmentation relation of the fragmentation base class and with tuples from reference relations.

Figure 3.1: Realization of a Fragmentation Schema

In the definition of a primary fragment class a fragment selection predicate $p_F$ and a tuple selection predicate $p_T$ can be used to narrow the set of fragments and tuples therein that are taken over from the fragmentation base class. For accurate presentation we introduce the notion of selection predicate $p$, which is defined as $p = p_F \wedge p_T$.

New content relation $FC\_G\_KL(X)$ then stores all fragments of the primary fragment class. It holds tuples from $FC\_F\_L(X)$ for which selection predicate $p$ applies. Its primary key attributes equal the ones of the content relation of the fragmentation base class.

⊙ *Example 18.* Content relation FC_PremWineries_region realizes primary fragment class PremWineries<region> (which is created as shown in Example 5). It holds tuples from FC_Wineries for which tuple selection predicate category="premium" applies. A fragment is retrieved by selecting all wineries of a certain region, e.g., fragment PremWineries_rioj is retrieved by evaluating $\sigma_{region="rioj"}$(FC_PremWineries_region). Attribute id is the primary key attribute of both of the aforementioned content relations.

New fragmentation relation $FR\_G\_KL(KL)$ stores the domain of the fragment class' parameters $KL$, $dom(KL) = dom(K) \times dom(L)$, and thus defines the fragments. For each parameter value the fragment selection predicate $p_F$ must apply. Because $dom(K) = \prod_{i=1}^{|K|} K_i$ is defined by a reference relation $RV\_E\_K_i$ for each parameter $K_i \in K$ (with primary key attribute $\phi_{RV\_E\_K_i}$[1]) and $dom(L)$ is defined by $FR\_F\_L$, data held by $FR\_G\_KL$ is defined by $dom(KL) = \{t \in ((\prod_{i=1}^{|K|} \pi_{\phi_{RV\_E\_K_i}}(RV\_E\_K_i)) \times FR\_F\_L) | p_F(t)\}$.

⊙ *Example 19.* Fragmentation relation FR_PremWineries_region(region) defines the fragments of PremWineries<region>. Its data is defined by RV_-Wineries_region only, because the fragmentation base class is a root fragment class and no parameters other than region are introduced. All primary key attribute values from RV_Wineries_region are taken over to FR_Prem-Wineries_region because no fragment selection predicate is defined.

To ensure data consistency, foreign key constraints are defined on both the content relation, referencing the fragmentation relation, and the fragmentation relation, referencing fragmentation and reference relations it depends on.

---

[1] In the following expression $\phi_R$ is used to denote the primary key attributes of relation $R$.

### 3.1.3   Derived Fragment Classes

When derived fragment class $G{<}KL{>}$ is defined upon fragmentation base class $F{<}L{>}$ and derivation base class $H{<}K{>}$, a new content relation and a new fragmentation relation are created. The content relation is initialized with tuples from the content relation of the fragmentation base class, while the fragmentation relation is initialized with tuples from the fragmentation relations of both the fragmentation and derivation base class.

New content relation $FC\_G\_KL(Z)$ stores all fragments of the derived fragment class. It holds tuples from $FC\_F\_L(X)$ that join with tuples from $FC\_H\_K(Y)$, where the join is defined by an equi-join $FC\_F\_L[J_F{=}J_H]FC\_H\_K$. Derived fragmentation is employed whenever fragmentation by properties of attributes $X$ of $FC\_F\_L$ is not possible, i.e., $K{\cap}X{=}\emptyset$. To store additional parameters $K$, $FC\_F\_L$'s schema $X$ is extended by $K$, i.e., $Z{=}X{\cup}K$. A tuple $t{\in}FC\_F\_L$ may qualify for several derived fragments (i.e., it may join with multiple tuples $s{\in}FC\_H\_K$). Therefore tuple $t$ from the fragmentation base class results in set $V$ of tuples in the derived fragment class, where $V{=}\pi_Z(t[J_F{=}J_H]FC\_H\_K)$. Consequently the primary key attributes of $FC\_G\_KL$ consist of the primary key attributes of $FC\_F\_L$ extended by $K$.

⊙ *Example 20.*   Content relation FC_PremWines_region of derived fragment class PremWines<region> (which is created as shown in Example 10) contains tuples from the fragmentation base's content relation FC_Wines that join with the derivation base's content relation FC_PremWineries_region on FC_Wine[producedBy=id]FC_PremWineries_region. The schema of FC_Prem-Wines_region comprises attribute region additionally to attributes from FC_Wines, its primary key attributes comprise attributes id and region.

New fragmentation relation $FR\_G\_KL(KL)$ stores the domain of parameter set $KL$, $dom(KL){=}dom(K){\times}dom(L)$. Because $dom(L)$ is defined by $FR\_F\_L$ and $dom(K)$ by $FR\_H\_K$, data held by $FR\_G\_KL$ is defined by $dom(KL){=}FR\_F\_L{\times}FR\_H\_K$.

⊙ *Example 21.*   Fragmentation relation FR_PremWines_region defines the fragments of PremWines<region>. Its data is defined by FR_Prem-Wineries_region only, because the fragmentation base class is not parameterized.

To ensure data consistency, foreign key constraints are defined on both the content relation, referencing the fragmentation relation, and the fragmentation relation, referencing the fragmentation relations of the fragmentation and derivation base.

The presented realization of fragment classes is not optimal with respect to disk space because it stores fragment class' contents redundantly. However, unlike main memory, disk capacity is usually not a limiting resource that must be managed wisely [99, 127]. Nevertheless, optimizations of disk space usage are possible at the expense of processing time. E.g., one possibility is not to store complete tuples in content relations but only their primary key attributes and attributes representing parameters, and to construct fragments by joining such tuples with the appropriate application relation. The employed configuration has to be determined when deploying the SMWP approach, either being optimized with respect to disk space usage or processing time.

### 3.1.4   Storing Page Classes

When page class $P<K>$ is defined upon foundation fragment class $F<L>$, its pages are pre-generated and stored inside the database in attributes of type CLOB. Pages are pre-generated as follows: for each fragment $f<l>$ of the foundation fragment class, a page $p<k>$ is generated (the mapping between parameter names $l$ and $k$ is specified with the definition of a page class, e.g., see Example 11). Furthermore, all fragments that are mapped to a page are read through SQL queries from the content relations of mapped fragment classes and transformed to XML, using a simple generic mapping (cf. [133, 134]).

⊙ *Example 22.*        A page of page class PremWines<region> is pre-generated for each fragment of its foundation fragment class PremWineries<region> and stored inside the database. For each page mapped fragments are read from content relations representing fragment classes. The fragments and tuples contained therein are transformed to XML, forming the page's content. An exemplary tuple contained in fragment premWineries<rioj> is shown below.

```
<fragment id="premWineries<rioj>" ..>
   <tuple ..>
      <attribute name="id">4</attribute>
      <attribute name="name">Finca Allende</attribute>
      <attribute name="region">rioj</attribute> ..
   </tuple> ..
</fragment>
```

## 3.2   Maintaining Fragments and Pages

Once fragments and pages have been created, later fragment modifications need to be propagated to dependent fragment classes and affected page classes to modify their data accordingly. A fragment class is dependent on another fragment class if the former contains a tuple of the latter as a replica or if a tuple's membership in the former depends on the existence of a "joining" tuple in the latter (i.e., in the derivation base class). Furthermore, by mapping fragment classes to page classes, modifications on fragments need to be propagated to affected pages (i.e., those pages that contain the tuple serialized as XML).

In this section, we show how the algorithms presented in [133, 134] at the conceptual level for the maintenance of fragments' content can be implemented in a relational database setting by database triggers maintaining content relations. In addition we also present triggers for the maintenance of fragment relations. These triggers are extenders (in terms of [39]) since they are generated automatically according to declaratively defined fragment classes to maintain their contents according to modifications of application relations.

With respect to the different kinds of relations that realize fragment classes and parameters, different kinds of triggers are distinguished as follows. *Content triggers* are defined on content relations to propagate modifications on fragments' content. *Serialization triggers* deal with update propagation from modified fragment classes to affected pages. They are defined on both content and fragmentation relations of fragment classes that are mapped to page classes. *Fragmentation triggers* are defined on both reference and fragmentation relations to propagate modifications on parameters and fragments to dependent fragmentation relations.

To provide for easy definition and maintenance of triggers, they are modularly designed in that each trigger propagates updates to a single dependent relation only. Such a relation is either a content or fragmentation relation of a fragment class that is defined upon the fragment class of the updated content relation (via its fragmentation or derivation base). The benefits of this approach are as follows: (a) triggers are easily created by using a template, thus a small set of algorithm templates can be used irrespective of an actual fragmentation schema, (b) triggers are easily added or deleted in case of new fragment classes are created or existing fragment classes are deleted.

Furthermore, to be able to use the same algorithm templates for fragmentation triggers on reference and fragmentation relations, their algorithms abstract from differences in structure and usage between reference and fragmentation relations.

To avoid a fragmentation schema's realization to influence legacy applications, newly created reference relations (such as RV_Regions_state, cf. Example 17) need particular attention. With the creation of a reference relation, a foreign key constraint is introduced on the according application relation referencing the reference relation to ensure data consistency. This brings up the problem of foreign key violations when legacy applications modify the application relation. Therefore, with the definition of a parameter that creates a new reference relation, a *synchronization trigger* is created on the application relation that keeps the contents of the reference relation synchronized.

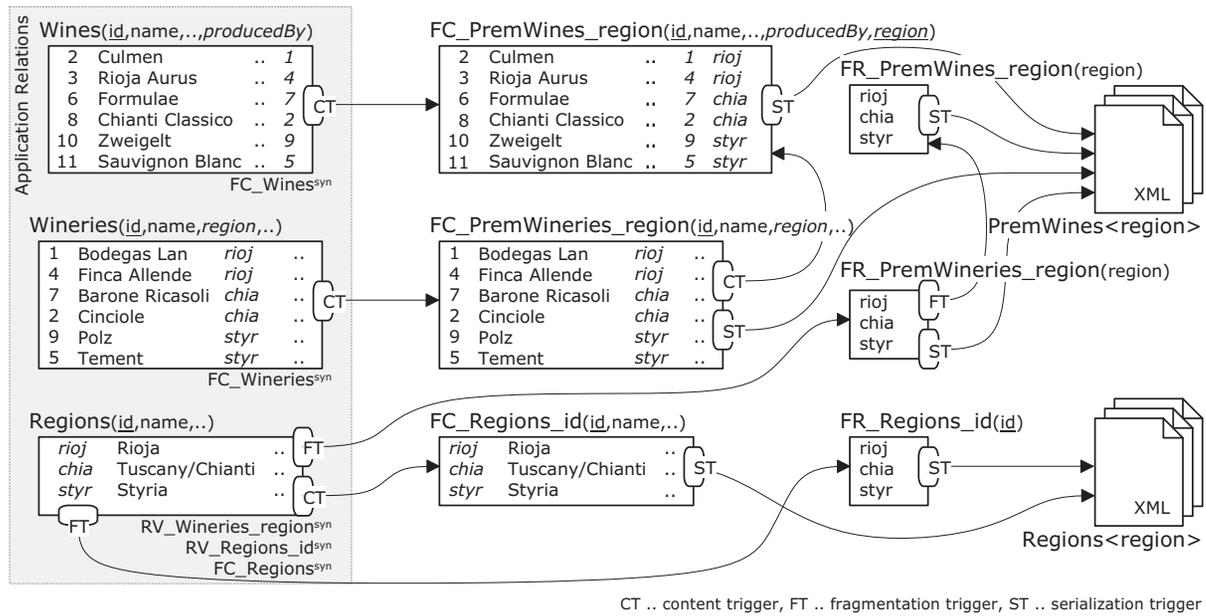

Figure 3.2: Realization of Incremental Push-Based Data Delivery

In short, modifications on content, fragmentation, and parameters are propagated incrementally by database triggers to maintain actual fragmentation and web pages. For each primary fragment class one content trigger (on the fragmentation base class' content relation) and several fragmentation triggers (one on the fragmentation base class' fragmentation relation and one on each introduced parameter's reference relation) propagate modifications. For each derived fragment class two content triggers (one on each of the fragmentation and derivation base class' content relation) and two fragmentation triggers (one on each of the fragmentation and derivation base class' fragmentation relation) propagate modifications. To maintain web pages according to database modifications, two serialization triggers are created (one on the fragment class' content relation and one on its fragmentation relation) for each fragment class that is mapped to at least one page class. They propagate modifications to affected pages.

⊙ *Example 23.* Figure 3.2 depicts the realization of incremental push-based data delivery of the wine shop's fragmentation and page content schema by content, fragmentation and serialization triggers.

### 3.2.1 Maintaining Primary Fragment Classes

With the definition of primary fragment class $G<KL>$ upon fragmentation base class $F<L>$, using selection predicate $p$, and parameters $KL$, which are defined upon root fragment class $E<>$, a new content trigger on $FC\_F\_L$, a new fragmentation trigger on $FR\_F\_L$, and a new fragmentation trigger on $RV\_E\_K_i$ for each $K_i \in K$ are created as described in the following.

#### Content Trigger

First, a content trigger is created on the content relation of the fragmentation base class (i.e., on $FC\_F\_L$) to propagate modifications thereon to the content relation of the primary fragment class (i.e., $FC\_G\_KL$), considering the selection predicate. Its algorithm template is described in this section.

Algorithm templates are depicted by pseudo-code mixed with SQL statements throughout the chapter. For better readability placeholders and variables are denoted in italics. While placeholders are replaced by actual values when the template is instantiated to implement a trigger, actual values of variables are determined at run time. Variables *old* and *new* are used in all templates. They are bound to the old and new tuple values as defined by SQL and are thus called bindings. For a concise presentation of the templates, we make use of the following abbreviations: (a) $X=Y$ is used to test two sets of attributes $X$ and $Y$ that both comprise the same attributes $a_1..a_n$ for equality of their respective attribute values (i.e., to test whether $X.a_1=Y.a_1 \wedge .. \wedge X.a_n=Y.a_n$), (b) analogously the same applies for testing inequality by $X \neq Y$, (c) in SQL Update statements "UPDATE ... SET $X=Y$" is used to set attributes $X.a_1=Y.a_1..X.a_n=Y.a_n$, and (d) whenever a variable that comprises a set of attributes (e.g., binding $new$) is used without further specification of attributes it refers to all attributes (e.g., $new$ refers to $new.a_1,..,new.a_n$).

The algorithm template for content triggers for primary fragment classes distinguishes the following three cases: (a) If a tuple is inserted (line *4*) and the tuple selection predicate applies ($X'$ denotes $X \backslash KL$) as well as the fragment selection predicate applies (lines *5*), the tuple is inserted into the primary fragment class (line *6*). (b) If a tuple is updated (line *7*), selection predicates for the old (line *8*) and new tuple (line *9*) are determined. If both the old and new tuple qualify for fragments of the primary fragment class (line *10*), the affected tuple is either updated if the fragment it is contained in remains identical (lines *11* and *12*), or deleted and inserted if the fragment changes (lines *13* to *15*). If the tuple does not qualify for any fragment of the primary fragment class any longer, it is deleted (lines *16* and *17*). Vice versa, if the tuple newly qualifies for a fragment, it is inserted (lines *18* and *19*). (c) If a tuple is deleted (*20*) and it qualified for a fragment of the primary fragment class (line *21*), it is deleted therefrom as well (line *22*).

*1* TRIGGER CT_FragBaseToPrimaryFC-$FC\_G\_KL$
*2* AFTER INSERT OR UPDATE OR DELETE ON $FC\_F\_L$
*3* FOR EACH ROW
*4*   **if** INSERTING
*5*     **if** $p_T(new.X')$ AND $p_F(new.KL)$
*6*       INSERT INTO $FC\_G\_KL$ VALUES $new$;
*7*   **if** UPDATING
*8*     $p_{old}:=p_T(old.X')$ AND $p_F(old.KL)$;
*9*     $p_{new}:=p_T(new.X')$ AND $p_F(new.KL)$;
*10*     **if** $p_{old}$ AND $p_{new}$
*11*       **if** $old.K=new.K$
*12*         UPDATE $FC\_G\_KL$ AS G SET G.$X=new.X$
          WHERE G.$\phi_G=old.\phi_G$;
*13*       **if** $old.K \neq new.K$
*14*         DELETE FROM $FC\_G\_KL$ AS G WHERE G.$\phi_G=old.\phi_G$;
*15*         INSERT INTO $FC\_G\_KL$ VALUES $new$;
*16*     **if** $p_{old}$ AND NOT $p_{new}$
*17*       DELETE FROM $FC\_G\_KL$ AS G WHERE G.$\phi_G=old.\phi_G$;
*18*     **if** NOT $p_{old}$ AND $p_{new}$
*19*       INSERT INTO $FC\_G\_KL$ VALUES $new$;
*20*   **if** DELETING
*21*     **if** $p_T(old.X')$ AND $p_F(old.KL)$
*22*       DELETE FROM $FC\_G\_KL$ AS G WHERE G.$\phi_G=old.\phi_G$;

⊙ *Example 24.* With the definition of primary fragment class Prem-

Wineries<region>, which is defined upon fragmentation base class Prem-
Wineries<>, a content trigger is created on FC_Wineries to propagate modi-
fications to FC_PremWineries_region.

### Fragmentation Trigger

Second, a fragmentation trigger is created on each relation $S_i \in S$, where $S$
is the set of reference and fragmentation relations, $FR\_G\_KL$ depends on.
$S$ comprises reference relations $RV\_E\_K_i$ for each parameter $K_i \in K$ and
fragmentation relation $FR\_F\_L$ if $|L| > 0$. By abstracting from differences in
structure and usage between reference and fragmentation relations, the same
algorithm template is used for fragmentation triggers on reference relations
as well as fragmentation triggers on fragmentation relations.

In the following the algorithm template for fragmentation triggers for
primary fragment classes is described. Insert, Update, and Delete SQL
Statements on reference and fragmentation relations reflect creation, modi-
fication, and deletion of fragments, activating fragmentation triggers. Such
a trigger is created on each relation $S_i \in S$. Variable $S'$, which is used in the
template, is defined by $S \backslash S_i$. In case of fragment modification or deletion
(line *4*), fragments of the primary fragment class that show corresponding
parameter values are deleted (line *5*). In case of fragment modification or
creation (line *6*), new fragments are determined by building the cartesian
product of the new parameter value by the domains of the remaining pa-
rameters and inserted if the fragment selection predicate applies (line *7*).
Note that the modification of a fragment's parameters causes the creation
and deletion of fragments.

*1* TRIGGER FT_FragBase-$S_i$-ToPrimaryFC-$FR\_G\_KL$
*2* BEFORE INSERT OR UPDATE OR DELETE ON $S_i$
*3* FOR EACH ROW
*4*   **if** UPDATING OR DELETING
*5*     DELETE FROM $FR\_G\_KL$ WHERE $\phi_{S_i} = old.\phi_{S_i}$;
*6*   **if** UPDATING OR INSERTING
*7*     INSERT INTO $FR\_G\_KL$
            (SELECT $new.\phi_{S_i}$, $S'_1.\phi_{S'_1}$, .., $S'_n.\phi_{S'_n}$ FROM $S'_1$, .., $S'_n$
                WHERE $p_F(new.\phi_{S_i}, S'_1.\phi_{S'_1}, .., S'_n.\phi_{S'_n})$));

⊙ *Example 25.* Fragmentation relation FR_PremWineries_region of pri-
mary fragment class PremWineries<region> depends on reference relation
RV_Wineries_region (i.e., application relation Regions) only. It does not de-
pend on any fragmentation relation, because its fragmentation base class
Wineries<> is not parameterized. Therefore, a fragmentation trigger is cre-
ated on RV_Wineries_region only to propagate modifications to the domain
of parameter region.

### 3.2.2   Maintaining Derived Fragment Classes

With the definition of derived fragment class $G<KL>$ upon fragmentation base class $F<L>$ and derivation base class $H<K>$, a new content trigger on $FC\_F\_L$, a new content trigger on $FC\_H\_K$, a new fragmentation trigger on $FR\_F\_L$, and a new fragmentation trigger on $FR\_H\_K$ are created as described in the following.

### Content Triggers

Because the derived fragment class is defined upon, and thus dependent on, its fragmentation and derivation base class, a content trigger is created on $FC\_F\_L(X)$ and $FC\_H\_K(Y)$ to propagate modifications to $FC\_G\_KL(Z)$ (where $Z{=}X{\cup}K$). While the fragmentation base class provides contents to be held by the derived fragment class, the derivation base class defines how this data is to be further partitioned. Due to these different roles, content triggers on $FC\_F\_L$ and $FC\_H\_K$ have different algorithms.

First, the algorithm template for the content trigger that is created on the content relation of the fragmentation base class is described in the following. It distinguishes the following three cases: (a) If a tuple is inserted (line *4*) it is inserted into all fragments of the derived fragment class it qualifies for as well (line *5*). (b) If a tuple is updated (line *6*) and the join attribute values did not change (line *7*), tuples of the derived fragment class can be updated as well (line *8*). If the join attribute values changed (line *9*), fragments the tuple qualifies for change as well. Thus it is deleted from fragments it is contained in but no longer qualifies for (line *10*), updated in fragments it is contained in and still qualifies for (line *11*), and inserted into fragments it newly qualifies for (line *12*). Remember that $old.L$ and $new.L$ never differ, because updates to $L$ would have been translated to according delete and insert statements by another trigger before (cf. Section 3.2.1). (c) If a tuple is deleted (line *13*) it is deleted from all fragments of the derived fragment class it is contained in (line *14*).

*1* TRIGGER CT\_FragBaseToDerivedFC-$FC\_G\_KL$
*2* AFTER INSERT OR UPDATE OR DELETE ON $FC\_F\_L$ AS F
*3* FOR EACH ROW
*4*   **if** INSERTING
*5*     INSERT INTO $FC\_G\_KL$
          (SELECT DISTINCT $new$, H.$K$ FROM $FC\_H\_K$ AS H
            WHERE H.$J_H{=}new.J_F$);
*6*   **if** UPDATING
*7*     **if** $old.J_F{=}new.J_F$
*8*       UPDATE $FC\_G\_KL$ AS G SET G.$X{=}new.X$
            WHERE G.$\phi_F{=}old.\phi_F$;
*9*     **if** $old.J_F{\neq}new.J_F$

*10*     DELETE FROM $FC\_G\_KL$ AS G
            WHERE G.$\phi_F$=$old.\phi_F$ AND G.$K$ NOT IN (
                SELECT $K$ FROM $FC\_H\_K$ WHERE $J_H$=$new.J_F$);
*11*     UPDATE $FC\_G\_KL$ AS G SET G.$X$=$new.X$
            WHERE G.$\phi_F$=$old.\phi_F$;
*12*     INSERT INTO $FC\_G\_KL$
            (SELECT DISTINCT $new$, $H.K$ FROM $FC\_H\_K$ AS H
                WHERE H.$J_H$=$new.J_F$ AND NOT EXISTS (
                SELECT * FROM $FC\_G\_KL$ AS G
                    WHERE G.$\phi_F$=$new.\phi_F$ AND G.$K$=H.$K$));
*13*  **if** DELETING
*14*     DELETE FROM $FC\_G\_KL$ AS G
            WHERE G.$\phi_F$=$old.\phi_F$;

⊙ *Example 26.*     A content trigger is created on FC_Wines to propagate modifications thereon to FC_PremWines_region of derived fragment class PremWines<region> with fragmentation base Wines<> and derivation base PremWineries<region>. For example, the insertion of a new wine $w$ into FC_Wines entails the insertion of tuple $v$=$\pi_Z(t$[producedBy=id]FC_Wineries_region) into FC_PremWines_region(Z), thus enriching the new wine with information about the region it is produced in.

Second, the algorithm template for the content trigger that is created on the content relation of the derivation base class is described in the following. It needs to detect upon a tuple modification whether another tuple contained in the same fragment has the same join attribute values as the tuple being modified. If not, tuples from the fragmentation base do no longer qualify for the derived fragment class and have to be removed. Thus this trigger needs to access bindings *old* and *new* as well as the content relation it is defined upon. However, while the bindings can only be accessed from row-level triggers (created as "FOR EACH ROW"), the content relation can only be accessed from statement-level triggers (created as "FOR EACH STATEMENT"), because for a row-level trigger the content relation is mutating and thus not accessible. Since a trigger can be either row-level or statement-level but not both, this problem needs to be solved specifically to SMWP.

The solution to the problem of mutating content relations is the following: (a) The content trigger is defined as a statement-level trigger (opposite to the other two content triggers CT_FragBaseToPrimaryFC-$FC\_G\_KL$ and CT_FragBaseToDerivedFC-$FC\_G\_KL$) to be able to query the content relation, (b) To achieve that the content trigger is triggered once for each modified tuple (i.e., to be able to use statement-level triggers instead of row-level triggers), all content triggers use cursors to modify several tuples by issuing a single SQL statement for each tuple being modified (and not a single SQL statement to modify several tuples as used in their algorithm templates for presentational purposes) and root fragment classes cannot be used as derivation base classes, (c) An auxiliary trigger is used to make the bindings available to the content trigger. It is a row-level trigger that is triggered before a tuple modification and stores the bindings in an auxiliary relation, named **SMWP_Worktrace** (which is in $NF^2$). By these means the content trigger can read bindings *old* and *new* from **SMWP_Worktrace** as well as the modified content relation $FC\_H\_K$.

The algorithm template of the auxiliary trigger, which is triggered before each row is shown below. It inserts the following data into **SMWP_Worktrace**: the name of the derived fragment class to distinguish between potentially several derived fragment classes that have $H{<}K{>}$ as derivation base class (needed by algorithm template **CT_DerivBaseToDerivedFC-**$FC\_G\_KL$, see later in this section) and bindings *old* and *new* where applicable.

*1* TRIGGER AT_DerivBaseToDerivedFC-$FC\_G\_KL$
*2* BEFORE INSERT OR UPDATE OR DELETE ON $FC\_H\_K$
*3* FOR EACH ROW
*4*   **if** INSERTING
*5*     INSERT INTO SMWP_Worktrace(fcName,old,new)
         VALUES ("$FC\_G\_KL$", NULL, *new*);
*6*   **if** UPDATING
*7*     INSERT INTO SMWP_Worktrace(fcName,old,new)
         VALUES ("$FC\_G\_KL$", *old*, *new*);
*8*   **if** DELETING
*9*     INSERT INTO SMWP_Worktrace(fcName,old,new)
         VALUES ("$FC\_G\_KL$", *old*, NULL);

The algorithm template of the content trigger, which is triggered after each statement, is shown below. Before performing any action, bindings *old* and *new* are read from **SMWP_Worktrace** where applicable (lines *5*, *10*, and *19*). The algorithm distinguishes the following three cases: (a) If a tuple is inserted into the derivation base (line *4*) and no other tuple with the same join attribute values already exists in the same fragment (lines *6*, *7*), the tuples from the fragmentation base that additionally qualify for the derived fragment class are determined and inserted (line *8*). (b) If a tuple is updated (line *9*) and the values of the join attributes have changed (line *11*), tuples from the fragmentation base may no longer or may newly qualify for fragments of the derived fragment class. In case no other tuple with the same join attribute values as the old ones exists in the same fragment (line *12*, *13*), the tuples that do no longer qualify are determined and deleted (line *14*). In case no other tuple with the same join attribute values as the new ones existed in the same fragment before (lines *15*, *16*), tuples that additionally qualify are determined and inserted (line *17*). Remember that *old.K* and $new.K$ never differ, because updates to $K$ would have been translated to according delete and insert statements by another trigger before (cf. Section 3.2.1 and earlier in this section). (c) If a tuple is deleted (line *18*) and no other tuple with the same join attribute values exists in the same fragment (lines *20*, *21*), tuples from the fragmentation base do no longer qualify for fragments of the derived fragment class and are thus deleted (line *22*). Finally, temporary buffered bindings *old* and *new* are removed from relation SMWP_Worktrace (line *23*).

*1* TRIGGER CT_DerivBaseToDerivedFC-$FC\_G\_KL$
*2* AFTER INSERT OR UPDATE OR DELETE ON $FC\_H\_K$
*3* FOR EACH STATEMENT
*4* **if** INSERTING
*5* SELECT new INTO *new* FROM SMWP_Worktrace
        WHERE fcName=$"FC\_H\_K"$;
*6* SELECT COUNT(*) INTO $count^i$ FROM $FC\_H\_K$ AS H
        WHERE H.$\phi_H \neq new.\phi_H$ AND H.$J_H = new.J_H$ AND H.$K = new.K$;
*7* **if** $count^i = 0$
*8* INSERT INTO $FC\_G\_KL$
        (SELECT *, $new.K$ FROM $FC\_F\_L$ WHERE $J_F = new.J_H$);
*9* **if** UPDATING
*10* SELECT old, new INTO *old*, *new* FROM SMWP_Worktrace
        WHERE fcName=$"FC\_H\_K"$;
*11* **if** $old.J_H \neq new.J_H$
*12* SELECT COUNT(*) INTO $count_1^u$ FROM $FC\_H\_K$ AS H
        WHERE H.$\phi_H \neq old.\phi_H$ AND H.$J_H = old.J_H$ AND H.$K = old.K$;
*13* **if** $count_1^u = 0$
*14* DELETE FROM $FC\_G\_KL$
        WHERE $K = old.K$ AND $J_F = old.J_H$;
*15* SELECT COUNT(*) INTO $count_2^u$ FROM $FC\_H\_K$ AS H
        WHERE H.$\phi_H \neq new.\phi_H$ AND H.$J_H = new.J_H$ AND H.$K = new.K$;
*16* **if** $count_2^u = 0$
*17* INSERT INTO $FC\_G\_KL$
        (SELECT *, $new.K$ FROM $FC\_F\_L$ WHERE $J_F = new.J_H$);
*18* **if** DELETING
*19* SELECT old INTO *old* FROM SMWP_Worktrace
        WHERE fcName=$"FC\_H\_K"$;
*20* SELECT COUNT(*) INTO $count^d$ FROM $FC\_H\_K$ AS H
        WHERE H.$\phi_H \neq old.\phi_H$ AND H.$J_H = old.J_H$ AND H.$K = old.K$;
*21* **if** $count^d = 0$
*22* DELETE FROM $FC\_G\_KL$
        WHERE $K = old.K$ AND $J_F = old.J_H$;
*23* DELETE FROM SMWP_Worktrace WHERE fcName=$"FC\_G\_KL"$;

⊙ *Example 27.* To propagate modifications on PremWineries<region>
to derived fragment class PremWines<region>, an auxiliary trigger and a
content trigger are created on FC_PremWineries_region. While the auxiliary
trigger is triggered before each row and insert bindings *old* and *new* into rela-
tion SMWP_Worktrace, the content trigger is triggered after each statement
and accesses content relation FC_PremWineries_region and auxiliary relation
SMWP_Worktrace.

### Fragmentation Triggers

Furthermore, one fragmentation trigger is created on the fragmentation re-
lations of the fragmentation base class (i.e., $FR\_F\_L$ if $|L|{>}0$) and one on
the derivation base class (i.e., $FR\_H\_K$). These triggers propagate creations
and deletions of fragments to $FR\_G\_KL$. They do not propagate modifi-
cations of fragments, because fragmentation triggers on reference relations
(cf. the algorithm in Subsection 3.2.1) translate updates on reference re-
lations to equivalent creations and deletions of fragments as soon as they
occur. Notice that derived fragment classes do not depend on any reference
relation, because they do not introduce new parameters.

The algorithm template for fragmentation triggers for derived fragment
classes is described in the following. It differs from the one presented in
Subsection 3.2.1 in that $|S|{\leq}2$, because $S$ comprises $FR\_H\_K$ and possibly
$FR\_F\_L$ only, and in that it does not need to treat updates to fragmenta-
tion relations for the reasons mentioned before. A fragmentation trigger is
created on each relation $S_i{\in}S$. Variable $S'$, which is used in the template, is
defined by $S{\setminus}S_i$ and may be empty. The algorithm template distinguishes
the following two cases: (a) If a fragment is deleted (line *4*), fragments of
the derived fragment class that have corresponding parameter values are
deleted (line *5*). (b) If a fragment is created (line *6*), new parameter values
are determined by forming the cartesian product of the new parameter value
by the domain of the other parameters and fragments are created (line *7*).

*1* TRIGGER FT_Frag/DerivBase-$S_i$-ToDerivedFC-$FR\_G\_KL$
*2* BEFORE INSERT OR DELETE ON $S_i$
*3* FOR EACH ROW
*4*   **if** DELETING
*5*     DELETE FROM $FR\_G\_KL$
         WHERE $\phi_{S_i}{=}old.\phi_{S_i}$;
*6*   **if** INSERTING
*7*     INSERT INTO $FR\_G\_KL$
         (SELECT $new.\phi_{S_i}$, $S'.\phi_{S'_1}$ FROM $S'_1$);

⊙ *Example 28.* Fragmentation relation FR_PremWines_region of de-
rived fragment class PremWines<region> depends on fragmentation relation

FR_PremWineries_region of the derivation base class only, because fragmentation base class Wines<> is not parameterized. Therefore the only fragmentation trigger created is the one on FR_PremWineries_region to propagate modifications thereon to FR_PremWines_region.

### 3.2.3   Maintaining Page Classes

Pre-generated pages contain contents drawn from a database and have therefore to be synchronized with it. For automatic synchronization, modifications to fragment's contents are propagated to pages, entailing equivalent modifications on XML serializations of tuples. Analogously modifications to fragmentation are propagated to pages, entailing equivalent modifications of XML serializations of fragments as well as the creation and deletion of pages.

A *fragmentation serialization-trigger* translates creations and deletions of fragments to equivalent operations on pages. It is created on fragmentation relation $FR\_F\_L$ if $F<L>$ is mapped to at least one page class. Such a trigger iterates over all page classes the modified fragment class is mapped to. If the modified fragment class is the current page class' foundation fragment class, a page is either created (comprising existing empty fragments of mapped fragment classes) or deleted. Otherwise all existing pages the fragment is mapped to are determined by comparing parameter values, taking parameter maps (e.g., cf. Example 11) into account, and a fragment is added (if not yet existent) or deleted thereon.

Besides modifications to fragmentation also modifications to fragments' content have to be propagated to pages. A tuple insert, update, or delete entails the according modification of affected pages. A *content serialization-trigger* is created on content relation $FC\_F\_L$ if $F<L>$ is mapped to at least one page class. It propagates modifications to fragment's contents to affected pages. Because the logic of both serialization triggers basically corresponds to the algorithms presented in [133], the interested reader is referred to there for their algorithms.

⊙ *Example 29.*    A fragmentation serialization-trigger is created on FR_PremWineries_region to propagate modifications on fragmentation to pages. Since fragment class PremWineries<region> is the foundation fragment class of page class PremWines<region>, pages are created or deleted whenever a tuple is inserted into or deleted from the fragmentation relation. Furthermore, a content serialization-trigger is created on FC_PremWineries_region to propagate modifications on fragments' content to according pages.

Serialization triggers comprise extensive generic logic for translating operations on fragments to equivalent operations on pages, and are thus implemented in Java. They are loaded into and executed within the database and use Oracle's XML Parser [121] and Xalan's Serializer [62] to perform operations on pages. To efficiently translate operations, serialization triggers need efficient access to fragmentation and page content schemas. Therefore auxiliary meta data relations are used to store this information.

Operations on pages are carried out by a SAX [111] filter in our prototype. The page to be modified is parsed by a SAX parser, which generates events for each recognized XML element. The filter listens to the event stream and passes events on to Xalan's Serializer, which constructs an XML document from its incoming event stream. If an event concerns an element that is to be modified, the filter modifies the event before passing it on. To efficiently identify elements that are to be modified, Oracle's rowid attribute, which is a database wide unique identifier of a tuple, is stored along with each XML element that represents a tuple. After parsing the original page is replaced with the newly constructed one, thus reflecting the fragment modification.

Summarizing the presented triggers, the following example shows the effects of modifying an application relation. The activation sequence of triggers is ordered by ascending numbers, however, the order may vary since the execution order of triggers created on the same relation may be undefined (in particular, step 1 and 2 may be interchanged).

⊙ *Example 30.* When region "South Australia" is *added* to the parameter domain of region, i.e., when a tuple with "saus" as the value of attribute id is inserted into application relation Regions (cf. Figure 3.2), (1) one of the two fragmentation triggers defined on Regions inserts the region into FR_PremWineries_region, (1.1) the fragmentation trigger defined on FR_Prem­Wineries_region inserts the region into FR_PremWines_region, (1.1.1) the serialization trigger defined on FR_PremWines_region does not perform any action since the page the fragment is mapped to does not yet exist, (1.2) the serialization trigger defined on FR_PremWineries_region creates page prem­Wines<saus> containing empty fragments premWineries<saus> and prem­Wines<saus>, (2) the other one of the two fragmentation triggers defined on Regions inserts the region into FR_Regions_id, (2.1) the serialization trigger defined on FR_Regions_id creates page regions<saus> containing empty fragment regions<saus>, (3) the content trigger defined on Regions inserts the region into FC_Regions_id, and finally (3.1) the serialization trigger defined on FC_Regions_id inserts the region into the page's fragment created in step 2.1. When a region is *deleted*, the activation sequence above applies as well, except that pages, fragments, and tuples are deleted rather than newly created or inserted. When a region is *updated*, a deletion and insertion of a region is performed by the fragmentation triggers defined on relation Regions.

The number of triggers that are involved in incremental propagation upon a modification of an application relation by a legacy application depends on the kind of data being modified (i.e., contents, fragmentation, or parameters), the fragmentation schema, the page content schema, and the modification itself (i.e., in which respect data is modified). Thus only a general statement can be made about the number of triggers involved as follows (for an empirical performance evaluation see [133]):

- Upon modification of a fragment class $F{<}L{>}$, (a) one content trigger is involved per dependent fragment class $G{<}KL{>}$, possibly causing multiple modifications of $G{<}KL{>}$, e.g., when join attribute values in a derivation base class are modified, and (b) one serialization trigger is involved if $F{<}L{>}$ is mapped to at least one page class.

- Upon modification of parameter $P$'s domain $dom(P)$, one fragmentation trigger is involved per fragment class $F{<}L{>}$ that introduces parameter $P$, causing maximally $2*|dom(L{\setminus}P)|$ modifications of $FR\_F\_L$ (since an update of a parameter is translated to an according deletion and insertion of fragments).

- Upon modification of the domain of fragment class $F{<}L{>}$'s parameter set $L$, (a) one fragmentation trigger is involved per dependent fragment class $G{<}KL{>}$ causing maximally $|dom(K)|$ modifications of $FR\_G\_KL$, and (b) one serialization trigger is involved if $F{<}L{>}$ is mapped to at least one page class.

## 3.3 Predicate Based Parameters

Experience has shown that sometimes a fragment class cannot be parameterized as desired, since it would need a parameter that is not part of the class' schema. In such cases one would fall back to on the fly generation or use other less flexible page pre-generation techniques. However, in many cases it is just sufficient to define a parameter on a derived attribute (i.e., an attribute whose value is derived from existing attributes). To provide for this kind of parameters, we introduce in this section the concept of *predicate based parameters* by extending our previous work and refer to parameters mentioned so far as *value based parameters*. Like the latter, predicate based parameters are defined upon root fragment classes to provide for reusable parameter definitions featuring consistent semantics.

Predicate based parameter $Q$ can be introduced for primary fragmentation and is defined by a set of label-predicate pairs, $Q{=}(l, q)$. If predicate $q_i$ applies for a given tuple, it is "tagged" with the according label $l_i$, thus enriching it with derived information. To determine the label of predicate based parameter $Q$ for tuple $t$, function $getLabel(Q, t)$ is used. For each tuple exactly one predicate must apply. The domain of a predicate based parameter is determined by the set of its labels.

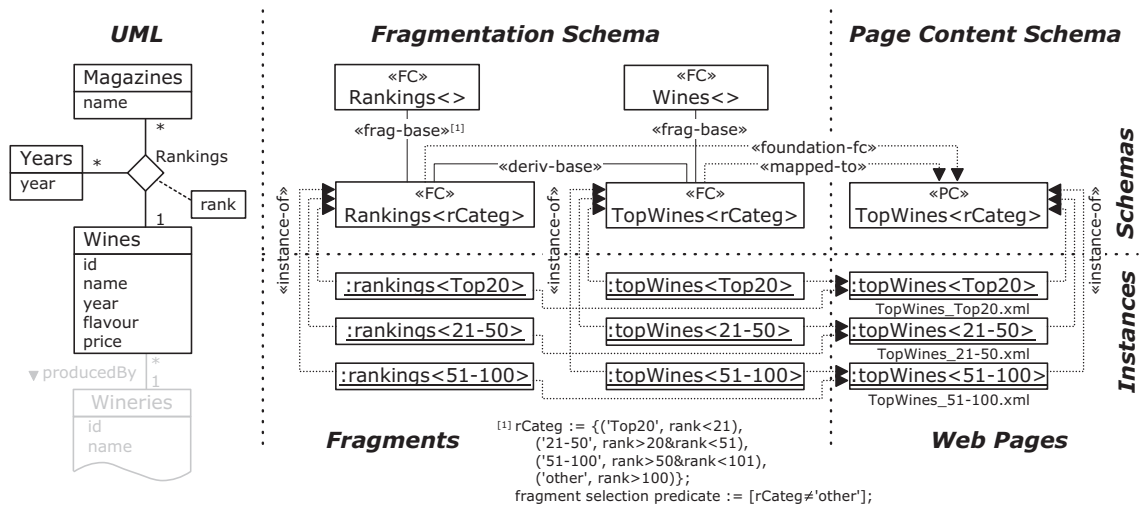

Figure 3.3: Parameterized Fragmentation using Predicate Based Parameters

⊙ *Example 31.* The wine shops stores, additionally to wines, wineries, and regions, contents about wine magazines (class **Magazine**), which publish annual rankings of newly available wines (ternary association **Rankings**), as depicted on the left of Figure 3.3. It wants to provide web pages enlisting wines that have been ranked in given ranking categories Top20, 21-50, and 51-100. However, the schema of application relation **Rankings**(magazine,year,wine,rank) (implementing UML association **Rankings**) does not contain an attribute for ranking categories. To derive the latter from attribute **rank**, a predicate based parameter **rCateg** is defined upon **Rankings**<> (as shown below).

CREATE PREDICATE BASED PARAMETER rCateg ON Rankings<>
    PREDICATES {("Top20", {rank<21}), ("21-50", {rank>20∧rank<51}),
      ("51-100", {rank>50∧rank<101}), ("other", {rank>100})};

The introduction of predicate based parameters slightly affects the definition of primary fragment classes and their content relations as presented in Subsection 3.1.2 as follows. Primary fragment class $G<KLM>$ with $FC\_G\_KLM(Y)$ is defined upon fragmentation base $F<L>$ with $FC\_F\_L(X)$ using selection predicate $p$. The primary fragment class introduces, additionally to value based parameters $L$ (where $|L|{\geq}0$ and $L{\in}X$), predicate based parameters $M$ (where $|M|{\geq}0$ and $M{\cap}X{=}\emptyset$).

⊙ *Example 32.* Predicate based parameter **rCateg** is used to define primary fragment class **Rankings**<rCateg>, which partitions rankings into fragments according to ranking categories as defined by parameter **rCateg** using fragmentation selection predicate **rCateg**≠"other" (as shown below).

CREATE PRIMARY FRAGMENT CLASS Rankings<rCateg>
    FRAGMENTATION BASE CLASS Rankings<>
    FRAGMENT SELECTION PREDICATE {rCateg≠"other"};

To smoothly integrate predicate based parameters into the presented realization of the SMWP approach, a predicate based parameter's label is stored, like a value based parameter, as part of a fragment's tuple to define the fragment it is contained in. Thereby fragment classes that do not introduce predicate based parameters can handle their parameters equally irrespective of whether they are value or predicate based.

However, when primary fragment class $G<KLM>$ introduces predicate based parameters $M$, they are not part of the schema of the fragmentation base class' content relation $FC\_F\_L(X)$. To hold these parameters, schema $X$ is extended by $M$. Thereby the primary fragment class' content relation $FC\_G\_KLM$ has schema $Z{=}X{\cup}M$. While tuple selection predicate $p_T$ may still reference $X \backslash KL$, fragmentation selection predicate $p_F$ may only reference introduced predicate based parameter set $M$ additionally to $KL$.

⊙ *Example 33.* Content relation FC_Rankings_rCateg of primary fragment class Rankings<rCateg> comprises attribute rCateg additional to attributes of fragmentation base class Rankings<>. Finally, to provide web pages enlisting wines that have been ranked in a ranking category as defined by Rankings<rCateg>, derived fragment class Wines-<rCateg> is defined upon fragmentation base Wines<> and derivation base Rankings<rCateg> (cf. Figure 3.3). Thereby fragment class Wines<>, which comprises a single fragment, is partitioned according to the fragmentation of Rankings<rCateg>. Thereafter, fragment class TopWines<rCateg> is mapped to page class TopWines<rCateg>, which is defined upon foundation fragment class Rankings<rCateg>.

Only the algorithm template CT_FragBaseToPrimaryFC-$FC\_G\_KL$ presented in Subsection 3.2.1 needs to be slightly adapted to support predicate based parameters. It needs to be modified to determine the label of the tuple whose modification is to be propagated. The remaining algorithm templates can be left unchanged for the following reasons: a) a content trigger in derived fragmentation can handle predicate based parameters like value based ones, b) fragmentation triggers are robust to the introduction of predicate based parameters, because they abstract from differences between reference and fragmentation relations, c) serialization triggers are concerned with incrementally modifying pages, thus they are not affected at all.

The adapted algorithm template for content triggers on the content relation of the fragmentation base class of a primary fragment class is shown below. The line numbers are identical with the line numbers of the original algorithm presented in Subsection 3.2.1. Lines that are modified or newly inserted are marked by the symbol "*", and opposed to original or modified lines, newly inserted lines are numbered alphabetically starting from "*a*" (e.g., *a**). Basically, the algorithm considers predicate based parameters $M$ additionally to parameters $KL$. The actual label of $M$ for a given tuple is determined by function *getLabel* (lines *a** through *d**). These labels are considered when dealing with parameters, e.g., when testing the selection predicate (line *5**), or when inserting a tuple (line *6**).

*1* TRIGGER CT_FragBaseToPrimaryFC-$FC\_G\_KL$
*2* AFTER INSERT OR UPDATE OR DELETE ON $FC\_F\_L$
*3* FOR EACH ROW
*4*     **if** INSERTING
*a**         $m_{new} := \prod_{i=1}^{|M|} getLabel(M_i, new)$
*5**     **if** $p_T(new.X')$ AND $p_F(new.KL \times m_{new})$
*6**         INSERT INTO $FC\_G\_KL$ VALUES $new, m_{new}$;
*7*     **if** UPDATING

**b\***   $m_{old} := \prod_{i=1}^{|M|} getLabel(M_i, old)$
**c\***   $m_{new} := \prod_{i=1}^{|M|} getLabel(M_i, new)$
**8\***   $p_{old} := p_T(old.X')$ AND $p_F(old.KL \times m_{old})$
**9\***   $p_{new} := p_T(new.X')$ AND $p_F(new.KL \times m_{new})$
**10**   **if** $p_{old}$ AND $p_{new}$
**11\***     **if** $old.K = new.K$ AND $m_{old} = m_{new}$
**12**        UPDATE $FC\_G\_KL$ AS G SET G.$X = new.X$
             WHERE G.$\phi_G = old.\phi_G$;
**13\***     **if** $old.K \neq new.K$ AND $m_{old} \neq m_{new}$
**14**        DELETE FROM $FC\_G\_KL$ AS G WHERE G.$\phi_G = old.\phi_G$;
**15\***        INSERT INTO $FC\_G\_KL$ VALUES $new, m_{new}$;
**16**   **if** $p_{old}$ AND NOT $p_{new}$
**17**     DELETE FROM $FC\_G\_KL$ AS G WHERE G.$\phi_G = old.\phi_G$;
**18**   **if** NOT $p_{old}$ AND $p_{new}$
**19\***     INSERT INTO $FC\_G\_KL$ VALUES $new, m_{new}$;
**20**   **if** DELETING
**d\***   $m_{old} = \prod_{i=1}^{|M|} getLabel(M_i, old)$
**21\***   **if** $p_T(old.X')$ AND $p_F(old.KL \times m_{old})$
**22**     DELETE FROM $FC\_G\_KL$ AS G WHERE G.$\phi_G = old.\phi_G$;

## 3.4   Performance Evaluation

This subsection presents a benchmark comparing the SWMP approach with triggered pull-based maintenance (cf., e.g., [137]). The benchmark description is preceded by a brief description of the SMWP prototype.

### 3.4.1   Overview of the SMWP Prototype

We implemented a prototype of the SMWP approach on top of Oracle9i DBS for storing and maintaining data in fragment classes as well as metadata describing frag-mentation schemas and page schemas. The prototype provides a generic framework in that it can be used with arbitrary relations, fragmentation schemas, and page schemas. Currently pages are stored in the file system, however the modular architecture allows to change the page storage system, e.g., using an XML database. Generated pages can be viewed with an arbitrary Web-browser that is capable of formatting XML documents by the means of CSS [146] or XSLT [151]. Processing XML is carried out by Java, which is loaded into and executed inside the DBS, and based solely on SAX [111] in order to minimize memory needs, which lead to performance problems when using DOM [164] in an earlier prototype. Employed tools are Oracle's XDK [121] (its XML parser and XSQL) and Apache's Xalan [62], which is used for serializing documents from SAX event streams, because to the best of our knowledge XDK does not support this functionality.

A modular architecture allows to change the employed XML tools to more efficient ones as they become available. The prototype is available on the Web[2] for download.

Pages of a page class $P<K>$ are pre-generated as follows: By reading the page content schema (e.g., Figure 7 in [133]) the foundation fragment class $F<K>$ as well as the other mapped fragment classes are determined. For each fragment $f<k>$ of the foundation fragment class a page p¡k¿ is generated. The content of a page is generated by read-ing the fragments through SQL queries and transforming the queries' results into an XML document (e.g., see Figure 10 for a sample document) according to the Page Content DTD (see Figure 9 in [133]).

Active behavior for maintaining fragment classes and page classes as described by the algorithms in Subsection 3.2 is implemented using PL/SQL database triggers which propagate a modification of a fragment class to dependent fragment classes and affected page classes. An incremental modification of a page that is affected by a fragment modification is carried out by a SAX filter. The affected page is parsed by a SAX parser, which generates events for each recognized XML element. The filter listens to the event stream and passes events to Xalan's SAX Serializer [62], which constructs an XML document from its incoming event stream. If an event concerns an element that is to be modified, the filter modifies the event before passing it on. The original page is replaced with the newly constructed one after parsing, thus reflecting the fragment modification.

For benchmarking the triggered pull-based approach, we extended the prototype by a simple mechanism that invalidates "dirty" web pages by inspecting modifications on relations.

### 3.4.2 Benchmark Architecture

The benchmark setting consisted of 28 fragment classes, 10 page classes (each com-prising up to 7 fragment classes), and 19,600 pages with page sizes between 2kB and 500kB. The Oracle database operates on an Ultra-60 machine (UltraSPARC-II 360MHz processor, 512MB of RAM) running Solaris 7.

The architecture of our benchmark is depicted in Figure 3.4. A modi-fication of a relation in the database (i.e., inserting, updating, or deleting a tuple) causes corresponding modifications of affected pages. We bench-marked three approaches for maintaining pre-generated web pages (cf. [133], Section 4.3):

(1) Pre-generation of web pages, using for maintenance

    a) incremental push-based data delivery via fragments ("core SMWP")

---

[2]at `http://www.dke.jku.at/smwp/`

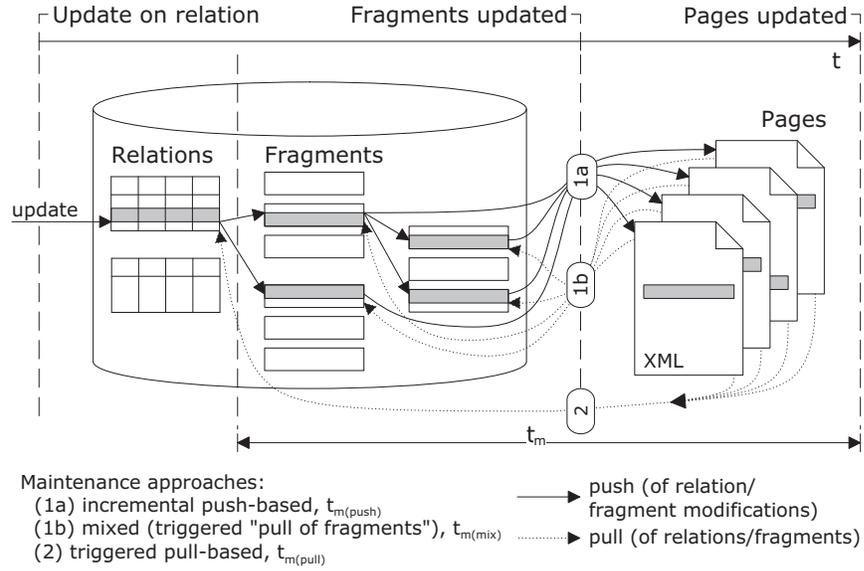

Figure 3.4: Benchmark Architecture

    b) a mixed approach in which triggers initiate complete regeneration of "dirty" web pages by pulling data from fragments (referred to as "triggered pull of fragments")

(2) triggered pull-based data delivery with complete regeneration of "dirty" web pages from base relations like [49], [99], and [137]. We have chosen to benchmark this approach with respect to SMWP, represented by options (1) a) and b). We did not benchmark the "on the fly generation" approach mentioned in [133] (Section 4.3), because its running time equals that one of the triggered pull-based approach. Both approaches differ only in whether page generation is triggered by a modification of a fragment or by a user's request respectively.

For comparing the running time of the different page maintenance approaches we are interested mainly in maintenance time $t_m$[3], which denotes the time needed for modifying affected pages once a base relation has been modified. Dependent on the employed maintenance approach, tm comprises the time needed for maintaining fragments (not effective in the case of triggered pull-based maintenance), the time needed for retrieving data from the database (not effective in the case of incremental push-based maintenance), and the time needed to modify affected pages accordingly. As $t_m$ does not include update time of relations, the benchmark times of the triggered pull-based approach basically coincide with the times needed for on the fly generation of a web page.

---

[3]An additional lower index put in brackets is used to distinguish between different maintenance approaches: "push" for incremental push-based, "mix" for mixed, and "pull" for triggered pull-based maintenance, e.g. $t_{m(pull)}$ or $t_{m(mixed)}$.

### 3.4.3  Benchmark Goal and Expectations

The goal of the benchmark was to compare the maintenance time of the core SMWP approach against other page maintenance approaches. We expected the core SMWP approach to outperform triggered pull-based maintenance or on the fly generation, as incremental propagation of changes to web pages via fragments does not involve complex queries and complete re-constructions of web pages. This expectation was nourished by Sindoni, who determined page generation to be the main cost factor in his approach, and Yagoub et al. [170], who experienced that retrieving web page content from the database takes up to 90% of the maintenance time.

The rationale behind considering the mixed approach was as follows: On the one hand, queries retrieving web page content are very simple (they just select all tuples from the according fragments) reducing query time compared to the triggered pull-based approach. On the other hand, parsing existing web pages for incremental modification might be more costly than simply re-generating the page from its fragments.

### 3.4.4  Benchmark Results

Figure 3.5 presents total maintenance times for a single page class with pages of different size . The page content schema of this page class is composed of 7 fragment classes which we experienced to be a typical number of fragment classes according to the common design rule "seven plus or minus two". To achieve representative results, each measurement point depicts the average of the maintenance times of a single page due to repeated changes to a base relation. Similar measurements with other page classes have been undertaken and show similar results: The graph (interpolation line) for incremental push-based maintenance re-mains identical, but the graphs for mixed and triggered pull-based maintenance vary in offset and gradient, although they never meet. Due to these variations, the intersection point of the graphs for incremental push-based and mixed maintenance varies between 5kB and 50kB, while the graphs for the triggered pull-based and the incremental push-based approach never meet.

Further measurements have shown that with decreasing database cache hit ratio the graphs for mixed and triggered pull-based maintenance increase in offset and gradient. Also, the intersection point between the graphs of their maintenance times moves, with respect to the graphs in Figure 3.5, towards larger page sizes (such as 250kB) with increased database load.

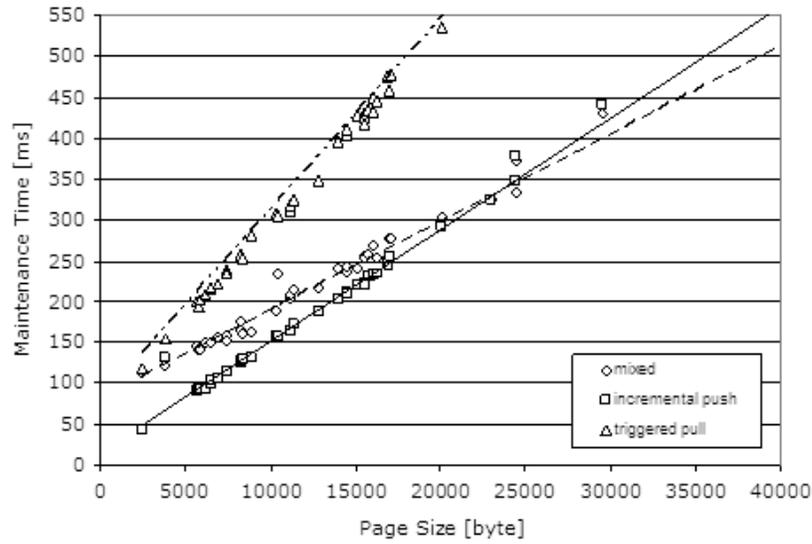

Figure 3.5: Maintenance Time

The benchmark results carried out for a single page class met our expectations. The incremental push-based (core SMWP) approach generally outperformed the other approaches. In the case of large web pages and high database cache hit ratios the mixed approach may be the better choice, because parsing a larger page by SAX can take longer than re-generating the page from fragments.

The advantage of incremental push-based and, also, mixed maintenance over triggered pull-based maintenance increases when more than one page is affected by a modification of a relation, since fragments need to be only modified once. With each additional affected page the graph for triggered pull-based maintenance moves, with respect to the graphs in Figure3.5, further up the y-axis than the graphs of the other approaches do.

A major goal of the SMWP approach was to come up with a declarative design architecture that simplifies the mapping between database entities and web pages and, thus, reduces design and maintenance complexity of pre-generated web pages. The competitive performance further justifies the use of the core SMWP approach. Should a particular application face performance problems with the core SMWP approach, one can flexibly switch to other page generation options as described in [133] (Section 4.3).

**Part II**

# Consistency within Data on the Web

# Chapter 4

# Active XML Schema (AXS)

## Contents

The chapter presents the basic Active XML Schema (AXS) approach. More advanced topics regarding the AXS approach are discussed in the following Chapters 5 and 6.

First, basic concepts of provided active behavior are presented in Section 4.1. They comprise event types, event classes, import and export of events, and rules. Second, advanced concepts of AXS are discussed in Section 4.2, comprising timestamping of events, scheduling events, composite and logical event classes, and exception rules. Third, the problem of defining an order on distributed events caused by possibly unsynchronized clocks in distributed environments and how it is dealt with in AXS is shown in Section 4.3. Fourth, Section 4.4 concludes the chapter by discussing briefly how active XML Schemas can be managed.

## 4.1   Active Extension of XML Schema

Adhering to an object-oriented approach, XML Schemas are enriched with additional components to provide for active and passive behavior of XML documents. While several schema languages have been proposed recently (e.g., [118, 89]) to replace XML 1.0 Document Type Definitions [147], we use XML Schema [157, 158].

An active XML Schema defines a distinguished global complex type, the *active document type*, that defines the structure and behavior of its instance documents. This is done by enriching a global complex type as defined by XML Schema with the definition and implementation of operations (i.e., passive behavior), with a set of definitions of event types and event classes, with a set of import and export statements defining the publication and subscription of event classes, and with a set of rules (i.e., active behavior).

To assure interoperability with legacy applications (i.e., applications that are not capable of interpreting active XML Schemas) the enrichments are transparent to those applications. Transparency is obtained by enriching complex type definitions via annotations that hold data defining active and passive behavior. XML schemas are instances of the XML Schema schema for XML Schemas (i.e., the metaschema [157]). To be able to validate an active XML Schema with a conventional XML Schema validator, this metaschema is enriched with metatype definitions for active and passive behavior, constituting the *Active XML Metaschema*. Like active XML schemas are instances of the active XML metaschema, active XML documents are instances of active XML schemas. They comprise, aside of static data, data representing events as well as subscriptions. These additions belong to a separate namespace, which is referenced by namespace-prefix act: in this chapter.

Along with the description of Active XML Schema concepts we repeatedly formalize the according part of the metaschema by an UML class diagram showing the contents of the dedicated UML package. The intention is to clarify details and to provide a concise representation of the metaschema. The datatypes of attributes are those defined by XML Schema Datatypes. For brevity we omit the namespace name when referring to a type, e.g., instead of xs:QName we write QName. The implementation of the metaschema in XML is discussed in Chapter 6.

In the examples for active XML Schemas throughout the chapter we use two notation to depict active XML Schemas: UML and XML. First, we use an extended UML [130] notation. This form of presentation is favored over object diagrams that model instances of the metaschema for conciseness. It is necessary to extend UML to model Active XML Schema concepts not originally present in UML. We introduce new compartments that an UML class can feature, and tag each compartment by a letter to reflect its semantics. While we do not use the compartments for structural attributes in the thesis, we tag the operation compartment by "O". The newly introduced event class compartment is tagged by "E", the exported event class compartment by "↑", the imported event class compartment by "↓", and the rule compartment by "R". Second, XML snippets show the serialization of the active XML Schema examples as part of XML Schema documents. In these, unprefixed elements are in the default namespace of the Active XML Metaschema[1], elements prefixed **actf** are in the namespace of the Active XML Framework[2] (cf. Section 6.6).

The metaschema for active document types is depicted in Figure 4.1, which shows package **axs**. It depicts active document type **ActiveDocTp** specifying definitions and implementations of operations by **psv::Interface** and **psv::Implementation**, event types by **evts::EventType**, event classes by **evts::EventClass**, and rules by **rule::Rule**. Regarding the import and export of event classes, **ixe::Proxy** is the only metaschema class of package **ixe** that is referenced in package **axs**, the other necessary classes are derived from **evts::EventType** and **evts::EventClass**. For detailed descriptions of these concepts and the metaschema packages **psv**, **evts**, **ixe**, **rule**, and **cle**, see the following sections. An active document type has a **name**, a **targetNamespace** similar to an XML schema, and import/include relationships to other active document types, which have the same semantics as import/include relationships in XML Schema. Moreover, class **NamespacePrefix** represents declared namespace prefixes that are used in defining the active document type, and **NativeCode** is an auxiliary datatype which describes an expression or source code in another, native language such as an XQuery expression or Java code. **NativeCode** is used in other packages.

In the following, we describe the basic behavioral components of Active XML Schema.

### 4.1.1 Passive Behavior

So far passive behavior for XML documents is specified procedurally, i.e. separately from the data, falling behind the prevailing object-oriented paradigm that integrates structure with behavior. This is the case with DOM, SAX, or XSLT [150]. However, XML databinding can be utilized to narrow the gap between definition of data and behavior. In particular Sun's

---

[1]which is `http://big.tuwien.ac.at/axs/metaschema/1.0`
[2]which is `http://big.tuwien.ac.at/axs/framework/1.0`

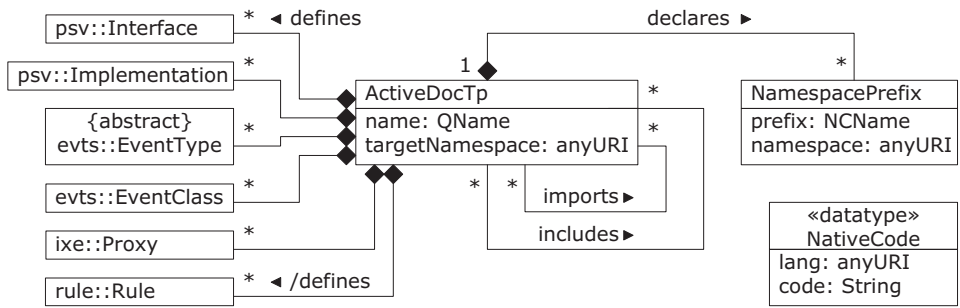

Figure 4.1: Metaschema for Active XML Schema (Package **axs**)

JAXB [139] can be used to define a mapping from XML to Java classes, behavior can be added by extending these classes.

Whereas XML databinding does not allow to integrate the definition of data and behavior, active XML Schemas define operations at XML document types in an object-oriented manner. They enrich XML types, which before described only structure, with passive behavior. Interfaces and implementations of operations are separated as in common practice in object oriented design.

In this chapter, implementations are illustrated by XSLT templates, which can access an operation's parameters via variables. They are, before being applied to an active XML document, embedded into a default stylesheet that performs a carbon copy (they override the default stylesheet's templates). The transformation's output document replaces the original active XML document, thus performing the modification as described by the operation's templates.

The metaschema for passive behavior is depicted in Figure 4.2, which shows package **psv**. An active document type may define several interfaces (class **Interface**) and implementations (class **Implementation**), where each implementation realizes a single interface, which is possibly defined by a different active document type. As in object-oriented programming languages, an interface defines a set of operations (class **Operation**), where each of them is described by its name, parameters (attribute **params** of datatype **ParamTp**), and return type. For every operation, up to two operation event types can be defined (see Subsection 4.1.2). Operations defined by interfaces have according implementations (class **OperationImpl**), an implementation is obliged to implement all operations defined by its interface (expressed by the constraint in the figure's bottom left).

⊙ *Example 34.* A job agency provides an active XML Schema defining active document type j:**JobAnnounce** and a document **academicJobs.xml** having that type, which comprises a list of current job offers (cf. Figure 4.3). A new job offer is announced by the invocation of operation **announce(j:Job)**, which adds a new job offer at the end of the list. The part of the active XML Schema document representing document type j:JobAnnounce that defines the interface of operation announce(j:Job) is depicted in snippet **1a**. The operation's implementation with XSLT is depicted in snippet **1b**. The type or object a snippet is part of, is represented by references to snippets in the extended UML model.

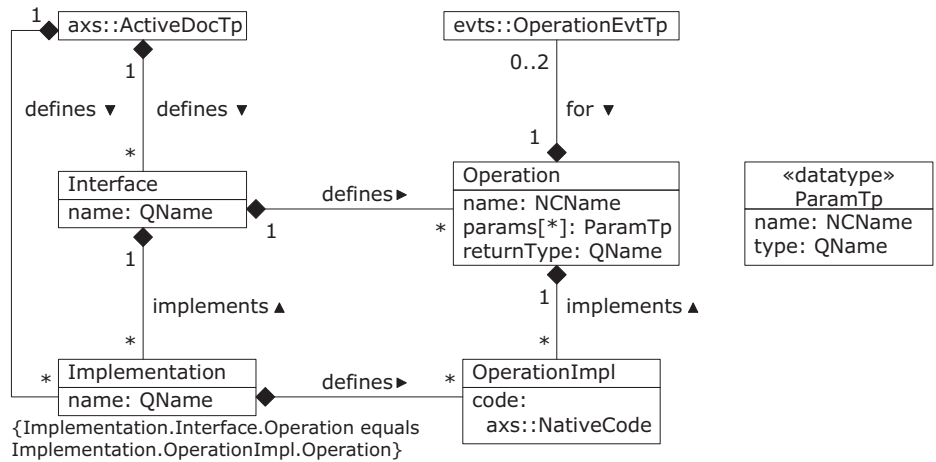

Figure 4.2: Metaschema for Passive Behavior (Package **psv**)

### 4.1.2 Event Types and Event Classes

Events are happenings of interest to a document. They are collected into event classes and are stored with documents as "first class XML elements" as any other XML element. Each event class has a member type that defines the structure of its member events.

Events can be distinguished into several *event categories*, which are: mutation events, operation events, calendar events, abstract events, and imported events.

- *Operation events* reflect the execution of an operation.

- *Mutation events* reflect changes to XML documents as according to the DOM Level 2 Event Module Specification. They are relevant in particular when a document is manipulated by means of legacy applications and not by predefined operations. Because the mutation events provided by the DOM Event Module may be too fine-grained fine-grained for directly defining rules on them, an approach has been developed to combine these events to so called composite mutation events, or short composite events (see Chapter 5).

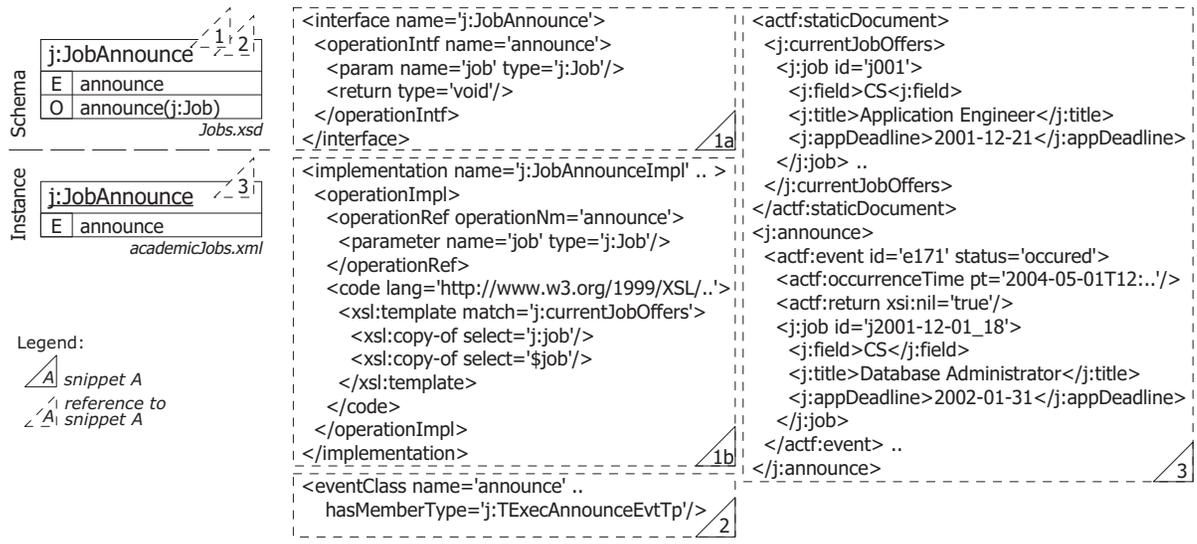

Figure 4.3: Operations and Event Classes at Schema and Instance Layer

- *Calendar events* constitute past, current, or future calendar entries.

- *Abstract events* comprise any other event explicitly raised or scheduled by users or rules.

- *Imported events* wrap remote events and are discussed in the next subsection.

Event types form a type hierarchy. The hierarchy at the metaschema layer (M2) is depicted in Figure 4.4 and is described in detail later in this section. Instances of the metaschema layer represent event types at the schema layer (M1). Instances of event types at M1 represent occurred events. Event types at M1 form a type hierarchy which is not shown for space limitations. Each event type at M1 defines various *parameters* reflecting the environmental setting in which it occurs. E.g., an operation event type will provide a parameter for each of the operation's parameters. The root event type defines the lowest common denominator of parameters (such as an identifier and an occurrence timestamp). For examples of event types at M1 see Subsection 6.3.3 and Section 6.6.

An event class is either a basic event class or a queried event class. The extension of a basic event class is defined by collecting every detected event of its member type. The extension of a queried event class is defined by a query over other event classes. A queried event class is either a composite event class or a logical event class which are discussed in Subsection 4.2.3.

Events are stored as "first class elements". As such they are part of a document and are thus persistent. This is different to most active database systems, whose events have a life span ending with the transaction in which they occurred. The extensions of all event classes of a document reflect the document's *event history*. Rules can query event classes and, thus, react according to the event history reflecting the document's past as well as the documents future as determined by scheduled events.

Event classes are modelled explicitly at the schema layer with document types and define the collections of events that are of interest at the instance layer. When an event class is defined with a document type, each document instance of that document type will maintain its extension of the event class.

The metaschema for event types and event classes is depicted in Figure 4.4, which shows package **evts**. An active document type defines several event types and event classes, represented by instances of classes **EventType** and **EventClass** respectively. Every event class has a name (attribute **name**), may be exported (indicated by attribute **exported**), and has an event type as its member type (association **hasMemberType**). An event class's name must be unique within all event classes of an active document type. Class **EventType** is the root of the type hierarchy of event types at the metaschema layer. Its name (attribute **name**) must be unique like the name of an event class. Its subclasses reflect the event categories mentioned above:

- Operation events are represented by the *operation event type*, Oper-ationEvtTp. For a single operation (class psv::Operation) up to two operation event types can be defined per active document type, one for events that occur before and one for events that occur after the execution of the operation.

- Mutation events are represented by the *mutation event type*, Muta-tionEvtTp. Events of a mutation event type occur in a certain part of the document defined by the event type's path type (attribute pathTp). Subclass PrimitiveMutEvtTp describes so called primitive mutation events, which reflect modifications of an XML document's contents, e.g., as defined by the DOM Event Module, while subclass CompositeMutEvtTp represents composite mutation events, which are detected according to the declarative specification of the accompanying composite event class. For details on composite events see Chapter 5, for details on composite event classes see Subsection 4.2.3.

- Calendar events are represented by the *calendar event type*, Calen-darEvtTp. Calendar events can be either periodical (class Periodic-CalEvtTp), meaning that they occur between a start time and an end time every given interval, or absolute (class AbsoluteCalEvtTp) mean-ing that the event occurs at a single point in time. Undefined calendar event types (class UndefinedCalEvtTp) are special in that their events are not specified at the schema layer. Instead, absolute events can be raised in event classes having this member type, e.g., by a rule's actions. Special control events, which again can be raised, e.g., by a rule's action, allow to initiate and cancel periodic events that occur in such event classes.

- Abstract events are represented by the *abstract event type*, Ab-stractEvtTp. The serialization of events of an abstract event type must accord to the schema defined in attribute instanceSpec which is expressed using the schema language identified by instanceSpec.lang.

- Finally, imported events are represented by the *imported event type*, ixe::ImportedEvtTp. For details see (the following) Subsection 4.1.3 and Figure 4.6 therein.

⊙ *Example 35.* Figure 4.3 shows the definition of event class announce with document type j:JobAnnounce, collecting all announce operation events representing invocations of operation announce (cf. snippet 2), and the ma-terialization of this event class in document instance acadamicJobs.xml (cf. snippet 3).

Storing occurred events with documents as first class XML elements causes the documents to grow in size with modifications to them. To control

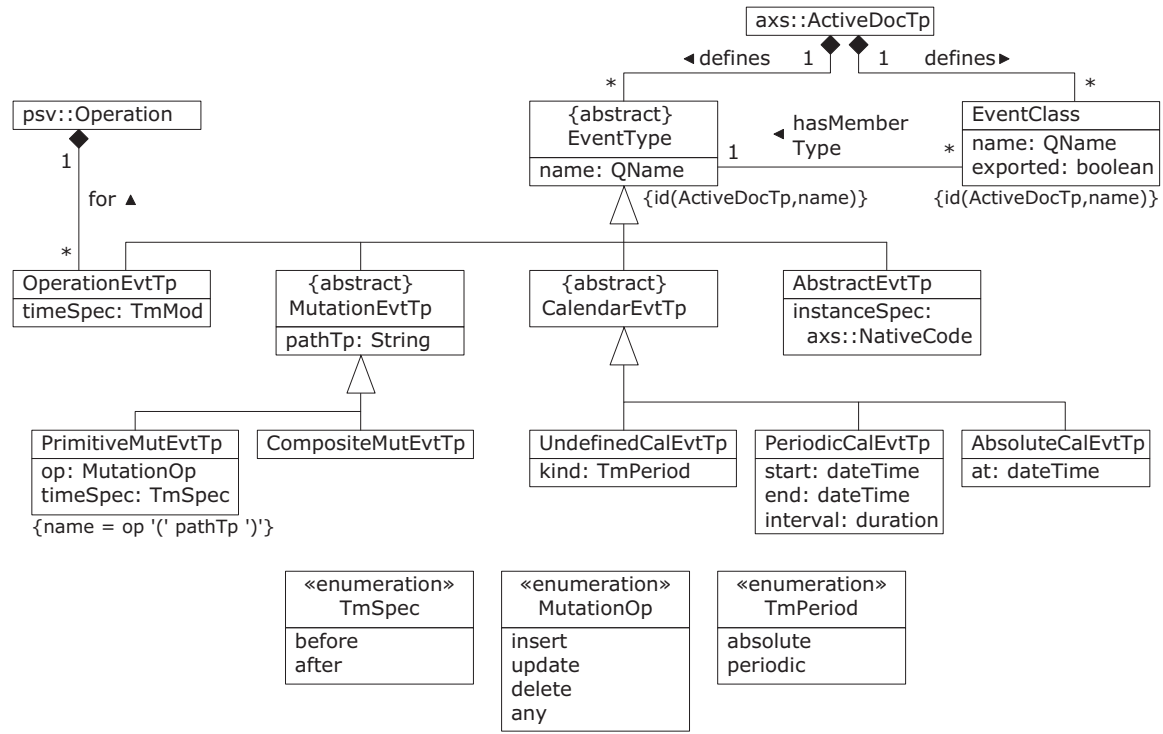

Figure 4.4: Metaschema for Event Types and Event Classes (Package evts)

a document's size, one can express size constraints on event classes and
the document as a whole, which are enforced by the reactive service the
document is stored at (see Subsection 4.4.1). If the constraints on event
classes are satisfied, but the one on the document is not, the reactive service
enforces the latter constraint by following some procedure to reduce the size
of event classes. A size constraint on an event class can be specified in
terms of maximum kilobytes, number of events, or age. The latter can only
be enforced if the location provides time service level tsl-2 or tsl-3. A size
constraint on a document can be specified in terms of maximum kilobytes.

### 4.1.3   Import and Export of Event Classes

Not only events that occur at a document locally may be of interest, but also
events that occur in another document. From a document's point of view,
one can distinguish between *local events* and *remote events*. A local event
happens at the document itself; a remote event happens at another, remote,
document. To be able to react to remote events locally, remote event classes
can be imported if they are exported by the document type of the remote
document. In such a case, a local event, referred to as imported event,
wraps a copy of the remote event. Replication ensures that the imported
event is available independently of the remote document, which is especially
of importance if the remote document is not under control of the owner of
the local document, a common setting in the Web environment.

Documents will often import several event classes from the same doc-
ument. Modelling import relationships between documents at the schema
level can support such a setting by separating the establishment of an im-
port relationship between document types from the definition of which event
classes are imported in that relationship. This separation is achieved by
defining a typed document *proxy* at the schema level that acts as an in-
termediary between the importing document type, which defines the proxy,
and a document of the proxy's type. A proxy's value specifies the document
the event classes are imported from, it is bound at the instance level. Thus,
to import an event class from another document the following steps are re-
quired: (1) at the schema level: (a) export the event class at the exporting
document type, (b) define a document proxy at the importing document
type, and (c) import the event class at the importing document type, and
(2) at the instance level: at the importing document, bind the document
proxy to the URI of a particular document,

⊙ *Example 36.* Figure 4.5 depicts an import scenario. As a courtesy to
its staff and students, a faculty posts relevant job offers supplied by a job
agency at its document of active document type u:Faculty. To provide others
access to newly announced jobs via an event class, the job agency exports
event class announce (cf. snippet 1). Active document type u:Faculty imports
this event class, using proxy jobSite to refer to the document from which the remote event class is imported (cf. snippet 2). Its value is bound in instance science.xml to document academicJobs.xml (cf. snippet 4). Announced job offers are now locally available within a faculty's page in the form of events.

Often a document needs to import event classes from several documents of the same document type, which corresponds to a "1:n" relationship in conceptual modelling terms, while we considered only "1:1" relationships so far. In such a case, a *set proxy* is defined at the schema level and bound to multiple URIs at the instance level. Imported event classes of the same kind are collected into an *event class family*. When querying event class families, member qualifiers are used for selecting specific event classes. An unqualified reference to an event class family refers to the union of all events of all family members.

⊙ *Example 37.*    A faculty comprises several departments, each one featuring its own document of type u:Department. A publication is added to such a document by invoking operation published(u:Pub). A faculty wishes to get notified whenever a publication is added to one of its departments' documents. Thus it imports event class family published, using set proxy depts to refer to its departments' documents, as depicted in Figure 4.10.

The metaschema for import and export of event classes is split over two figures. First, since any event class can be exported, the part of the metaschema dealing with the export of events is depicted in Figure 4.4. An event class is exported, if the flag EventClass.exported is set to true (aforementioned step 1.a). Second, the part of the metaschema that deals with the import of events is depicted in Figure 4.6. A proxy is defined by instantiating class Proxy (step 1.b). A proxy has a name, a type, which specifies whether it is a single or set proxy (enumeration ProxyTp), and acts as a placeholder for a specific document type. The import of an event class (step 1.c) is modelled by instantiating class ImportedEvtCs, which is a special event class. Its member type is an imported event type (class ImportedEvtTp), which wraps the remote event type. A filter expression (attribute filterExpr) allows to filter events based on their content. This can be used to minimize event traffic between documents, if the filter is evaluated at the publishing document, and/or to reduce the amount of events that are stored in the event history. The binding of a proxy to a concrete document takes place at the instance level and has thus no effects on the metaschema.

### 4.1.4   Rules

Active behavior is defined by *ECA rules* with a document type. Each rule is defined upon an event class, and consists of a condition and an action. If an event of the event class occurs, all rules defined on that event class are triggered. After a rule is triggered, its condition is evaluated, and if the condition applies, the action of the rule is executed.

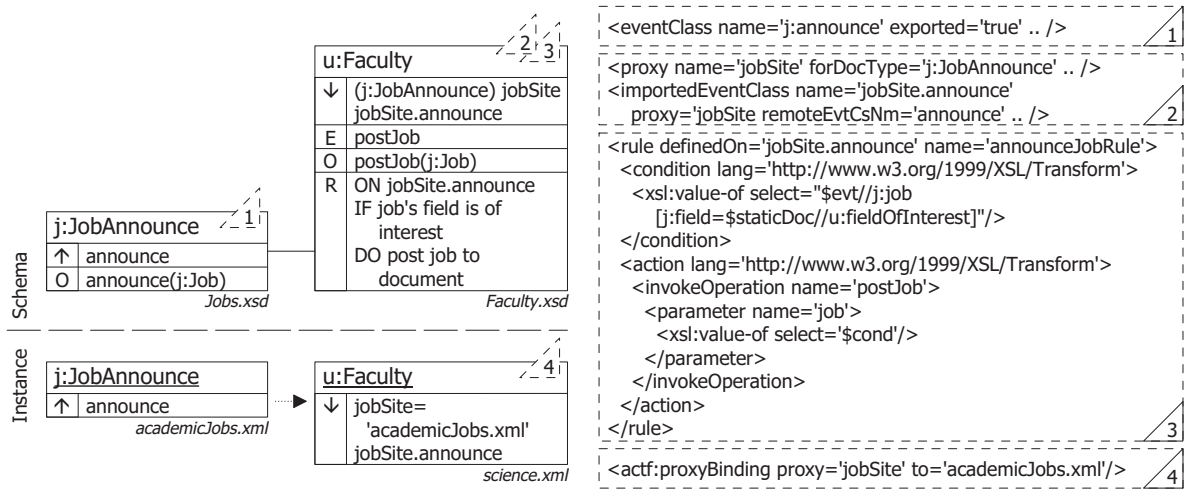

Figure 4.5: Export and Import of Event Classes and Rules

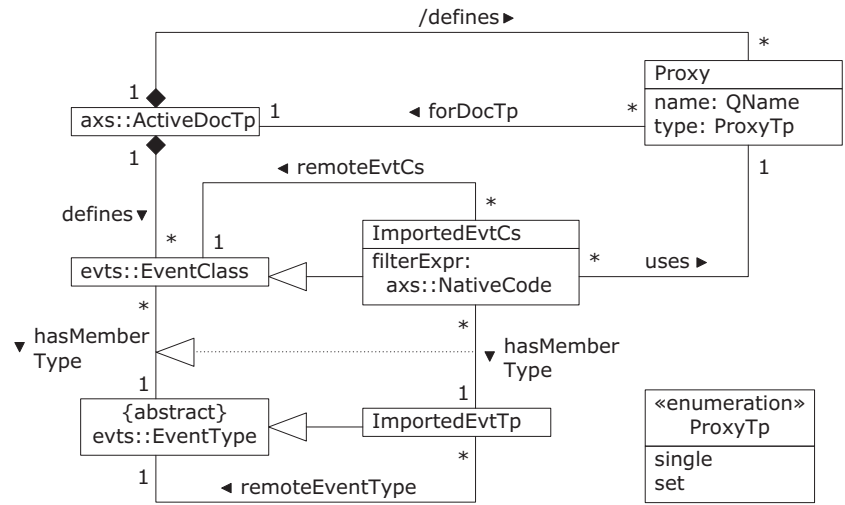

Figure 4.6: Metaschema for Im-/Export of Event Classes (Package ixe)

To ensure loosely coupling between documents, rules are restricted to have a *local scope*, which means that event class, condition, and action may refer only to the document with whose document type the rule is defined. This corresponds to the "Law of Demeter" [103], which suggests a restricted scope for type's operations in object-oriented design and is generally considered beneficial. If a rule shall be triggered upon events of a remote event class, the event class has to be imported. If a rule shall be triggered upon rule execution in remote documents, an event must be defined upon the rule's action, exported to and imported by the other document, which can react to this imported event by a local rule. Furthermore, conditions can only access data from the local document, in order to be completely independent from other documents, and actions of rules may only invoke operations that modify data of the local document.

Conditions are queries on the static data and the event history of a document. The latter two are available to the query as bindings **staticDoc** and **evtHistory** respectively. The event that triggered the rule is available as binding **evt**. Conditions apply when the query's result is not empty. We utilize XSLT for the definition of conditions because of the widespread support and knowledge of XSLT. Alternatively, XQuery can be used as well, if XQuery queries are translated to XSLT by an XQuery-to-XSLT translator [102].

Actions can invoke operations on the local document and raise or schedule events in a local event class (scheduling events is described in Subsec-tion 4.2.2). As is the case for conditions, the document's data, its event history, and the triggering event are available to the action as bindings. Moreover the result of the condition evaluation is available as binding cond.

Rules defined at the schema level with document types may need to react specifically depending on the document instance they are triggered at. Document-specific reactions can be achieved without introducing explicit rule parameters by specifying rule conditions that query the document's data the rule is triggered at. It is, however, reasonable from a design perspective, to group elements which are employed to implicitly parameterize rules into a separate part of the document.

The metaschema for rules is depicted in Figure 4.7, which shows package rule. A rule (class Rule) has a name and priority, and is defined upon a single event class. No two rules may have the same priority, rules with higher priority are executed before rules with lower priority. A rule comprises a condition (class Condition) and an action (class Action). Both are specified (attribute spec) using some language (identified by attribute language).

⊙ *Example 38.* Figure 4.5 shows the active document type for faculties, which defines rule announceJobRule on the imported event class announce (cf. snippet 3). Its condition tests whether the job's field is of interest to the specific faculty by querying elements fieldOfInterest, whose content is, e.g., 'CS' (computer science) and 'EE' (electrical engineering) for the science faculty science.xml. If the condition applies, the new job is posted to the faculty's document by invoking operation postJob(j:Job).

## 4.2  Advanced Concepts

In this section, we motivate and discuss advanced concepts of Active XML Schema: event timestamps, scheduling events, composite and logical events, and exception rules.

### 4.2.1  Event Timestamps

When an event occurs (at its *occurrence time*) that is stored in an exported event class, it is subsequently delivered to subscribing documents. Such a communication is characterized by two timestamps, the *publication time*, which records the time when the remote document publishes the event, and the *delivery time*, which records the time the event is delivered. These timestamps may be of different importance to different applications, like the time of the post mark (corresponding to publication time) is relevant to meet the deadline of a postal vote or the time of delivery to meet an application deadline.

To comply with these three notions of timestamps, every local event is described by an occurrence time while every imported event is described by all three timestamps. The occurrence time of an imported event reflects the time the event is stored in the imported event class opposed to the other two timestamps which describe the communication. An example given later shows how the distinction between publication time and delivery time can be used in an exception rule.

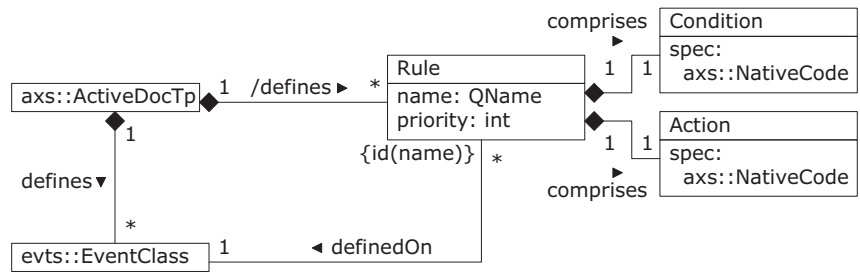

Figure 4.7: Metaschema for Rules (Package rule)

Timestamps are defined with event types at the schema layer, as already mentioned in Subsection 4.1.2. How this can be done is discussed in Chapter 6 in detail, which briefly describes how it has been done in Section 6.6. As will be shown in more detail in Subsection 4.3.2, a timestamp comprises several components, recording, e.g., a logical clock count and a physical clock count.

### 4.2.2   Scheduling Events

Comprehensive support for web content management requires not only to handle event occurrences that were caused by modifications to the document or were explicitly raised, but also to schedule events in the future. A typical application is to schedule an event in the future to trigger removal of then outdated content. With active XML Schemas, events can be scheduled by operations or rules. Such a *scheduled event* has an occurrence time in the future.

⊙ *Example 39.*   To remove a job offer from a faculty's document when its application deadline is over, rule scheduleJobRemovalRule is defined on event class postJob, which collects invocations of operation postJob(j:Job) (as depicted in Figure 4.8). The rule schedules the removal of a job offer by adding an event that is scheduled to occur when the application deadline ends to event class jobExpired. Rule removeJobRule defined on event class jobExpired removes an expired job offer, whose identifier was bound to the event, by invoking operation unpostJob(xs:ID).

### 4.2.3   Composite and Logical Events

Active database system provide *operational languages* for defining a composite event from other events [45, 48, 67, 70, 123, 173, 175] based on their occurrence time. E.g., sequence event C = A ; B occurs if event B occurs after event A has occurred. As events of types A and B may occur several times, the notion of *event context* has been introduced to indicate which of these occurrences are used to detect an occurrence of C. E.g., in the "recent" context the last occurrences of A and B are used, and in the "chronical" context the i-th occurrence of A is paired with the i-th occurrence of B. Various techniques based on Petri nets, state machines, and event graphs have been introduced to detect composite events incrementally. E.g., in the example of the above sequence event, an occurrence of A would start the incremental detection of an occurrence of composite event C.

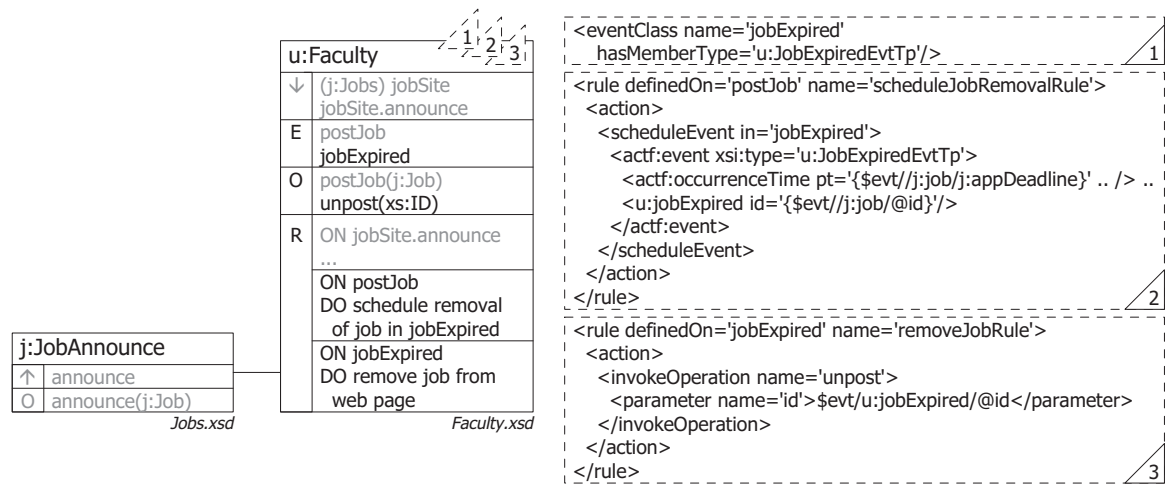

Figure 4.8: Scheduling an Event

Experience in modelling business rules at the conceptual level has shown that a declarative approach for defining composite events is better suited for end users (i.e., schema designers) than an operational approach, which is more appropriate for database internal usage. Therefore, active XML Schemas provide the functionality of composite events being defined by querying a document's event history, e.g., using XQuery. Such events are called *logical events*. Different to operational languages, which detect composite events incrementally by observing events as they occur from the past till "now", logical events query from the perspective of "now" the past history of events and may also query future scheduled events. Furthermore, this approach allows to waive consumption policies, whose effect is sometimes hard to grasp for end users, especially when several contexts need to be combined. The query approach can also cope very easily with multiple timestamps provided by active XML Schemas, avoiding a difficult adaption of the notion of "event context" to multiple timestamps. Finally, the goal of having a "light-weight" approach in the web setting suggests to reuse already available query languages rather than providing an additional composite event detector and cluttering web documents with many unfinished event detections.

Logical events are defined by logical event classes, which specify a query over the document's event history. A logical event class can be *event preserving*, in which case it selects members from other event classes, or *event generating*, in which case its member events are newly built from other events. The specification of a logical event class identifies next to its member type a set of terminating event classes and participating event classes. A logical event occurs if a member of a terminating event class occurs and the stated query over terminating and participating event classes is satisfied.

The metaschema for logical event classes is depicted in Figure 4.9, which shows package **loev**. The metaschema class representing logical event classes is **LogicalEvtCs**. It is derived from the abstract class **QueriedEvtCs**, which represents event classes that collect events that are derived from the event history by defining a query (attribute **querySpec**) in some language (attribute **language**) over the document's event history. **LogicalEvtCs** adds two associations, one for terminating event classes and one for participating event classes. For each of these classes, an alias can be specified that can be used in the query.

Depending on whether the logical event class is event preserving or generating, it may have different member types. If it is preserving, its member type must be one of the member types out of the union of terminating and participating event classes. In case it is event generating, its member type can be an arbitrary subclass of **evts::EventType** except imported event types and mutation event types. Events of the former are "generated" by importing it from remote documents while events of the latter are "generated" by composite mutation event classes, as described later in this section.

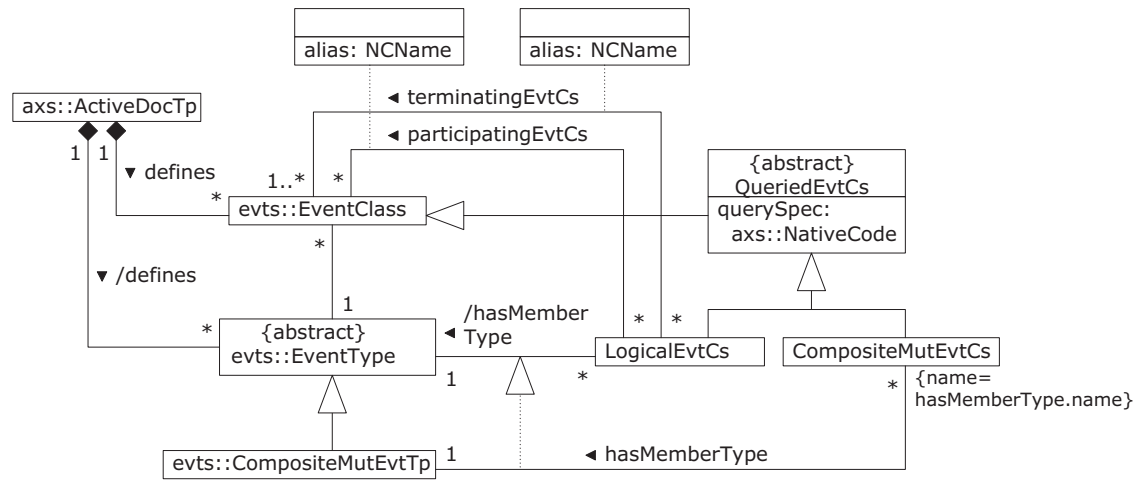

Figure 4.9: Metaschema for Composite and Logical Events (Package cle)

⊙ *Example 40.* A faculty publishes the active researcher status of its employed researchers at the beginning of each year, which depends on a researcher's number of publications he/she published the year before. Each publication is reflected by a **published** event that is imported by the faculty's document from its departments' documents. The active researcher status is determined by operation **publishActResStatus(u:Researcher, xs:gYear)** for a given researcher and year. Due to failures, a **published** event may be delivered after the active researcher status has already been determined, although the event was published in time (i.e., in the year before the current year). Logical event class **pubDeliveredLate** collects **published** events that should have been considered when determining a researcher's active researcher status. This is the case if a **published** event (terminating event) is delivered and a **publishActResStatus** event (participating event) exists such that (1) the **publishActResStatus** event occurs before the **published** event is delivered – notice that such a comparison is necessary, since in general a participating event class may also contain future events –, and (2) the **published** event occurred the year before and concerns a paper published by a researcher the year before, for whom the **publishActResStatus** operation has already been executed for that year (cf. Figure 4.10).

Defining composite events by querying the event history, however, is only an appropriate means of definition if the meaning of queried events is near to the application domain, i.e., if terminating and participating event classes collect other events than mutation events, such as operation and calendar events. To support the detection of composite events from mutation events, which are especially of interest when they occur in the course of manipulations by a legacy application, a dedicated operational language has been developed, which is presented in Chapter 5. Such events are called *composite mutation events* (or short *composite events*).

The construction of composite mutation events is restricted so that it does not result in the undesired properties mentioned before (in the course of motivating logical events). First, the complexity of choosing the right consumption policies, which are also referred to as contexts, does not apply, because the operational language presented in this thesis introduces a new context which is sufficient for most usage scenarios. Second, an adaption of contexts to multiple timestamps is not an issue, since the presented operational language is intended to compose events that occur within the same document. Finally, documents are not cluttered with unfinished event detections by waiving them when the manipulation of a document ends. Because the operational language is intended to compose mutation events only, the behavior described in the latter two arguments is reasonable.

The metaschema for composite event classes is depicted in Figure 4.9. An event class for composite mutation events is represented by an instance

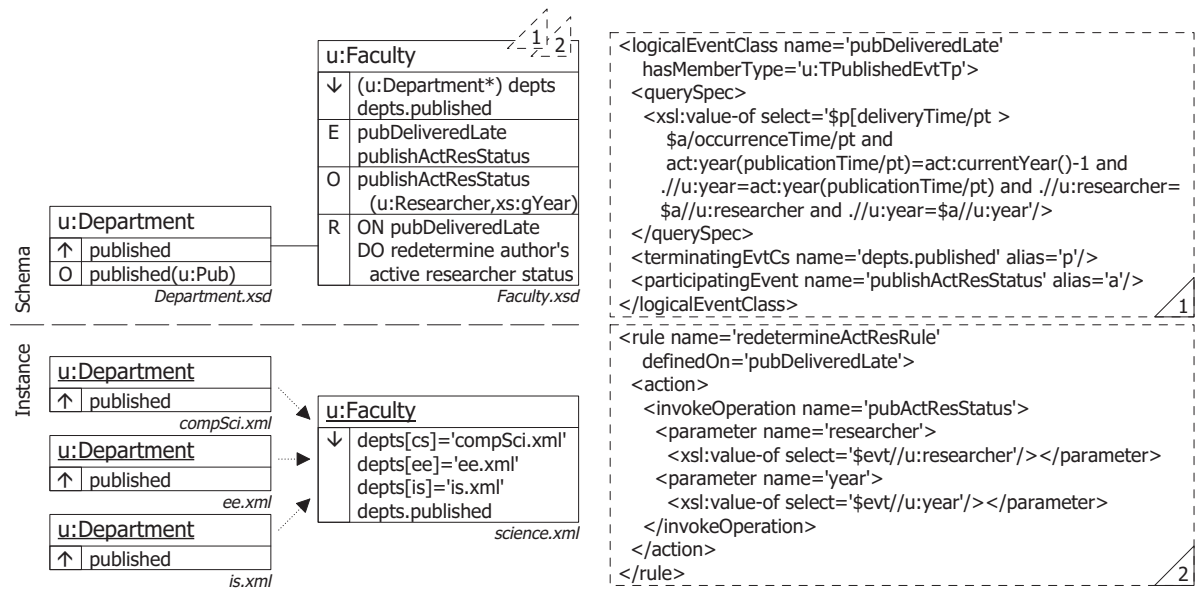

Figure 4.10: Logical Event Class and Exception Rule

of the metaschema's CompositeMutEvtCs class, the language attribute holds
the namespache of the operational language[3] and the querySpec attributes
defines the according event expression. Terminating and participating event
classes could be derived from the event expression, but are not since this
information is not explicitly required in the metaschema. The event class's
member type is a mutation event type (evts::CompositeMutEvtTp), refining
the association hasMemberType inherited from class QueriedEvtCs.

### 4.2.4   Exception Rules

Situations that do not constitute "standard" situations during deployment
of active XML Schemas may occur. They originate in effects of an active
XML Schemas' environment. In particular extraordinary behavior in com-
municating events between documents, such as the delayed delivery of a
subscribed event, builds the ground for these situations. Exceptional sit-
uations can be anticipated in schema design and should be dealt with by
exception rules, providing a meaningful design distinction. No additional
concepts are necessary to model exception rules. They react to specific log-
ical events, which determine exceptional situations, and trigger necessary
corrections.

⊙ *Example 41.*   Figure 4.10 shows event class pubDeliveredLate, which
collects exceptional situations, i.e. published events that are delivered late.
Exception rule redetermineActResRule (cf. snippet 2) defined on pubDeliv-
eredLate reacts to the late delivery by redetermining the researcher's active
researcher status by invoking operation publishActResStatus.

## 4.3   Distributed Events

AXS as a distributed system is inherently different from centralized systems,
in particular, two characteristics of distributed systems affect event detec-
tion. First, each location has its own clock, potentially being incompatible
with other clocks in the system or showing a drift. Second, messages sent
over a network can be delayed depending on transmission behavior of the
sender, receiver, and the network itself. In AXS these two characteristics
make it difficult to determine whether an event $e_i$ that occurred at location
$loc_1$, denoted as $e_{1i}$, occurred before or after event another $e_{2j}$.

This section describes the kinds of orders that are distinguished in dis-
tributed systems in general (Subsection 4.3.1), the infrastructure that sup-
ports event ordering in AXS (Subsection 4.3.2), and the orders that can
be established between events in AXS and the conditions that must apply
(Subsection 4.3.3).

---

[3]which is http://www.big.tuwien.ac.at/research/composite-events for the oper-
ational language presented in this thesis (cf. Chapter 5)

### 4.3.1 Causal and Temporal Order

In AXS it may be necessary to order events when testing a rule's condition, determining events stored in a logical event class, or forming composite events. Remember, only those events can be accessed thus ordered by queries that are contained in a document's event history. To make events that occur remotely available in a document's event history additional to local events, AXS provides for the import of event classes from remote documents. When an event occurs in a remote event class that is imported into a local document, the event that occurred remotely, the so-called remote event, is sent via a message to the local document. There it is wrapped in a so-called imported event which provides timestamps reflecting when the message was sent (publication time), when it was received (delivery time), and when it was stored in the document (occurrence time).

⊙ *Example 42.* An exemplary situation of event occurrences and their exchange between documents is depicted in Figure 4.11 as a time diagram, where time advances from left to right. Each horizontal line represents a document stored at a different location on the Web and each arrow represents a sent message, i.e., the import of a remote event. For instance, event $e_{12}$ is imported by document $d_2$ and is available therein as imported event $e_{22}$, which wraps the remote event $e_{12}$.

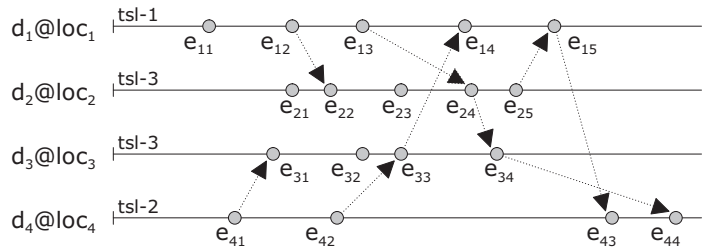

Figure 4.11: Time Diagram of Distributed Events

Between events occurring at different locations in a distributed environment only a *partial order* can be established, as will be described later in more detail. Thus it may be undecidable for two events $e_{1i}$ and $e_{2j}$, which occur at different locations, whether $e_{1i}$ occurred before $e_{2j}$ or vice versa. In case the order is undecidable, the events are said to have occurred concurrently, which is denoted as $e_{1i} \parallel e_{2j}$. Otherwise either $e_{1i}$ happened before $e_{2j}$, denoted as $e_{1i} < e_{2j}$, or $e_{1i}$ happened after $e_{2j}$, denoted as $e_{1i} > e_{2j}$. Two events can only be determined to have occurred at the same time if they have occurred at the same location, denoted as $e_{1i} = e_{1j}$. Between events occurring at the same location a *total order* can be established, i.e., concurrent events always occur at different locations.

Events occurring at different locations can be ordered based on causality or time. They are characterized as follows:

- *Causal order* reflects whether an event $e_{mi}$ may have causally affected the occurrence of another event $e_{nj}$, which is the case if knowledge about $e_{mi}$ is available in the document at location $loc_n$ before $e_{ni}$ occurs therein.

- *Temporal order* assumes a physical clock at each location with which the occurrence time of events is measured. Temporal order reflects whether an event $e_{mi}$ occurred before or after another event $e_{nj}$ solely by comparing their occurrence time. Temporal order assumes the clocks at distributed locations to be synchronized.

Lamport defines *causal order* using logical clocks in [100]. A logical clock means a counter that is incremented with every event occurrence. He defines a "happened before" relation between events, denoted as "$\rightarrow$". The relation satisfies the following conditions[4]: (1) If $e_{mi}$ and $e_{nj}$ are events that occur at the same location, i.e., $m = n$ and $e_{mi}$ occurs before $e_{nj}$, then $e_{mi} \rightarrow e_{nj}$. (2) If $e_{nj}$ is the imported event that encapsulates the remote event $e_{mi}$, then $e_{mi} \rightarrow e_{nj}$. (3) If $e_{mi} \rightarrow e_{nj} \wedge e_{nj} \rightarrow e_{ok}$ then $e_{mi} \rightarrow e_{ok}$. Two distinct events $e_{mi}$ and $e_{nj}$ occur concurrently if $\neg(e_{mi} \rightarrow e_{nj} \vee e_{nj} \rightarrow e_{mi})$. The happened before relation reflects causal order, i.e., $e_{mi} \rightarrow e_{nj}$ reflects that $e_{mi}$ may have causally affected $e_{nj}$. Viewed differently, when looking at Figure 4.11, an event happens before another if there exists a path from the former to the latter.

⊙ *Example 43.*      For events at different locations there must exist a path between the events to determine a causal order between them, e.g., $e_{12} \rightarrow e_{22}$, $e_{23} \rightarrow e_{14}$ (since $e_{23} \rightarrow e_{24} \wedge e_{24} \rightarrow e_{14}$), and $e_{11} \rightarrow e_{44}$. If a path does not exist, the events happen concurrently, e.g., $e_{11} \parallel e_{21}$, $e_{21} \parallel e_{33}$, and $e_{41} \parallel e_{25}$. Two events occurring at the same location never occur concurrently. Looking at Figure 4.11, e.g., this applies for $e_{11} \rightarrow e_{12}$ and $e_{21} \rightarrow e_{25}$.

Furthermore, Lamport derives conditions on the local logical clocks from the happened before relation. In [100] it is stated that the strongest reasonable condition that can be derived from $e_{mi} \rightarrow e_{nj}$ is that $e_{mi}$ should happen at an earlier time than $e_{nj}$, i.e., if $e_{mi} \rightarrow e_{nj}$ then $e_{mi}.occTime < e_{nj}.occTime$ should apply (where $e_{mi}.occTime$ denotes the occurrence time of event $e_{mi}$ as observed by the local logical clock at location $loc_m$). To satisfy this condition, logical clocks at different sites are synchronized in the course of exchanging messages between them. For details see Example 44. Lamport's main motivation for the clock condition is for introducing a relation that establishes a total order among all events that occur distributed.

[4]adapted from [100], p. 559

While a total order is not of particular interest in AXS, a side effect of the clock condition which is of interest is that it allows for checking the possibility of causal order simply by comparing occurrence times since if $\neg(e_{mi}.occTime < e_{nj}.occTime)$ then $\neg(e_{mi} \rightarrow e_{nj})$ (see Subsection 4.3.3 for details).

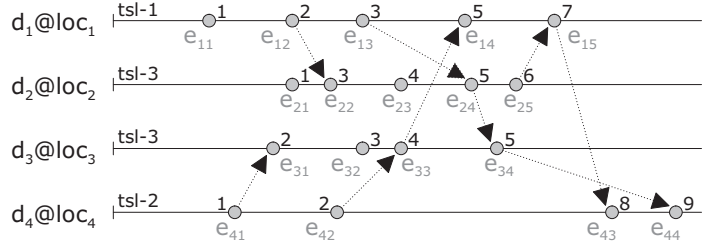

Figure 4.12: Time Diagram with Exemplary Timestamps According to Lamport

⊙ *Example 44.* Figure 4.12 exemplifies Lamport's clock condition. Since $e_{12} \rightarrow e_{22}$, $e_{12}.occTime < e_{22}.occTime$ must apply, thus in course of the import of $e_{12}$ as $e_{22}$ and storing it in $d_2$ the clock at $loc_2$ is increased from 1 to 3. Analogously, the import of $e_{14}$ at $loc_4$ as $e_{43}$ causes the clock at location 4 to be increased from 2 to 8 upon storing it in $d_4$.

The 2g-precedence model proposed by [96] defines a *temporal order* on distributed events. The model assumes synchronized physical clocks at each location, whereby any two local clocks may have a maximum deviation of $\pi$. A global occurrence time can then be determined by reducing the granularity of every local clock to the global clock granularity $g_g$, whereby $g_g > \pi$ must apply. A clock's granularity refers to the duration of a single clock tick. Two events that occur at different sites at the same time thus have timestamps that differ at maximum one clock tick from each other, measured by global time. Thus two events that are $\geq 2$ clock ticks apart can be temporally ordered.

In the 2g-precedence model, the global occurrence time of an event upon which an order is defined is determined as follows. Assuming that the reference clock's granularity $g_z$ is 1 millisecond (for presentation purposes), precision $\pi$, which is the maximum difference between two corresponding ticks of any two local clocks as observed by the reference clock, is measured in milliseconds too. For all locations a maximum granularity of local time $g_l$ can be determined, $g_l = Max(\{g_{l_i}\})$ (where $g_{l_i}$ denotes local clock granularity at location $i$). Since global clock granularity $g_g$ must be greater than both $g_l$ and $\pi$, $g_g = n * g_l$ such that $g_g > \pi$ and $g_g - g_l < \pi$ (if $g_l < \pi$). The global occurrence time of a remote event is its local occurrence time divided by granularity of global time $g_g$.

The 2g-precedence model was adapted to distributed active systems in closed networks by [136, 172]. It is employed by various approaches, such as EVE [71] which provides for event-driven workflow execution. It has been refined by [104] in that the network time protocol (NTP) and global positioning system (GPS) time servers have been used to guarantee local clocks to be synchronized with global reference time. They define groups of locations (called strata) where each group's clock shows a defined maximum deviation from the reference clock (called accuracy interval). The farer a group is located from a GPS time server, the larger is its accuracy interval. By introducing stratas, the authors claim to have modified the 2g-precedence model so that it can be used in loosely coupled distributed systems like the Web, for which they claim the 2g-precedence model does not scale.

Causal and temporal order differ in focusing on orthogonal aspects of ordering, being either causality or time. Regarding their weaknesses, causal order lacks expressing causal relationships that are established through other channels than the observed message passing and is less robust to long network partitionings while temporal order lacks causal relationships. Both orders have a notion of time and in both orders an event occurs before another one if its timestamp is smaller than the other one's. Note, however, that in causal order this is a necessary conditions while in temporal order this is a sufficient condition. Temporal order overcomes causal order's limitation of not representing real world time.

### 4.3.2   Time Service Levels

To provide maximum flexibility, AXS provides for both causal ordering and temporal ordering of events. This is in-line with the reasonable assumption that not all clocks at all locations in a Web environment can be synchronized and in-line with the expectation that not every location features a physical clock. Providing for both kinds of order is achieved by introducing *time service levels* that identify the kind of clock by which the occurrence time of an event is measured and thereby allow to determine the kind of order that can be established between any two events. A location determines occurrence times of events according to a single time service level. The levels differ in the kind of clock they feature:

- Level tsl-1: logical clock constrained by Lamport's clock condition (for establishing causal order);

- Level tsl-2: physical clock without any synchronization (for causal order);

- Level tsl-3: synchronized physical clock with a guaranteed maximum deviation from a reference clock (for temporal order).

Every time service level subsumes the functionality of clocks provided by lower levels. Thus, a tsl-2 clock provides for unsynchronized physical time and logical time according to tsl-1, and a tsl- 3 clock provides for synchronized physical time and logical time according to tsl-1. As one can easily see, a logical clock is provided by every time service level. The physical clocks of tsl-2 and tsl-3 measure time with a certain granularity, depending on the location's hardware and software. AXS defines 1 ms as the reference time's granularity $g_z$, because it is assumed that a smaller granularity in an internet setting is not reasonable and because AXS is not intended for applications that are in need of a higher granularity. Moreover many operating systems measure time in units of tens of milliseconds[5]. Thus, opposite to [136, 172], granularity of local time $g_l$ may be larger than granularity of reference time $g_z$.

AXS provides for defining groups of synchronized clocks because it cannot be assumed that all tsl-3 clocks are synchronized within a single precision $\pi$, or $\pi$ might become large otherwise. Within a group $r_i$ any two clocks have a maximum deviation of $\pi_{r_i}$ from each other as observed by the group's reference clock $g_{z_{r_i}}$. Thus two events that occur at different tsl-3 locations can be ordered temporally only if the locations they occurred at belong to the same group, otherwise they are ordered based on causality.

Grouping is favored over accuracy intervals as proposed in [104], because it provides for autonomously creating and maintaining groups within controlled environments on the Web. For a brief description of [104] see Subsection 4.3.1. Moreover, it seems unlikely that all tsl-3 clocks on the Web implement the protocol proposed in [104]. AXS can be extended, however, to allow a group to use the protocol described in [104], thereby separating the group into stratas with defined accuracy intervals and ordering events based on these. This would avoid the drawback of grouping that a temporal order cannot be established between events occurring at locations that are not member of the same group.

An event's *timestamp* records clock counts of all clocks the location the event occurs at provides. Having provided a logical clock by all time service levels, an event's occurrence timestamp, denoted as $e.occTime$, records the count of the logical clock, denoted as $e.occTime.lt$. Moreover, if the level is tsl-2 or tsl-3, the count of the physical clock is recorded as well, denoted as $e.occTime.pt$. From the perspective of ordering events that occur at different locations, ordering using tsl-2 equals tsl-1 because in both cases only causal order can be determined using the time stamps' logical clock counts. From the perspective of semantic expressiveness, however, tsl-2 differs significantly from tsl-1 in that the time between two events occurring at the same site can be determined, affecting the expressiveness of timestamps. Finally, an event's timestamp records the identifier of the location it occurred at, the time service level provided by the location, and the tsl-3 groups the location was member of denoted as $e.occTime.pid$, $e.occTime.tsl$, and $e.occTime.gids$ respectively.

---

[5]E.g., cf. `http://java.sun.com/j2se/1.4.2/docs/api/java/lang/System.html#currentTimeMillis()`

### 4.3.3   Ordering Distributed Events

Events are ordered differently depending on the locations they occur at and the time service levels provided by these locations. If two events occur at the same location, they can be ordered totally establishing a temporal order, otherwise they are ordered partially. In the latter case, i.e., if two events occur at different locations, a temporal order can be established by default if both time service levels are tsl-3, otherwise a causal order is established. The default order is established if operators $=$, $>$, $<$, and $\parallel$ are used when comparing events. In addition to these operators, subscripted variants exist that can be used to enforce causal order or temporal order (denoted by subscripts $c$ and $t$ respectively, e.g., $<_c$ and $<_t$). If one tries to enforce a temporal order, but at least one time service level is not tsl-3, the operator falls back and determines causal order.

First, two events $e_{mi}$ and $e_{nj}$ that occurred *at the same location*, i.e., where $m = n$, are ordered as follows:

- if $e_{mi}.tsl \neq$ tsl-3

  – $e_{mi} <_c e_{nj}$ iff $e_{mi}.occTime.lt < e_{nj}.occTime.lt$
  – $e_{mi} >_c e_{nj}$ iff $e_{mi}.occTime.lt > e_{nj}.occTime.lt$
  – $e_{mi} =_c e_{nj}$ iff $e_{mi}.occTime.lt = e_{nj}.occTime.lt$

- if $e_{mi}.tsl =$ tsl-3

  – $e_{mi} <_t e_{mj}$ iff $e_{mi}.occTime.pt < e_{mj}.occTime.pt$
  – $e_{mi} >_t e_{mj}$ iff $e_{mi}.occTime.pt > e_{mj}.occTime.pt$
  – $e_{mi} =_t e_{mj}$ iff $e_{mi}.occTime.pt = e_{mj}.occTime.pt$

Second, two events $e_{mi}$ and $e_{nj}$ that occurred *at different locations*, i.e., where $m \neq n$, are ordered as follows:

- if $e_{mi}.tsl \neq$ tsl-3 $\vee$ $e_{nj}.tsl \neq$ tsl-3 $\vee$ $e_{mi}.occTime.gids \cap e_{nj}.occTime.gids = \emptyset$

  – $e_{mi} <_c e_{nj}$ iff $e_{mi}.occTime.lt < e_{nj}.occTime.lt \wedge e_{mi} \rightarrow e_{nj}$[6]

---

[6]Note that the first condition ($e_{mi}.occTime.lt < e_{nj}.occTime.lt$) is a necessary condition while the second condition ($e_{mi} \rightarrow e_{nj}$) is a sufficient one. Thus an evaluation of the first condition may be detected obsolete if the second evaluates to true. However, it is expected that testing the first condition first pays off because it can be evaluated more easily and may make evaluating the second condition obsolete if it is false. The same applies for the next sub-bullet.

- $e_{mi} >_c e_{nj}$ iff $e_{nj}.occTime.lt < e_{mi}.occTime.lt \land e_{nj} \rightarrow e_{mi}$

- $e_{mi} \parallel_c e_{nj}$ iff $\neg(e_{mi} <_c e_{nj} \lor e_{mi} >_c e_{nj})$

- if $e_{mi}.tsl = \mathsf{tsl\text{-}3} \land e_{nj}.tsl = \mathsf{tsl\text{-}3} \land e_{mi}.occTime.gids \cap e_{nj}.occTime.gids \neq \emptyset$

  - $e_{mi} <_t e_{nj}$ iff $\lfloor e_{nj}.occTime.pt/g_g \rfloor - \lfloor e_{mi}.occTime.pt/g_g \rfloor \geq 2 * g_g$

  - $e_{mi} >_t e_{nj}$ iff $\lfloor e_{mi}.occTime.pt/g_g \rfloor - \lfloor e_{nj}.occTime.pt/g_g \rfloor \geq 2 * g_g$

  - $e_{mi} \parallel_t e_{nj}$ iff $\neg(e_{mi} <_t e_{nj} \lor e_{mi} >_t e_{nj})$.

Causal order between two events determined from the view of a document may differ from causal orders determined from other views. This is caused by using only events that are stored locally in a document to determine causal order between distributed events, which is in-line with the Web's loose coupling and in-line with redundantly storing events by importing them as presented in Section 4.1.3. For example, Figure 4.13 shows the events stored at document $d_3$ and $d_4$. Because the set of events a document stores, i.e., the set of events that occurred locally or were imported, is a subset of all globally occurring events, it may be the case that two events have a different order (a) from a document's view than they have from the global view, or (b) between two documents' views. For better results, all events stored in the documents at a given location could be used. This limitation does not apply for temporal order.

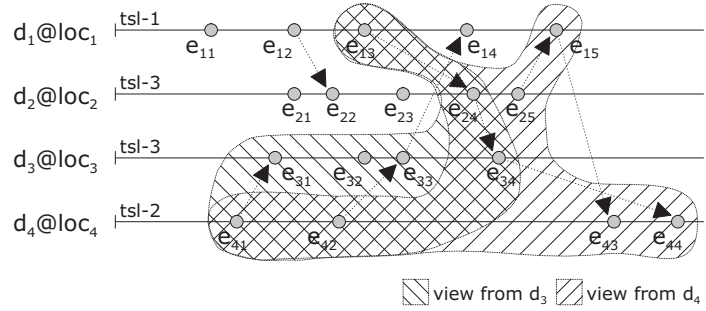

Figure 4.13: Different Views on Distributed Events

⊙ *Example 45.* The following examples show different causal orders depending on the point of view (see Figure 4.13): (1) the global view and the local view from $d_3$ equal each other in $e_{13} \parallel_c e_{42}$, but $e_{33} <_c e_{34}$ which are the respective imported events wrapping $e_{13}$ and $e_{42}$, and $e_{41} <_c e_{42}$; (2) the global view and a local view differ in $e_{42} \parallel_c e_{15}$ from the view of $d_4$, but $e_{42} <_c e_{15}$ from the global view; (3) two local views differ in $e_{41} \parallel_c e_{34}$ from the view of $d_4$, but $e_{41} <_c e_{34}$ from the view of $d_3$. The difference between causal and temporal order is exemplified when ordering $e_{25}$ and $e_{34}$ which is $e_{34} \parallel_c e_{25}$, but $e_{34} <_t e_{25}$.

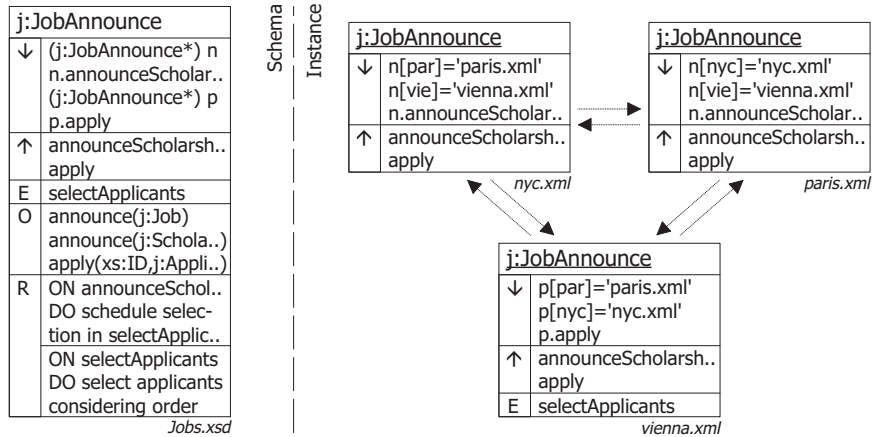

Figure 4.14: Distributed Events in AXS

⊙ *Example 46.*    An international corporation uses an extended version of active document type j:JobAnnounce to announce internal job offers and scholarships for training programs at their subsidiaries in New York, Paris, and Vienna (as depicted in Figure 4.14, where event classes without any events are not shown for presentation purposes). Offers for jobs and scholarships are published at the headquarters in Vienna (vienna.xml) and replicated in the other subsidiaries (paris.xml and nyc.xml) by importing event class announceScholarship. An employee can apply for a job or scholarship by locally invoking apply(xs:ID, j:Application) through some user interface. An according event is stored in event class apply which is imported by vienna.xml.

After the application deadline for a scholarship is over, rule selectApplicantsRule is executed to select the people to which the scholarship is granted. If the number of applicants is larger than the number of scholarships, a ranking is established based on a) the applications, and b) the order of their application if two applicants have the same qualification.

If all three locations provide time service level tsl-3, a temporal order between applications can be established, otherwise a causal order will be used.

## 4.4   Managing Active XML Schemas

Concluding this chapter, we briefly discuss the management of active XML Schema documents.

### 4.4.1   Reactive Service

Since XML documents are static by nature, they need to be complemented by a *reactive service* that handles the presented behavioral extensions (for passive behavior as well as active behavior). Various components of a reactive service, such as an event detector and a rule manager, can be built upon experience from active database systems. The management of simple active XML rules based on mutation events has already been discussed in [31]. Due to the higher-level approach taken with active XML Schemas, new issues arise, which we sketch in the following.

First, the occurrence of an event does not correspond to the addition of an event to an event class, but to the event's occurrence time matching wall clock time (system time), since event classes may contain scheduled events.

Second, the delivery time of a remote event needs not be past the event's occurrence time, because a remote event may be delivered to subscribing documents before its occurrence time. This is similar to employers delivering salary payment records ahead of the payment day to employees' bank accounts, which also list such transactions on account statements with a future effective date (occurrence time).

Third, events need not be added to a document's event classes when they occur. When a document is checked out from a reactive service to be edited in a legacy application, an edit script as a collection of mutation events (cf. [31]) reflecting the document changes is added to the document only when it is checked in again. In a similar way, subscribed remote events may be collected by a behavior service in an "in-box" before these events are added to a document. All these events, which are added delayed, should be added to a document's event classes before any rule is triggered.

Fourth, as documents may not be accessed for some time, event occurrences need not to trigger rules immediately (*eager mode*), but may trigger rules only before a document is accessed next time (*lazy mode*). In lazy mode or when events are added delayed, rules can inquire by querying event classes what other events are "pending". This approach may be compared to skimming all e-mails in a mail box before answering in order to avoid a reply that would have been different if an already delivered later e-mail had been read. This enables designers to realize a "net effect" policy as known from active databases [123].

Fifth, rules may also need to react according to events reflecting active behavior, i.e. rule executions, complementary to events reflecting passive behavior, i.e. operation executions. To inquiry the status of rule executions, for each event and each rule defined on the event's class a *rule trace* records whether the rule has not yet been processed, is being processed, or has been processed successfully. For easy checking whether all rules that are defined on an event class have been considered or processed for a given member event, each event provides information reflecting its *event status*.

The status of an event is "occurred" after its addition and it is "processed" after all rules defined on the event class the event is member of, have been considered. An event that occurs in the future, but is already added to an event class, shows status "scheduled".

### 4.4.2   Communication Transparency

Import and export of event classes between documents is specified declaratively, disregarding the delivery approach chosen. To provide for communication transparency, the delivery of subscribed events is delegated to a *publication handler* associated with the exporting document and a *subscription handler* associated with the importing document. The delivery approach may be specified separately for each document type or document and states whether a pull or pushed-based approach is taken, what the polling period is in the case of a pull-based approach, how delivered events are buffered, or whether events are pre-filtered by the publication handler if not all events of an exported event class are needed by the subscribing document (techniques outlined in [32] for pushing reactive services to the XML repositories may be employed for pre-filtering). Similar to physical data independence, which hides changes to the internal storage structure of a database from database applications, communication transparency hides changes from the delivery semantics from event classes and rules.

### 4.4.3   Event Source Transparency

Events that occur within legacy applications may as well as active XML documents give raise to active behavior. Therefore, and because an active XML Schema should be shielded from changes to its event sources, we provide for event source transparency by which changes to event sources are hidden from event classes and rules.

For example, an active XML document may contain data drawn from a database. Database changes need to be propagated to the document. In such a situation, an active XML Schemas adapter can be defined for the database, which exports appropriate event classes reflecting database updates.

# Chapter 5

# Composite Mutation Events

## Contents

The chapter discusses the functionality of the Active XML Schema (AXS) approach to compose occurred primitive mutation events to more complex ones, as already mentioned in Subsections 4.1.2 and 4.2.3. After introducing and motivating the problem of composite events in XML in Section 5.1, the refinement of an event algebra and a motivating example is shown in Section 5.2. The event algebra that is refined is one from active database literature, where event algebras have been commonly used to define the composition of primitive events. The main contribution of this chapter is made in Section 5.3, where the algebra's extension is presented. Finally, Section 5.4 briefly discusses the implementation of a proof-of-concept prototype.

Note that this chapter, opposite to Chapter 4, uses the notion of "type" to refer to both intensional and extensional aspects of events, because this is how the active database literature dealing with composite events uses it (e.g., cf. [48, 67, 70, 173, 175]). The reason may be that a clear distinction between intensional and extensional aspects is not required to describe the semantics of composite events. Moreover, it is expected that this loss in precision results in a presentation that is more concise and better readable. How the notion of "event type" as used in this chapter subsumes the notions of "event type" and "event class" from Chapter 4 is clarified in Subsection 5.2.2.

## 5.1 Introduction

Recently, several approaches for active XML, i.e., for enriching XML documents with active behavior have been proposed, comprising Active XML Schema (AXS, see Chapter 4) and the ones described in Subsection 1.4.5. All provide for the definition of ECA rules on primitive mutation events, which occur upon, e.g., element insertions and attribute modifications.

In parallel, the W3C standardized the Document Object Model (DOM) Event Module [161]. It provides for detection of events in DOM documents so that application programs can react accordingly. Defined events comprise among others mutation events, which are events that represent modifications of XML data. The Event Module may thus be used to provide the proprietary approaches for active XML and custom applications with mutation events.

When using any of the approaches for active XML, however, it is sometimes impossible to decide upon which event to react. The reason is that often not a single event but a combination of multiple events determines a situation where some action has to be executed. A potential work-around in such a situation is to use the event that always occurs at last of multiple events or to use another event that usually occurs after multiple events. Using such a work-around, however, makes rules dependent on applications which define the order of event occurrences.

Obviously, a technique is needed to detect occurrences of combinations of multiple events, i.e., to detect so called *composite events*. This has long been studied in the active database literature where *event algebras* have been proposed for the description of composite events (e.g., cf. [48, 67, 70, 173, 175]) and several techniques for realizing detection of composite events have been proposed, namely event graphs [46], state automata [70], and petri nets [68].

Events in XML, however, differ from the concept of events in literature as follows:

(1) XML events are not only ordered by time but also by hierarchical structure. It is mostly undesired to use hierarchically unrelated events to form composite events as previous approaches do.

(2) An XML schema may constrain the number of element and attribute occurrences in documents. Existing approaches do not support the detection of when such constraints are satisfied.

(3) Event types, which are descriptions of events at the schema level, are hierarchically related as their events are. This allows for more expressive and more reusable event type definitions than in previous approaches.

Due to the above peculiarities of XML events, existing approaches for detecting composite events, such as [48, 67, 70, 173, 175], cannot be reasonably employed for XML events because one encounters the following problems: (i) depending on the order of multiple modifications that all result in the same XML data different composite events are detected, (ii) most of the detected composite events are meaningless since they are not hierarchically related, they have to be filtered out by application code, (iii) it cannot be detected when multiplicity constraints defined by an XML schema are satisfied, and (iv) event types are unrelated and their extents are disjunct, limiting expressiveness and reusability of composite event type definitions. For a motivating example that shows problems i and ii when using a refined existing approach see Subsection 5.2.3.

The contribution of the chapter is to present an approach to detect composite events in XML that takes the above peculiarities of XML events into account. It *refines* an event algebra known from literature by defining the employed abstract model for XML data, XML events, and XML event types. Thereby it provides for more expressive and reusable event type definitions (addressing peculiarity 3). Moreover, it *extends* the semantics of the refined event algebra by introducing the hierarchical context to combine events according to hierarchy (addressing peculiarity 1), by introducing the multiplicity operator to detect when multiplicity constraints are satisfied (addressing peculiarity 2), and by introducing operator modifiers to provide for more expressive event type definitions.

In particular, the presented approach refines and extends the event algebra Snoop [46, 48], because it is both extensible and well suited for XML. Snoop is extensible because it uses contexts to define the semantics of an event expression, thus by defining a new context semantics can be extended. Snoop is well suited for XML because *event trees* are used to realize event expressions and demonstrate event detection. Event trees fit well for processing XML events because they are hierarchically ordered as well. Moreover, Snoop is used in the Sentinel active DBMS, is prominent among [67, 70, 173, 175] according to CiteSeer[1], and is still subject to active research [9].

---

[1] http://citeseer.nj.nec.com

## 5.2    Refined Event Algebra

This section shows how the event algebra Snoop is refined so that it can be
used with XML events. First it presents the employed abstract model for
XML data, a syntax for referrers to portions of XML data at the schema
and instance level, and operators on referrers in Subsection 5.2.1. Second,
an abstract model for events and event types is introduced in Subsection 5.2.2.
Finally, Subsection 5.2.3 briefly introduces Snoop and shows an example
that uses the refined event algebra with contexts from Snoop.

### 5.2.1    Path Types and Path Instances

An XML document is represented by a tree of nodes where an XML docu-
ment's elements, attributes and text is represented by the tree using element,
attribute, and text nodes respectively. As such it is a subset of the XML
Infoset [154]. Each node has an identifier.

A *path type* identifies a node of a tree by using type information, i.e.,
independently of concrete documents. A path type accords to a restricted
XPath expression [149] that refers to either element, attribute, or text nodes
in each of its steps via respective axis and node tests. A path type is absolute
or relative with respect to the root of the tree, e.g., /order/item/price denotes
an absolute path type while item/price denotes a relative path type to ele-
ment price. Path type $pt$ is a tuple comprising a $kind \in \{\mathsf{absolute}, \mathsf{relative}\}$
and an ordered set of steps, thus $pt = \langle kind, steps \rangle$ or $pt = null$. Two single
steps are equal if they equal in their respective axis (child or attribute) and
node test (test for an XML-QName or text()).

A *path instance* identifies a node of a tree representing a concrete docu-
ment. For node $\mathsf{price_1}$ its path instance comprises an ordered set of identifiers
that starts with the identifier of the tree's root node and ends with $\mathsf{price_1}$'s
identifier and is thus always absolute. A path instance is denoted similar to a
path type by using "/" to separate nodes, e.g., /$\mathsf{order_1}$/$\mathsf{item_1}$/$\mathsf{price_1}$ denotes
a path instance to node $\mathsf{price_1}$. Path instance $pi$ is a tuple comprising its
absolute path type and an ordered set of node identifiers, thus $pi = \langle pt, ids \rangle$
or $pi = null$.

To compare and operate on path types, operators for testing for equality
($=$), containment ($\subset$), ending ($\subset_{\mathsf{e}}$), and intersection ($\cap_{\mathsf{lb}}, \cap_{\mathsf{ab}}$) are defined.
The operators complement the ones defined by XPath which operate on path
instances only, such as $=$ [149] and intersection [163]. The result of applying
operators on path types are defined as follows (where $m = |pt_1.steps|$ and
$n = |pt_2.steps|$):

- $pt_1 = pt_2$
  Path type $pt_1$ equals $pt_2$ iff $pt_1.kind = pt_2.kind \land m = n \land \forall 1 \leq i \leq$

$m : pt_1.steps[i] = pt_2.steps[i]$.

⊙ *Example 47.* item/price = item/price,
/order = /order,
order ≠ /order.

- $pt_1 \subset pt_2$
Path type $pt_2$ uniquely contains path type $pt_1$ iff $\forall 1 \leq i \leq m :$ $pt_1.steps[i] = pt_2.steps[c+i]$ where $c$ is a constant offset and $m+c \leq n$. No $c' \neq c$ may exist for which the expression above applies as well. Additionally, if $pt_1$ and $pt_2$ are both absolute $c = 0 \wedge m < n$ must apply, if both are relative only $m < n$ must apply. A relative path type cannot contain an absolute one.

  ⊙ *Example 48.* item/price ⊂ /order/item/price,
  order/item ⊂ /order/item/price,
  /order ⊄ order/item.

- $pt_1 \subset_{\mathsf{e}} pt_2$
Relative path type $pt_1$ ends path type $pt_2$ if the end of $pt_2.steps$ contains $pt_1.steps$ and $pt_2$ is more special than $pt_1$, i.e., $pt_1 \subset_{\mathsf{e}} pt_2$ iff $pt_1.kind = \mathsf{relative} \wedge (m < n \vee (pt_2.kind = \mathsf{absolute} \wedge m = n)) \wedge \forall 1 \leq i \leq m : pt_1.steps[i] = pt_2.steps[n - m + i]$.

  ⊙ *Example 49.* item/price ⊂$_{\mathsf{e}}$ /order/item/price,
  price ⊂$_{\mathsf{e}}$ item/price,
  order/item ⊄$_{\mathsf{e}}$ /order/item/price.

- $r := pt_1 \cap_{\mathsf{lb}} pt_2$
The left-bound intersection operator is commutative and determines for path types $pt_1$ and $pt_2$ equal steps at the beginning of $pt_1.steps$ and $pt_2.steps$. If $pt_1.kind = pt_2.kind$, $r.kind := pt_1.kind$, otherwise it is absolute. Resulting $r.steps := \{pt_1.steps[i] | pt_1.steps[i] = pt_2.steps[i]\}$ where $1 \leq i \leq j$ where $j$ is either the largest index for which $pt_1.steps[j] = pt_2.steps[j]$ applies or the minimum out of $m$ and $n$. The result is $null$ if $pt_1 = null \vee pt_2 = null \vee pt_1.steps[1] \neq pt_2.steps[1]$.

  ⊙ *Example 50.* order/item ∩$_{\mathsf{lb}}$ /order/billTo = /order,
  item/@partnum ∩$_{\mathsf{lb}}$ item = item,
  order ∩$_{\mathsf{lb}}$ item = $null$.

- $r := pt_1 \cap_{\mathsf{ab}} pt_2$
The absolute-path intersection is not commutative and makes path type $pt_1$ absolute according to absolute path type $pt_2$. If $pt_1 \subset pt_2 \vee pt_1 = pt_2$, $r$ is defined by $r.kind := \mathsf{absolute}$ and $r.steps :=$ $\{pt_2.steps[i] | 1 \leq i \leq j\}$ where $j$ is the last index where $pt_2$ contains $pt_1$. The result is $null$ if $pt_1 = null \vee pt_2 = null \vee (pt_1 \not\sqsubset pt_2 \wedge pt_1 \neq pt_2)$.

⊙ *Example 51.* item $\cap_{\mathsf{ab}}$ /order/item/price = /order/item,
item $\cap_{\mathsf{ab}}$ /order = $null$.

To compare and operate on path instances, operators for testing for equality ($=$) and projection ($\pi$) are defined. The result of applying operators on path instances are defined as follows:

- $pi_1 = pi_2$
  Two path instances $pi_1$ and $pi_2$ equal iff $pi_1.pt = pi_2.pt \wedge \forall 1 \leq i \leq |pi_1.ids| : pi_1.ids[i] = pi_2.ids[i]$.

  ⊙ *Example 52.* /order$_1$/item$_1$ = /order$_1$/item$_1$.

- $r := \pi_{pt}(pi)$
  A projection of path instance $pi$ on path type $pt$ is a path instance iff $pt \sqsubset pi.pt \vee pt = pi.pt$. Then $r.pt := pt$ and $r.ids := \{pi.ids[i] | j \leq i \leq k\}$ where $j$ and $k$ are the indexes between which $pi.pt.steps$ contains $pt.steps$. The result is $null$ if $pi = null \vee pt = null \vee (pt \not\sqsubset pi.pt \wedge pt \neq pi.pt)$.

  ⊙ *Example 53.* $\pi_{\mathsf{item}}$(/order$_1$/item$_1$/price$_1$) = item$_1$,
  $\pi_{\mathsf{/order}}$(/order$_1$/item$_1$) = /order$_1$,
  $\pi_{\mathsf{item}}$(/order$_1$) = $null$.

## 5.2.2 Event Types and Events

The DOM Event Module defines among others event types for mutation events, which reflect modifications of DOM documents' data. They basically comprise one event type for the insertion of nodes, one for the deletion of nodes, one for manipulation of attributes, and one for manipulations of text nodes.

While the DOM event types are sufficient for a procedural handling of occurred events, they are too coarse grained for a declarative handling by an event algebra. Hence, for *every* path type $pt$ the presented approach distinguishes three *primitive event types*, denoted as $\mathsf{ins}(pt)$, $\mathsf{upd}(pt)$, and $\mathsf{del}(pt)$. Like in the DOM Event Module, $\mathsf{ins}$ and $\mathsf{del}$ events reflect insertions and deletions of element, attribute, and text nodes, while $\mathsf{upd}$ events reflect modifications of text nodes and attribute nodes. Moreover, instead of an operation wildcard "$*$" can be used. The path type defines where events of that type occur. It can be relative or absolute. Primitive event type $et$ is a tuple comprising an operation, which is one of $\{\mathsf{ins}, \mathsf{upd}, \mathsf{del}, *\}$, and a path type, thus $et = \langle op, pt \rangle$.

A *primitive event* occurs whenever a node is manipulated. Primitive event $e$ is represented by a tuple comprising its identifier $id$, timestamp $ts$, event type $et$, and path instance $pi$ which identifies the manipulated node, thus $e = \langle id, ts, et, pi \rangle$. The event type's operation does not equal wildcard "$*$", its path type is absolute, and the path instance's path type $pi.pt$ equals $et.pt$.

A *composite event type*, i.e., the event type of a composite event, is a tuple that comprises a unique name and a path type, thus $et = \langle name, pt \rangle$. Wildcard "$*$" can be used instead of a name, referring to any composite event type having the path type. A composite event type is denoted as $\mathsf{name}(pt)$ analogously to a primitive one. The path type of a composite event type defines, like the path type of a primitive event type, where events of that type occur. It can be relative or absolute.

A *composite event* is formed by combining primitive and other composite events, which are referred to as constituent events. Composite event $e^c$ is represented by a tuple comprising its identifier $id$, composite event type $et$, path instance $pi$, which identifies where the event occurred, and a set of constituent events $cevts$, thus $e^c = \langle id, et, pi, cevts \rangle$. The composite event type's name does not equal wildcard "$*$", its path type is absolute, and the path instance's path type $e^c.pi.pt$ equals $e^c.et.pt$.

The notion of "event type" as used in this Chapter subsumes the notions of "event type" and "event class" from Subsection 4.1.2 as follows:

1. A tuple representing a primitive event type denotes an instance of metaschema class $\mathsf{evts::PrimitiveMutEvtTp}$. The instance's attribute $\mathsf{timeSpec}$, which defines whether the event should be detected before or after the respective manipulation takes place, is not of relevance in the context of detecting composite events and is thus not dealt with in this chapter. The same applies for the instance's attribute $\mathsf{name}$, which is derived as the concatenation of operation $\mathsf{op}$ and path type $\mathsf{pathTp}$ (cf. Figure 4.4).

2. A tuple representing a composite event type denotes an instance of metaschema class $\mathsf{cle::CompositeMutEvtTp}$. The tuple's $name$ correspond to the instances $\mathsf{name}$, and $pt$ to $\mathsf{pathTp}$.

3. An event expression denotes an instance of metaschema class $\mathsf{cle::CompositeMutEvtCs}$. Conceptually, the expression specifies a query over event classes with mutation events, thereby defining the extension of the composite event class. It thus corresponds to the instance's attribute $\mathsf{querySpec}$. When an event expression refers to another event expression via its assigned composite event type (see Subsection 5.3.1), it uses the latter type to refer to the class's extension (thus using the notion "type" to refer to both intensional and extensional aspects). To assure that an event class with the composite event type's name exists, a constraint is specified with metaschema class cle::CompositeMutEvtCs (see Figure 4.9).

The logical clock of time service level tsl-1 (see Subsection 4.3.2) is sufficient for detecting composite events. Primitive events occur at distinct points in time and for simplicity it is assumed that the detection of composite events takes no time. Therefore one primitive and multiple composite events may be detected at a single point in time.

Primitive as well as composite event types are hierarchically related via their path type and lead to *more expressive* composite event type definitions than in Snoop. Most important this allows to constrain the combination of event types by an operator to related event types. For the constraints on operator nodes see Subsection 5.3.1.

An event can be an instance of more than one event type, providing for *more reusable* composite event type definitions than in Snoop. Event $e$ is a direct instance of its type $e.et$ and an indirect instance of event types $et_i \neq e.et$ to which the event's type is compatible to, denoted by $e.et \succeq et_i$ (not commutative). Primitive event type $et_1$ is *compatible to* primitive event type $et_2$, i.e., $et_1 \succeq et_2$ iff $((et_1.op = et_2.op) \vee (et_2.op = \text{``*''})) \wedge ((et_2.pt \subset_e et_1.pt) \vee (et_2.pt = et_1.pt))$. Analogously, for two composite event types $et_1^c \succeq et_2^c$ iff $((et_1^c.name = et_2^c.name) \vee (et_2^c.name = \text{``*''})) \wedge ((et_2^c.pt \subset_e et_1^c.pt) \vee (et_2^c.pt = et_1^c.pt))$. For more details on reuse see Subsection 5.3.1, especially Examples 59 and 60.

⊙ *Example 54.*    Primitive event $e_{p_1}$ reflecting an insertion in path type $e_{p_1}.et.pt = $ /order/item/price is an instance of event types such as $E_1 = \text{ins(/order/item/price)}$ and $E_2 = *(\text{price})$ since $e_{p_1}.et \succeq E_1$ and $e_{p_1}.et \succeq E_2$. Analogously, composite event $e_{h_1}^c$ with path type $e_{h_1}^c.et.pt = $ /order/item and name Nm is an instance of composite event types such as $E_1^c = \text{Nm(/order/item)}$ and $E_2^c = *(\text{item})$ since $e_{h_1}^c.et \succeq E_1^c$ and $e_{h_1}^c.et \succeq E_2^c$.

Like in Snoop, composite event types are defined by event expressions according to an event algebra. An expression combines events by the algebra's operators. The presented approach uses operators $\triangle$, $\triangledown$, and ; from Snoop to form conjunction, disjunction, and sequence of events. An expression is realized by an event tree, e.g., Figure 5.2 depicts the event tree that realizes the expression described in Subsection 5.2.3.

An event tree comprises event type nodes and operator nodes. An *event type node* is a tuple $\langle et, evts \rangle$ which stores a set of events *evts*. All events in *evts* are a direct or indirect instance of event type $et$. An *operator node* combines events from child nodes *nds* to events of composite event type $et$ according to operator *opr* and stores them in a set of composite events *evts*, and is thus a tuple $\langle opr, nds, et, evts \rangle$. Leaf nodes in an event tree are event type nodes while inner nodes are operator nodes.

In the presented approach the event type of an event tree node (i.e., of an event type node or operator node) can be compared to the *static type* of a variable in strongly typed object-oriented programming languages (e.g., Java), while the event type of an event can be compared to the *dynamic type* of an expression, e.g., an object. An occurred event is stored in all leaf nodes that have a compatible event type. By using the ending operator in the definition of type compatibility a leaf node only stores events that occur directly in the node's path type and not in descendants.

⊙ *Example 55.* Continuing Example 54, when primitive event $e_{p_1}$ with (dynamic) event type $e_{p_1}.et$ occurs, it is stored in event type nodes that have compatible (static) event types such as $E_1$ and $E_2$. Analogously, when composite event $e_{h_1}^c$ is raised, it is stored in event type nodes that have event types such as $E_1^c$ and $E_2^c$.

### 5.2.3 Motivating Example

This section exemplifies the need for composite events in XML and demonstrates the application of the refined event algebra. Consider an XML document that represents a purchase order (document element **order**) comprising items to be ordered (element **item**), each in turn described by a price (element **price**) and a quantity (element **quantity**). When defining a rule that reacts on the insertion of an item and re-calculates the overall order value by multiplying price by quantity of each item and summing it up, one encounters the problem to decide which primitive mutation event to react upon. Upon the insertion of element **item** it does not comprise any of the necessary child elements, and upon the insertion of element **price** the **quantity** element may not be available and vice versa.

The problem can be overcome by using composite events. A composite event is raised according to its definition after certain primitive events have occurred. In the example, a composite event should occur after the occurrence of events reflecting insertions of an **item**, a **price**, and a **quantity** element (where the latter two are children of the first) so that a rule can be defined on it. This can be achieved by event expression "$E_i$ ; ($E_p \triangle E_q$)", where $E_i := \text{ins(item)}$, $E_p := \text{ins(item/price)}$, and $E_q := \text{ins(item/quantity)}$.

Figure 5.1 shows the example's active XML schema. It defines composite event class **InsertItem** whose query is set to the mentioned event expression. Thereby detected composite events are stored in the event class upon their occurrence. Upon occurrence of a such a composite event, rule **recalculate-OrderValueRule** is executed, recalculating the overall order value by invoking operation **calculateTotal()**.

The event tree realizing the example's event expression is depicted in Figure 5.2. The event tree's behavior when using contexts from Snoop, i.e., its processing of four sequences of primitive events $S_1..S_4$ is shown in

Table 5.1: Raised Composite Events when using Contexts from Snoop

| - | $t_1$ | $t_2$ | $t_3$ | $t_4$ | $t_5$ | $t_6$ |
|---|---|---|---|---|---|---|
| | | | | Cumulative Context | | |
| $S_1$ | $e_{i_1}$ | $e_{p_1}$ | $e_{q_1}, \{e_{i_1}e_{p_1}e_{q_1}\}^c$ | $e_{i_2}$ | $e_{p_2}$ | $e_{q_2}, \{e_{i_2}e_{p_2}e_{q_2}\}^c$ |
| $S_2$ | $e_{i_1}$ | $e_{i_2}$ | $e_{p_1}$ | $e_{p_2}$ | $e_{q_1}, \{e_{i_1}e_{i_2}e_{p_1}e_{p_2}e_{q_1}\}^c$ | $e_{q_2}$ |
| $S_3$ | $e_{i_1}$ | $e_{i_2}$ | $e_{p_1}$ | $e_{q_1}, \{e_{i_1}e_{i_2}e_{p_1}e_{q_1}\}^c$ | $e_{p_2}$ | $e_{q_2}, (\{e_{q_2}e_{q_2}\}^c)$ |
| $S_4$ | $e_{i_1}$ | $e_{i_2}$ | $e_{p_1}$ | $e_{q_2}, \{e_{i_1}e_{i_2}e_{p_1}e_{q_2}\}^c$ | $e_{p_2}$ | $e_{q_1}, (\{e_{q_2}e_{q_1}\}^c)$ |
| | | | | Chronicle Context | | |
| $S_1$ | $e_{i_1}$ | $e_{p_1}$ | $e_{q_1}, \{e_{i_1}e_{p_1}e_{q_1}\}^c$ | $e_{i_2}$ | $e_{p_2}$ | $e_{q_2}, \{e_{i_2}e_{p_2}e_{q_2}\}^c$ |
| $S_2$ | $e_{i_1}$ | $e_{i_2}$ | $e_{p_1}$ | $e_{p_2}$ | $e_{q_1}, \{e_{i_1}e_{p_1}e_{q_1}\}^c$ | $e_{q_2}, \{e_{i_2}e_{p_2}e_{q_2}\}^c$ |
| $S_3$ | $e_{i_1}$ | $e_{i_2}$ | $e_{p_1}$ | $e_{q_1}, \{e_{i_1}e_{p_1}e_{q_1}\}^c$ | $e_{p_2}$ | $e_{q_2}, \{e_{i_2}e_{p_2}e_{q_2}\}^c$ |
| $S_4$ | $e_{i_1}$ | $e_{i_2}$ | $e_{p_1}$ | $e_{q_2}, \{e_{i_1}e_{p_1}e_{q_2}\}^c$ | $e_{p_2}$ | $e_{q_1}, \{e_{i_2}e_{p_2}e_{q_1}\}^c$ |
| | | | | Recent Context | | |
| $S_1$ | $e_{i_1}$ | $e_{p_1}$ | $e_{q_1}, \{e_{i_1}e_{p_1}e_{q_1}\}^c$ | $e_{i_2}$ | $e_{p_2}, \{e_{i_2}e_{p_2}e_{q_1}\}^c$ | $e_{q_2}, \{e_{i_2}e_{p_2}e_{q_2}\}^c$ |
| $S_2$ | $e_{i_1}$ | $e_{i_2}$ | $e_{p_1}$ | $e_{p_2}$ | $e_{q_1}, \{e_{i_2}e_{p_2}e_{q_1}\}^c$ | $e_{q_2}, \{e_{i_2}e_{p_2}e_{q_2}\}^c$ |
| $S_3$ | $e_{i_1}$ | $e_{i_2}$ | $e_{p_1}$ | $e_{q_1}, \{e_{i_2}e_{p_1}e_{q_1}\}^c$ | $e_{p_2}, \{e_{i_2}e_{p_2}e_{q_1}\}^c$ | $e_{q_2}, \{e_{i_2}e_{p_2}e_{q_2}\}^c$ |
| $S_4$ | $e_{i_1}$ | $e_{i_2}$ | $e_{p_1}$ | $e_{q_2}, \{e_{i_2}e_{p_1}e_{q_2}\}^c$ | $e_{p_2}, \{e_{i_2}e_{p_2}e_{q_2}\}^c$ | $e_{q_1}, \{e_{i_2}e_{p_2}e_{q_1}\}^c$ |
| | | | | Continuous Context | | |
| $S_1$ | $e_{i_1}$ | $e_{p_1}$ | $e_{q_1}, \{e_{i_1}e_{p_1}e_{q_1}\}^c$ | $e_{i_2}$ | $e_{p_2}, \{e_{i_2}e_{p_2}e_{q_1}\}^c$ | $e_{q_2}, (\{e_{p_2}e_{q_2}\}^c)$ |
| $S_2$ | $e_{i_1}$ | $e_{i_2}$ | $e_{p_1}$ | $e_{p_2}$ | $e_{q_1}, \dagger_1$ | $e_{q_2}$ |
| $S_3$ | $e_{i_1}$ | $e_{i_2}$ | $e_{p_1}$ | $e_{q_1}, \dagger_2$ | $e_{p_2}, (\{e_{p_2}e_{q_1}\}^c)$ | $e_{q_2}, (\{e_{p_2}e_{q_2}\}^c)$ |
| $S_4$ | $e_{i_1}$ | $e_{i_2}$ | $e_{p_1}$ | $e_{q_2}, \dagger_3$ | $e_{p_2}, (\{e_{p_2}e_{q_2}\}^c)$ | $e_{q_1}, (\{e_{p_2}e_{q_1}\}^c)$ |
| $\dagger_1$: $\{e_{i_1}e_{p_1}e_{q_1}\}^c, \{e_{i_1}e_{p_2}e_{q_1}\}^c, \{e_{i_2}e_{p_1}e_{q_1}\}^c, \{e_{i_2}e_{p_2}e_{q_1}\}^c$ $\dagger_2$: $\{e_{i_1}e_{p_1}e_{q_1}\}^c, \{e_{i_2}e_{p_1}e_{q_1}\}^c$ $\dagger_3$: $\{e_{i_1}e_{p_1}e_{q_2}\}^c, \{e_{i_2}e_{p_1}e_{q_2}\}^c$ | | | | | | |

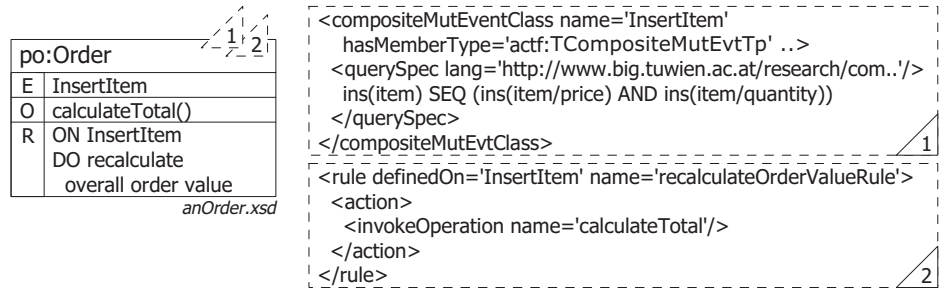

Figure 5.1: Recalculate Order Value upon Insertion of an Item

Table 5.1. The sequences reflect the insertions of **item** (abbreviated as $i_n$), **price** ($p_n$), and **quantity** ($q_n$) elements. Numerical index $n$ of an element represents its hierarchical position, meaning that elements with the same numerical index are hierarchically related, e.g., $p_1$ is a child of $i_1$. In case two numerical indices are separated by a dot, the first number represents the element's hierarchical position and the second one the time of its insertion, e.g., the insertion of $q_{1.1}$ occurs before the insertion of $q_{1.2}$.

Briefly and informally introducing the contexts from Snoop, a composite event is raised by a conjunction operator node as soon as a child node's set of stored events is modified, i.e., an event is added (the so called "terminator") and every child node contains at least one event. Note that only the conjunction operator is described here, however, sequence and disjoint operator are defined analogously. A raised composite event's constituent events are defined as follows:

- in *cumulative context* they comprise all events from every child node wherefrom they are removed.

- in *chronicle context* they comprise the oldest event from every child node wherefrom they are removed, i.e., all constituent events are consumed in chronological order of occurrence.

- in *recent context* they comprise the most recent event from each child node. All events that cannot be the earliest constituent event of subsequently raised composite events (i.e., that cannot be an "initiator") are removed from child nodes.

- in *continuous context* its constituent events comprise the terminator and the most recent event from every child node except the one of the terminator. Subsequently all constituent events are removed from child nodes except the terminator. The procedure is repeated until there are no events left to be combined with the terminator. Finally the terminator is removed if it cannot be an initiator of subsequently raised composite events.

The rationale of Table 5.1 is to introduce the unfamiliar reader to Snoop's contexts and to exemplify that incorrect composite events are raised when any of Snoop's contexts is used (for which one falsifying event sequence would suffice). Incorrect events are raised because events are selected only by their occurrence time and not their hierarchical position. Where under a correct composite event it is referred to a composite event whose constituent events are hierarchically related. Composite events raised by the root of the tree are shown by their constituent events, e.g., $\{e_{i_1}e_{p_1}e_{q_1}\}^c$. Unconsumed composite events that remain in the tree after $t_6$ are shown at the time they are raised, but in brackets, e.g., $(\{e_{p_2}e_{q_2}\}^c)$. The table does not show the unrestricted context which basically forms the cartesian product of all events. Naturally, it raises even more incorrect composite events.

Summarized, the refined event algebra as presented in this section is still not applicable to detect composite events in XML when used with contexts from Snoop. The reason is that the requirements on applications using such an event algebra are inadequate, which would have to use the unrestricted context and filter out huge amounts of incorrect events or manipulate XML data in a defined temporal order so that some context only detects correct events. Still, the satisfaction of multiplicity constraints cannot be detected using the refined event algebra.

## 5.3   Extended Event Algebra

This section presents an extension to the refined event algebra. The extension comprises the hierarchical context presented in Subsection 5.3.1, the multiplicity operator in Subsection 5.3.2, and operator modifiers in Subsection 5.3.3. Event trees which are generated from event expressions are used for presentational purposes throughout the section.

### 5.3.1   Hierarchical Context

The hierarchical context is introduced since it is necessary to combine events according to their hierarchical position, which is not supported by existing contexts. It raises only correct composite events, i.e., composite events whose constituent events are hierarchically related.

To combine events according to their hierarchical position an event tree maintains data concerning hierarchy. Therefore, as mentioned earlier, every node $n$ in an event tree has an assigned path type $n.et.pt$. Naturally, an event type node specifies a path type, however, the path type of an operator node, if not specified by the event expression, has to be derived from its child nodes $c_1, c_2, ..., c_n$ by evaluating $c_1.et.pt \cap_{\mathsf{lb}} c_2.et.pt \cap_{\mathsf{lb}} ... \cap_{\mathsf{lb}} c_n.et.pt$ (left-bound intersection is used for simplicity). The derivation of path types for all operator nodes is done bottom up. For an event tree to be valid, every node $n$ must have a non-*null* path type whose steps are not empty and do not violate the constraints on child nodes (cf. later in this section).

⊙ *Example 56.*    Figure 5.2 shows an exemplary event tree on the left defining composite event type $\mathsf{E}_i^c := \mathsf{E}_i$ ; $(\mathsf{E}_p \triangle \mathsf{E}_q)$ for the insertion of item elements and the derivation of path types for operator nodes on the right. The derivation is done bottom up, step 1 determines the path type of operator node $\triangle$ by evaluating item/price $\cap_{\mathsf{lb}}$ item/quantity = item. Subsequently, step 2 determines the path type of operator node ; by evaluating item $\cap_{\mathsf{lb}}$ item = item.

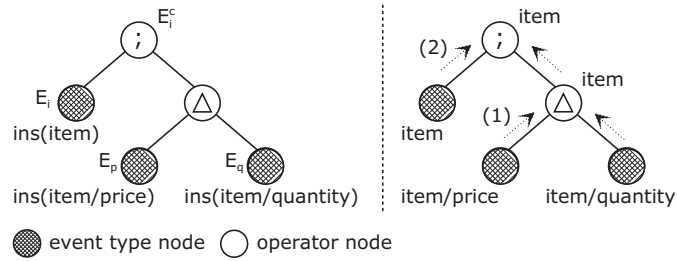

Figure 5.2: Derivation of Operator Nodes' Path Types

An operator node raises a composite event by selecting events from its child nodes that satisfy certain conditions. Basically, (a) at least one child node of a disjunction operator node must hold an event, (b) all child nodes of a conjunction operator node must hold an event, and (c) all child nodes of a sequence operator node must hold an event and they must have occurred in the specified order.

If operator node $o$ combines multiple events from child nodes they must all have the same ancestor node, i.e., $\forall 1 \leq i \leq m-1 : \pi_{o.et.pt}(e_i.pi) = \pi_{o.et.pt}(e_{i+1}.pi)$ where $m$ is the number of events in $o$'s child nodes that are to be combined. Events for which the projection evaluates to *null* are not combined. Raised composite event $e^c$'s path instance, which must be absolute, is derived from constituent events by $\pi_{o.et.pt \cap_{\mathsf{ab}} e_1.pi.pt}(e_1.pi)$. Event $e_1$ is the first constituent event, however, any other constituent event $e_i$ could be used instead because if all constituent events $e_i$ equal in $\pi_{o.et.pt}(e_i.pi)$ they equal in $\pi_{o.et.pt \cap_{\mathsf{ab}} e_i.pi.pt}(e_i.pi)$ as well.

⊙ *Example 57.*    How the event tree in Figure 5.2 forms composite events in hierarchical context is shown in Figure 5.3. Its status after the occurrence of the event sequence $e_{i_1}, e_{i_2}, e_{p_1}, e_{p_2}, e_{q_1}$ is shown on the left. In path instances and event indices, $o$ abbreviates order, $i$ abbreviates item, $p$ abbreviates price, $q$ abbreviates quantity, and the numerical index indicates the hierarchical position, e.g., $p_1$ is a child of $i_1$. Events $e_{p_1}$ and $e_{q_1}$ are combined by operator $\triangle$ to form composite event $e_{h_1}^c$, since $\pi_{\mathsf{item}}(e_{p_1}.pi) = \pi_{\mathsf{item}}(e_{q_1}.pi)$ $(= \mathsf{i}_1)$. With the creation of $e_{h_1}^c$ the two primitive events are consumed (step 1). Subsequently, $e_{i_1}$ and $e_{h_1}^c$ are combined by operator $;$ since $\pi_{\mathsf{item}}(e_{i_1}.pi) = \pi_{\mathsf{item}}(e_{h_1}^c.pi)$ $(= \mathsf{i}_1)$ (step 2). Upon the occurrence of $e_{q_2}$ later on the same combination process starts over again.

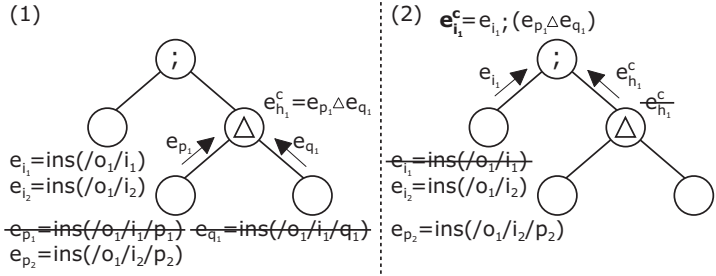

Figure 5.3: Event Selection and Event Consumption in Hierarchical Context

*Constraints* on operator node $o$'s path type are enforced so that the projection of an event's path instance on $o$'s path type is likely to return a non-*null* value, i.e., so that for child event $e_j$ expression $\pi_{o.et.pt}(e_j.pi) \neq null$ applies. The constraints seek a compromise between the operator's expressiveness for the reuse of event trees (see later in this section) and restrictions on child nodes and do thus not guarantee a non-*null* value. They are as follows (where $E_j$ refers to a child node's event type): (a) if both $o.et.pt$ and $E_j.pt$ are absolute, $o.et.pt \cap_{\mathsf{lb}} E_j.pt \neq null$ must apply, (b) if only $E_j.pt$ is absolute, $o.et.pt \subset E_j.pt$ must apply. If $E_j.pt$ is relative no constraints must apply so that it may contain only a single step that is not contained in $o.et.pt$. In such cases one must rely that an event's (dynamic) event type is compatible to the operator's (static) event type so that the event's path instance can be projected on the operator's path type. This can be compared to a type-cast in strongly typed object-oriented programming languages where an object's dynamic type must be compatible to the static casted type which can only be determined at runtime. For examples see Example 59 and 60.

Operators combine events with *interval-based semantics* not detection-based semantics, since composite events detected with the latter are not always exactly as, presumably, intended (cf. [9, 66]). Basically, this means that operators do not combine events according to their detection time but take the intervals during which the events occurred into account. A primitive event detected at $t_d$ occurs in interval $[t_d, t_d]$, and a composite event starts at the beginning of the smallest interval and ends at the end of the latest one. Comparing intervals, $[t'_a, t''_a] < [t'_b, t''_b]$ iff $t''_a < t'_b$. Interval-based semantics only affect the sequence operator, so that expression $E_i\,;E_j$ combines two hierarchically matching events $e_i$ and $e_j$ only if the occurrence interval of the former is smaller than the other.

The hierarchical context is *orthogonal* to existing contexts from Snoop, because selection by XML hierarchy is orthogonal to selection by time. Thus it can be combined with any existing context, acting like a filter. For operator node $o$, first the hierarchical context groups all events from child nodes according to $o$'s path type, and second $o$ selects and consumes events within each group according to its context from Snoop.

⊙ *Example 58.* To detect when both price and quantity information of an order item are modified, event expression ∗(item/quantity) △ upd(item/price) can be used. Table 5.2 shows raised composite events, represented by their constituent events for the above expression in the hierarchical variants of Snoop's contexts. Events that remain in the event graph after $t_5$ are shown in the rightmost column for completeness.

A composite event type definition can *reuse* other, existing event type definitions, i.e., and event tree can reuse other event trees. All event trees together form the *event graph*. A composite event $e_i^c$ of type $E_i^c$ raised by an event tree is reused in another tree by an event type node with a compatible event type, i.e., if $e_i^c.et \succeq E_j^c$, presuming that $E_i^c.name = E_j^c.name$. It may be the case that the node's parent operator node $o$'s path type $o.et.pt$ cannot be derived from its child nodes, because the left-bound intersection of $E_j^c.pt$ with the path types of $o$'s other child nodes is *null*, e.g., if their path types are of different kind or $E_j^c.pt$ contains only a single step. In either case $o$'s path type must be specified explicitly.

⊙ *Example 59.* The order element has aside from item elements a billTo and a shipTo child element. When two composite event types $E_b^c$ and $E_s^c$ for complete insertions of the latter two elements are defined in addition to $E_i^c$ (from Example 56), a fourth composite event type can use the three to define a composite event that occurs after both addresses and at least one item have been inserted by $E_i^c \triangle E_s^c \triangle E_b^c$. Since, e.g., $E_i^c.pt \cap_{lb} E_s^c.pt \cap_{lb} E_b^c.pt =$ item $\cap_{lb}$ shipTo $\cap_{lb}$ billTo $= null$, the conjunction operator's path type must be specified explicitly, e.g., as order. Because the (dynamic) event types of occurred events have absolute path types, the projection of their path instances on order is not *null*.

Reuse of event type definitions is facilitated by using type compatibility instead of type equality for storing occurred events in event type nodes, as mentioned before and in Subsection 5.2.2. The reason is that the more leaf nodes with "general" event types an event tree has (i.e., nodes with event types that have relative path types and/or wildcard "∗" as operator or name), the more events will be stored in it because an event's "special" event type (i.e., one with absolute path type and given operation) may be compatible to a general one but not to another special one. The more events are stored, the more composite events are raised, and the more the tree, i.e., the event definition is reusable.

Table 5.2: Combination of the Hierarchical Context with Contexts from Snoop

| - | $t_1$ | $t_2$ | $t_3$ | $t_4$ | $t_5$ | unconsumed events |
|---|---|---|---|---|---|---|
| | Hierarchical Cumulative Context | | | | | |
| $S_5$ | $e_{q_{1.1}}$ | $e_{q_2}$ | $e_{q_{1.2}}$ | $e_{p_2}, \{e_{p_2}e_{q_2}\}^c$ | $e_{p_1}, \{e_{p_1}e_{q_{1.1}}e_{q_{1.2}}\}^c$ | – |
| | Hierarchical Chronicle Context | | | | | |
| $S_5$ | $e_{q_{1.1}}$ | $e_{q_2}$ | $e_{q_{1.2}}$ | $e_{p_2}, \{e_{p_2}e_{q_2}\}^c$ | $e_{p_1}, \{e_{p_1}e_{q_{1.1}}\}^c$ | $e_{q_{1.2}}$ |
| | Hierarchical Recent Context | | | | | |
| $S_5$ | $e_{q_{1.1}}$ | $e_{q_2}$ | $e_{q_{1.2}}$ | $e_{p_2}, \{e_{p_2}e_{q_2}\}^c$ | $e_{p_1}, \{e_{p_1}e_{q_{1.2}}\}^c$ | $e_{p_1}, e_{q_{1.2}}, e_{p_2}, e_{q_2}$ |
| | Hierarchical Continuous Context | | | | | |
| $S_5$ | $e_{q_{1.1}}$ | $e_{q_2}$ | $e_{q_{1.2}}$ | $e_{p_2}, \{e_{p_2}e_{q_2}\}^c$ | $e_{p_1}, \{e_{p_1}e_{q_{1.1}}\}^c, \{e_{p_1}e_{q_{1.2}}\}^c$ | $e_{p_1}, e_{p_2}$ |

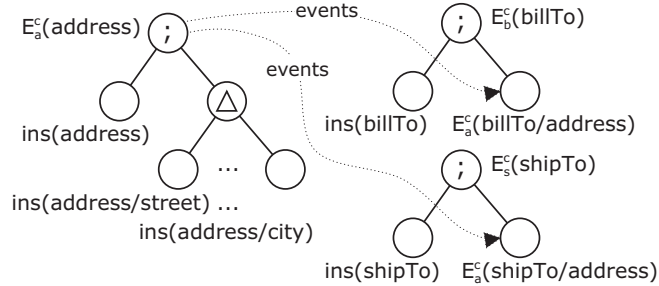

Figure 5.4: Reuse of Event Trees

⊙ *Example 60.* The order's shipTo and billTo elements both have an address child element that comprises other elements such as street and city. By defining composite event type $E_a^c$ for the insertions of address elements and using only relative path types that start with address, the according event tree in Figure 5.4 raises composite events for complete insertions of address elements as childs of both shipTo and billTo. Thus $E_a^c$ can be reused by both $E_b^c$ and $E_s^c$ from Example 59.

## 5.3.2 Multiplicity Operator

Because an XML schema can define multiplicity constraints on XML elements and attributes it is desirable to be able to detect when multiplicity constraints are satisfied by observing occurred events. A multiplicity constraint is defined by lower bound $l$ and upper bound $u$ ($l \leq u$), meaning that between $l$ and $u$ child elements or attributes with the same name may occur as child of a parent element. After $l$ events reflecting insertions the multiplicity constraint is satisfied (if the parent element did not contain any such child elements before). The operator from Snoop that closest resembles the required functionality is the ANY operator. It detects a fixed number $> 0$ of events of distinct event types, however, in XML the required number is $\geq 0$, since 0 events reflect optional elements, and the events are of the same event type.

To detect when multiplicity constraints are satisfied unary multiplicity operator "×" is introduced, denoted as $\times [l, u] E_i$. It raises a composite event as soon as $l$ events of $E_i$ occurred, thereby indicating the constraint's satisfaction. The consumption of the composite event, however, may take place after other events of $E_i$ occurred. To provide composite events with most extensive sets of constituent events, the multiplicity operator has *integrative semantics.*

Event integration starts after composite event $e_1^c$ is raised upon the occurrence of the $l^{th}$ event of $E_i$. A subsequently occurring event gives rise to the new composite event $e_2^c$ which integrates $e_1^c$'s constituent events. Event $e_1^c$ is waived because it has been integrated and is thus assumed not to be of interest any longer. If $u$ is reached or the integrating composite event is consumed, integration is suspended and starts over after the $l^{th}$ occurrence of $E_i$.

When employed in a hierarchical context, the multiplicity operator shows the above behavior for every distinct path instance of its path type. If not specified explicitly, a multiplicity operator's path type is set to the child node's path type omitting the last step. The same constraints on a multiplicity operator's path type must apply as on other operator's path types (see Subsection 5.3.1). Multiplicity operator $o$ raises a composite event $e_1^c$ as soon as there exist $l$ child events that equal in $\pi_{o.et.pt}(e_{i_j}.pi)$ for $1 \leq j \leq l$. Thereafter it raises new composite event $e_2^c$ upon the occurrence of child event $e_{i_k}$ where $\pi_{o.et.pt}(e_{i_k}.pi) = \pi_{o.et.pt}(e_1^c.pi)$ if $e_1^c$ has not been consumed yet, where $e_2^c.cevts$ is defined by the union of $e_1^c.cevts$ and $e_{i_k}$.

⊙ *Example 61.* The event tree depicted in Figure 5.5 allows multiple insertions of `quantity` elements as childs of element `item`. For two insertions of `quantity` elements $e_{q_{1.1}}$ and $e_{q_{1.2}}$ the multiplicity operator first raises composite event $e_{q_1}^c$ upon the occurrence of $e_{q_{1.1}}$. If $e_{q_1}^c$ has not been consumed upon the occurrence of $e_{q_{1.2}}$ new composite event $e_{q_2}^c$ is raised, where $e_{q_2}^c.cevts = \{e_{q_{1.1}}, e_{q_{1.2}}\}$, and $e_{q_1}^c$ is waived. Otherwise, i.e., if $e_{q_1}^c$ has been consumed new composite $e_{q_2}^c$ is raised as well, however, with $e_{q_2}^c.cevts = \{e_{q_{1.2}}\}$.

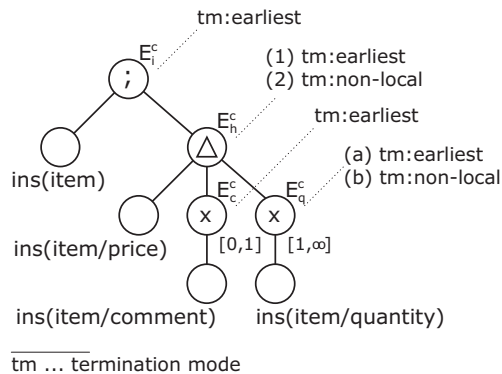

Figure 5.5: Event Tree using a Multiplicity Operator

A multiplicity operator in hierarchical context with a lower bound of zero raises a composite event without the occurrence of a child event. Instead, composite event $e_1^c$ is raised with the event representing the creation of (absolute) path instance $pi$ that satisfies the constraint of multiplicity operator $o$ by having $o.et.pt \sqsubset_e pi.pt \vee o.et.pt = pi.pt$. It does not comprise any constituent event, its path type is set to $pi.pt$, and its path instance to $pi$. Analogously, such composite events are raised when an event graph is registered with a document for every path instance $pi$ in the document that satisfy the condition above. When event $e_i$ with $\pi_{o.et.pt}(e_i.pi) = \pi_{o.et.pt}(e_1^c.pi)$ subsequently occurs in $o$'s child node before $e_1^c$ is consumed, new composite event $e_2^c$ which integrates $e_1^c$ is raised. If composite event $e_2^c$ is consumed later on and $e_2^c.pi$ still exists, new composite event $e_3^c$ is raised since $o$'s constraints are still satisfied.

⊙ *Example 62.*    The event tree depicted in Figure 5.5 allows optional element **comment** as child of element **item**. With the insertion of **item** element $i_1$ composite event $e_{c_1}^c$ occurs with $e_{c_1}^c.et.pt = $ /order/item and $e_{c_1}^c.pi = $ /$o_1$/$i_1$. Thereafter, the conjunction operator raises composite event $e_{h_1}^c$ as soon as matching events are stored in its other child nodes, i.e., as soon as $p_1$ and $q_1$ are inserted as childs of $i_1$. If matching **comment** element $c_1$ is inserted before both elements $p_1$ and $q_1$ are inserted, event $e_{c_2}^c$ integrating $e_{c_1}^c$ is raised which becomes a part of $e_{h_1}^c$ later on.

### 5.3.3   Operator Modifiers

To enrich the expressiveness of event type definitions, operator nodes are parameterized by two modifiers to exactly define the points in time when composite events are raised. When an operator node detects events in child nodes that satisfy the operator's semantics, a "potential composite event", which is not stored in the tree, is detected. A composite event, which is stored in the tree, is raised with the detection of the potential event or is deferred to a later date. This is of importance, e.g., when a multiplicity operator or the hierarchical cumulative context is used, because the later a composite event is raised the more constituent events it will possibly have. Thus applications can react to deferred composite events comprising a rich set of constituent events and can determine the net-effect [123], i.e., overall effect of multiple events more easily.

First, the operator's *termination mode* determines when a composite event is raised relative to the detection of a potential event. If the termination mode is **earliest**, composite event $e_i^c$ is raised with the detection of potential composite event $e_p^c$. If it is **non-local**, $e_i^c$ is raised after the detection of $e_p^c$ and the first occurrence of event $e_j$, where $\pi_{o.et.pt}(e_p^c.pi) \neq \pi_{o.et.pt}(e_j.pi)$. This means that operator node $o$ waits until an event occurs that reflects a manipulation in another subtree of the document, i.e., it assumes that the manipulation of a document is done hierarchically. If the termination mode is custom, composite events are raised upon flushing or closing the event tree (cf. later in this section).

⊙ *Example 63.*     For the event tree depicted in Figure 5.5 and four exemplary event sequences $S_6..S_9$ Table 5.3 shows when composite events are raised depending on the termination mode of (i) the conjunction operator node and (ii) the multiplicity operator node of the quantity element. The other two operator nodes have termination mode earliest. To clearly point out the consequences of termination modes the table shows composite events raised by any operator node. A composite event is denoted by $e^c$ with an alphabetical index indicating its event type and a numerical index indicating its hierarchical position. Note that due to the order of occurred events in $S_6..S_9$ the same composite events are raised irrespective of the context from Snoop that is combined with the hierarchical context.

Second, the operator's *termination condition* must be fulfilled for a composite event to be raised. It is a condition on the subtree of the XML document with root node $e_p^c.pi$ in the form of an arbitrary XPath expression which is evaluated relative to the subtree's root.[2] The XPath expression must evaluate to boolean. The termination condition differs from operators and termination modes in that it is used to test the instance document, not events. Termination conditions can be arbitrarily used. E.g., they can ease the handling of events representing modifications of text nodes in case an element's text content may consist of multiple adjacent text nodes as in DOM (for which we have implemented a prototype, see Section 7.3). In this case, termination conditions can be used, e.g., to test if an element's child text nodes contain any or certain text such as a number. In certain situations, a termination condition can also be used to resemble the NOT operator from Snoop by using an XPath expression with a negation and a node test.

⊙ *Example 64.*   Instead of adding event type nodes and operator nodes that test for the insertion of text nodes to the event tree depicted in Figure 5.5, termination condition "item/price $\geq$ 0 $\wedge$ item/quantity $>$ 0" can be defined on the root node so that a composite event is only raised when both price and quantity element contain a value that is a number. This has also the advantage that text nodes can be arbitrarily inserted, updated, and deleted because its their value that matters, not the operations that lead to it.

The presented approach does not provide for termination conditions that test potential events themselves for two reasons. First, this requires a defined binding of data to events that is more detailed than the one presented in Subsection 5.2.2, a defined representation of the binding and/or a defined syntax for expressing conditions. This is, however, outside the focus of this work. Second, related approaches for detecting composite events in the active database literature (e.g., cf. [48, 67, 70, 173, 175]) do not provide this capability too. There usually the environment in which composite event detection is employed provides this capability, i.e., the condition of an ECA rule may usually test the composite event in active databases [123]. The second reason also applies for termination conditions that test the document (provided by the approach), however, only partially because there are differences in semantics: While a termination condition that tests events and resembles a rule's condition is intended to avoid an action's execution by avoiding raising a composite event, a termination condition that test the document may be also intended to defer raising a composite event.

---

[2]The restriction of querying the subtree only can be waived easily so that the condition is an arbitrary XPath expression which is evaluated relative to node $e_p^c.pi$. This, however, negatively affects performance because it is necessary to evaluate the termination condition in processTree($t,E$) with every modification to the document, not only with every modification to the subtree.

Table 5.3: Raised Composite Events in Hierarchical Context when different Termination Modes are used

| - | $t_1$ | $t_2$ | $t_3$ | $t_4$ | $t_5$ | $t_6$ |
|---|---|---|---|---|---|---|
| **(1a) $\triangle$ : earliest, $\times[1,\infty]$ : earliest** | | | | | | |
| $S_6$ | $e_{i_1}, e^c_{c_1}$ | $e_{p_1}$ | $e_{q_1}, e^c_{q_1}, e^c_{h_1}, \mathbf{e^c_{i_1}}$ | $e_{i_2}, e^c_{c_2}$ | $-$ | $-$ |
| $S_7$ | $e_{i_1}, e^c_{c_{1.1}}$ | $e_{p_1}$ | $e_{c_1}, e^c_{c_{1.2}}$ | $e_{q_1}, e^c_{q_1}, e^c_{h_1}, \mathbf{e^c_{i_1}}$ | $e_{i_2}, e^c_{c_2}$ | $-$ |
| $S_8$ | $e_{i_1}, e^c_{c_{1.1}}$ | $e_{p_1}$ | $e_{q_{1.1}}, e^c_{q_{1.1}}, e^c_{h_1}, \mathbf{e^c_{i_1}}$ | $e_{c_1}, e^c_{c_{1.2}}$ | $e_{q_{1.2}}, e^c_{q_{1.2}}$ | $e_{i_2}, e^c_{c_2}$ |
| $S_9$ | $e_{i_1}, e^c_{c_1}$ | $e_{p_1}$ | $e_{i_2}, e^c_{c_2}$ | $e_{p_2}$ | $e_{q_1}, e^c_{q_1}, e^c_{h_1}, \mathbf{e^c_{i_1}}$ | $e_{q_2}, e^c_{q_2}, e^c_{h_2}, \mathbf{e^c_{i_2}}$ |
| **(1b) $\triangle$ : earliest, $\times[1,\infty]$ : non-local** | | | | | | |
| $S_6$ | $e_{i_1}, e^c_{c_1}$ | $e_{p_1}$ | $e_{q_1}$ | $e_{i_2}, e^c_{c_2}, e^c_{q_1}, e^c_{h_1}, \mathbf{e^c_{i_1}}$ | $-$ | $-$ |
| $S_7$ | $e_{i_1}, e^c_{c_{1.1}}$ | $e_{p_1}$ | $e_{c_1}, e^c_{c_{1.2}}$ | $e_{q_1}$ | $e_{i_2}, e^c_{c_2}, e^c_{q_1}, e^c_{h_1}, \mathbf{e^c_{i_1}}$ | $-$ |
| $S_8$ | $e_{i_1}, e^c_{c_{1.1}}$ | $e_{p_1}$ | $e_{q_{1.1}}$ | $e_{c_1}, e^c_{c_{1.2}}$ | $e_{q_{1.2}}$ | $e_{i_2}, e^c_{c_2}, e^c_{q_1}, e^c_{h_1}, \mathbf{e^c_{i_1}}$ |
| $S_9$ | $e_{i_1}, e^c_{c_1}$ | $e_{p_1}$ | $e_{i_2}, e^c_{c_2}$ | $e_{p_2}$ | $e_{q_1}$ | $e_{q_2}, e^c_{q_1}, e^c_{h_1}, \mathbf{e^c_{i_1}}$ |
| **(2a) $\triangle$ : non-local, $\times[1,\infty]$ : earliest** | | | | | | |
| $S_6$ | $e_{i_1}, e^c_{c_1}$ | $e_{p_1}$ | $e_{q_1}, e^c_{q_1}$ | $e_{i_2}, e^c_{c_2}, e^c_{h_1}, \mathbf{e^c_{i_1}}$ | $-$ | $-$ |
| $S_7$ | $e_{i_1}, e^c_{c_{1.1}}$ | $e_{p_1}$ | $e_{c_1}, e^c_{c_{1.2}}$ | $e_{q_1}, e^c_{q_1}$ | $e_{i_2}, e^c_{c_2}, e^c_{h_1}, \mathbf{e^c_{i_1}}$ | $-$ |
| $S_8$ | $e_{i_1}, e^c_{c_{1.1}}$ | $e_{p_1}$ | $e_{q_{1.1}}, e^c_{q_{1.1}}$ | $e_{c_1}, e^c_{c_{1.2}}$ | $e_{q_{1.2}}, e^c_{q_{1.2}}$ | $e_{i_2}, e^c_{c_2}, e^c_{h_1}, \mathbf{e^c_{i_1}}$ |
| $S_9$ | $e_{i_1}, e^c_{c_1}$ | $e_{p_1}$ | $e_{i_2}, e^c_{c_2}$ | $e_{p_2}$ | $e_{q_1}, e^c_{q_1}$ | $e_{q_2}, e^c_{h_1}, \mathbf{e^c_{i_1}}$ |
| **(2b) $\triangle$ : non-local, $\times[1,\infty]$ : non-local** | | | | | | |
| $S_6$ | $e_{i_1}, e^c_{c_1}$ | $e_{p_1}$ | $e_{q_1}$ | $e_{i_2}, e^c_{c_2}, e^c_{q_1}, e^c_{h_1}, \mathbf{e^c_{i_1}}$ | $-$ | $-$ |
| $S_7$ | $e_{i_1}, e^c_{c_{1.1}}$ | $e_{p_1}$ | $e_{c_1}, e^c_{c_{1.2}}$ | $e_{q_1}$ | $e_{i_2}, e^c_{c_2}, e^c_{q_1}, e^c_{h_1}, \mathbf{e^c_{i_1}}$ | $-$ |
| $S_8$ | $e_{i_1}, e^c_{c_{1.1}}$ | $e_{p_1}$ | $e_{q_{1.1}}$ | $e_{c_1}, e^c_{c_{1.2}}$ | $e_{q_{1.2}}$ | $e_{i_2}, e^c_{c_2}, e^c_{q_1}, e^c_{h_1}, \mathbf{e^c_{i_1}}$ |
| $S_9$ | $e_{i_1}, e^c_{c_1}$ | $e_{p_1}$ | $e_{i_2}, e^c_{c_2}$ | $e_{p_2}$ | $e_{q_1}$ | $e_{q_2}, e^c_{q_1}, e^c_{h_1}, \mathbf{e^c_{i_1}}$ |

An event tree/graph can be opened, flushed, and closed by an application. After opening (or "registering" with a document) it stores occurring primitive and composite events. If it is flushed, remaining composite events are raised by operator nodes with **non-local** or **custom** termination mode. Closing an event tree/graph first flushes it and afterwards takes it out of order, e.g., when an XML document is closed after manipulation. An event tree/graph is automatically closed if the document is unloaded[3].

## 5.4 Implementation

The implemented proof-of-concept prototype extends the DOM event module of Apache's Xerces [63] and thus provides Java applications with composite events. Detected composite events are dispatched to the DOM node identified by the event's path instance. This has the advantage that the API to react to composite events is the same as for DOM mutation events (comprising event listeners etc.). This section briefly describes the prototype's execution model.

Every occurred primitive event is inserted into the event graph, which is processed to detect composite events. To determine the order in which the event trees of the graph are processed, which remains the same as long as neither the graph nor the trees are modified, the notion of *event tree dependency* is introduced. Event tree $t_b$ directly depends on event tree $t_a$, denoted as $t_a \rightarrow t_b$, iff $t_b$ uses composite event type $et_b$ to which event type $et_a$ defined by $t_a$ is compatible to, i.e., if $et_a \succeq et_b$. Naturally, direct dependency is not transitive. The transitive closure of a tree $t_a$ refers to the set of trees that directly and indirectly depend on it and is denoted by $(t_a)^+$. An event tree may not be in its own closure, i.e., it may not depend directly or indirectly on itself.

---

[3]cf. DOM event `http://www.w3.org/2001/xml-events#unload` [161])

Algorithm `processGraph` for processing event graph $G$ upon the occurrence of primitive event $e$ is shown below. It processes every event tree of the graph exactly once. Line **2** determines the ordered set of event trees that do not directly depend on any other event tree and assigns it to $T$. Subsequently, the set of occurred and raised events $E$ is initialized to $e$ (line **3**). While there are event trees left that need to be processed (line **4**), every event tree $t_i \in T$ is processed (line **5**–**6**, see later in this section), where $t_i$ denotes the $i^{th}$ element of $T$. Subsequently, all event trees that directly depend on any $t \in T$ and thus need to be processed are determined and assigned to $T$ (line **7**). Finally, $T$ is purged (line **8**, see later in this section).

```
1 processGraph(e)
2     T := {t ∈ G| ∄u ∈ G : u → t}
3     E := {e}
4     while T ≠ ∅ do
5        for i = 1 to |T| do
6           E := processTree(tᵢ, E)
7        T := {t' ∈ G|∃t ∈ T : t → t'}
8        T := purge(T)
```

Algorithm `processTree(t,E)` processes event tree $t$ with the set of events $E$ as follows. The tree is traversed in postorder (a form of depth-first traversal) during which (a) every visited event type node $n$ is tested for type compatibility with the event type of every event $e \in E$ and if they are compatible, i.e., $e.et \succeq n.et$, $e$ is stored in $n.evts$ (but not taken out of $E$), and (b) every visited operator node $o$ is evaluated if (i) an event was stored in a (direct) child node, (ii) if $o$ specifies a termination condition and for the primitive event $e \in E$ it applies that $o.et.pt \subset e.pt \lor o.et.pt = e.pt$, (iii) if $o$ has **non-local** termination mode, or (iv) if $o$ is a multiplicity operator node with $l = 0$ and for the primitive event $e \in E$ it applies that $o.et.pt \subset_e e.pt \lor o.et.pt = e.pt$. Finally, composite events that are raised in the root node of event tree $t$ are added to $E$ to be processed by dependent event trees later on.

Algorithm `purge(T)` removes every event tree $t_k \in T$ from $T$ if it is contained in the transitive closure of another event tree $t_j \in T$. Executing `purge(T)` in `processGraph` above guarantees that an event tree is processed exactly once at the latest time possible.

```
1 purge(T)
2     T' := {}
3     for i = 1 to |T| do
4        if ∄tⱼ ∈ T, j ≠ i : tᵢ ∈ (tⱼ)⁺
5           T' := T' ∪ tᵢ
6     return T'
```

The prototype including additional examples is available on the Web[4]. Naturally the approach for composite mutation events is independent of an implementation and can be used by any application in need of composite events in XML, presumably, such as [12] and [32].

---

[4]at `http://www.big.tuwien.ac.at/research/prototypes/composite-events/`

# Chapter 6

# Realizing the Metaschema

## Contents

The chapter discusses the realization of the metaschema of the Active XML Schema (AXS) approach, which was presented in Chapter 4. As described in Section 4.1, AXS defines active behavior within XML schemas along metadata, and stores traces of active behavior such as occurred events within XML documents along data. The semantic expressiveness of XML Schema, the schema language recommended by the W3C, however, is not sufficient to define the active semantics of Active XML Schema concepts. The contribution of this chapter is to identify, explore, and evaluate approaches to implementing AXS with XML Schema, discussing the trade-off between semantic expressiveness and interoperability. Assuming that AXS may be seen as representative for tailored schema languages, the findings of this chapter can be applied for arbitrary tailored schema languages.

The chapter is organized as follows. After giving a brief overview of the problem in Section 6.1, Section 6.2 defines criteria by which the approaches are evaluated, which are presented in Section 6.3. Subsequently, the four approaches are ranked according to the criteria. Section 6.4 summarizes the approaches by directly comparing them to each other. For a better understanding of the presented approaches, Section 6.5 discusses examples
from theory and practice. Finally, Section 6.6 gives an in-depth example of
applying a mixed approach for realizing AXS.

## 6.1   Introduction

Tailored schema languages define domain concepts thus semantics once and
for all across schemas. In relational databases for example, the schema lan-
guage defines concepts such as tables and foreign keys, constituting mod-
elling primitives for database schemas. Applications exhibiting event driven,
active behavior are another example where the use of a dedicated schema
language is favorable. Such a tailored schema language defines the seman-
tics of active concepts such as event-condition-action (ECA) rules and event
types independent of individual application schemas.

As the usage of XML increases, the need for tailored XML schema lan-
guages, which go beyond the semantic expressiveness of XML Schema, arises.
This goes in parallel with the emerging practice to define an XML syntax
both for schemas and instances (e.g., as RDF does).

For instance, the Active XML Schema (AXS) approach provides an XML
syntax both for schemas and instances. Its schema language[1] allows to
define circumstances having *intensional aspects* and/or *extensional aspects*.
While the former refers to circumstances that only affect a schema, such as
ECA rules, the latter refers to circumstances that only affect instances of a
schema, such as the structure of occurred events.

Using XML for schemas and instances instead of using other data for-
mats is beneficial with respect to *interoperability*, *openness*, and *integration*.
This means that schemas and instances described with XML syntax are ac-
cessible under various platforms and environments, they can be extended by
employing XML namespaces, and they can be integrated with other XML
standards such as XLink, XSLT, and RDF.

The contribution of this chapter is to identify, explore, and evaluate
approaches to implementing tailored metaschemas with XML Schema. In
particular, four approaches with distinct characteristics are presented. They
are explored and applied to Active XML Schema, giving insight into their
respective implications. Furthermore, the approaches are evaluated with
respect to criteria that have been identified to be relevant for the quality of
a tailored metaschema's implementation.

Since Active XML Schema comprises concepts that have intensional and
extensional aspects, it can be assumed to be a representative for tailored
metaschemas. Thus the chapter generalizes statements about Active XML

---

[1]For the purpose of readability, we use the term metaschema instead of schema language
throughout the rest of the chapter. If we talk about a schema expressed in XML Schema,
we concisely call it XML schema.

Schema to statements about tailored metaschemas. However, keep in mind that the findings presented in this chapter, except for the approaches and evaluation criteria themselves, are based on experiences in implementing the metaschema of Active XML Schema.

The running example is described in Example 65, which summarizes Examples 34, 35, and 36 from Chapter 4.

⊙ *Example 65.* A job agency provides an Active XML schema defining active document type j:JobAnnounce and a document academicJobs.xml having that type, which comprises a list of current job offers. A new job offer is announced by an invocation of operation announce(j:Job), which is defined by the active document type. It adds new job offer j at the end of the list. A university's science faculty posts, as a courtesy to its staff and students, relevant job offers supplied by the job agency at its document science.xml of active document type u:Faculty. The latter imports event class announce, using proxy jobSite to refer to the document from which the remote event class is imported. The proxy's value is bound in instance science.xml to document academicJobs.xml. Announced job offers are now locally available within a faculty's page in the form of events contained in the imported event class.

## 6.2   Evaluation Criteria

This section presents criteria to determine the quality of the different approaches. These quality criteria, which we have identified to be relevant in the narrow context of implementing tailored metaschemas, are related to quality factors proposed in literature [29, 88] to facilitate a better understanding of their implications. This is only done informally since it is not the focus of this chapter.

*Semantic expressiveness* describes how much semantics is expressed formally and concisely by a schema. Since semantics is defined by the tailored metaschema, a schema expressed therein is most expressive. When using XML Schema instead, semantics of the tailored metaschema need to be mapped to XML Schema. Because usually not all semantics can be mapped, schemas expressed in XML Schema are less expressive. The more semantics is explicit in a schema, the better the schema can be verified against an explicit metaschema such that errors can be detected at design time. Semantic expressiveness influences quality factors such as understandability, maintainability, testability, and reusability.

*Schema interoperability* in general describes the ability of a system to exchange schemas with other systems and interpret them. Interoperability also affects design time of a system in that it allows to reuse schema components from interoperable schemas and to reuse software components implemented for interoperable schemas and metaschemas. In the case of AXS, schema interoperability describes the ability of standard XML software to process schemas that have been created following each of the presented approaches. Thus it directly affects the extent of reusing standard XML software when writing applications that process such schemas. This criteria influences quality factors such as interoperability, flexibility, and portability.

*Locality of change* describes the self-containedness of a schema such that a change in one schema component does not require subsequent changes in dependent schema components in the same or other schemas. It is negatively affected by redundantly modelled schema components (i.e., components that model the same circumstance by different concepts) and non-atomic schema components (i.e., a circumstance is modelled by several dependent components). Locality of change is influenced by two aspects: First by the schema's environment (i.e., the employed metaschema and its imposed usage), and second, by the design of a given schema. Because the second aspect is application specific and thus independent of the presented approaches, we focus on the first aspect. Locality of change influences quality factors such as understandability, maintainability, and integrity.

## 6.3   Approaches

This section describes four approaches to implementing a tailored metaschema and ranks them with respect to the evaluation criteria presented in Section 6.2.

### 6.3.1   Proprietary Schema Approach

This approach expresses schemas directly in terms of the tailored metaschema, constituting the most "natural" approach with respect to schema design. As shown on the left of Figure 6.1, schema s (an XML document) is created by instantiating tailored metaschema m (e.g., an XML schema). Instance data in turn is created by proprietarily instantiating schema s.

For intensional aspects, i.e., aspects that apply for all instances but have no corresponding materialization at the instance level, it is not necessary to consider an XML syntax for M0.

⊙ *Example 66.*    The tailored metaschema and the exemplary proprietary schema below show how rules are defined and expressed by element actm:rule. A rule is identified by its name and is defined upon an event class (attributes name and definedOn), it comprises a condition and an action (elements actm:condition and actm:action). Since ECA rules only have intensional aspects, no XML syntax needs to be considered for M0.

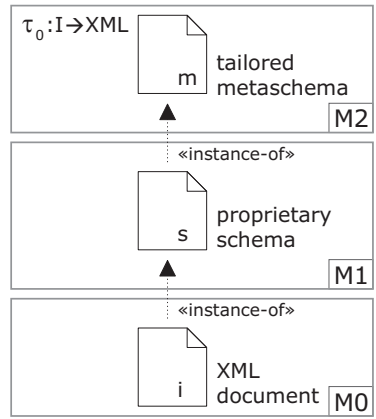

Figure 6.1: Proprietary Schema Approach

---

*(M2) Tailored metaschema with target namespace actm:*
```
01   <xs:element name="rule" ..> ..
02     <xs:sequence>
03       <xs:element name="condition" ../>
04       <xs:element name="action" ../>
05     </xs:sequence>
06     <xs:attribute name="name" type="xs:QName" ../> ..
07     <xs:attribute name="definedOn" type="xs:QName" ../> ..
08   </xs:element>
```

*(M1) Exemplary proprietary schema:*
```
09   <actm:rule definedOn="u:jobSite.announce" name="announceJobRule" ..>
10     <actm:condition>..</actm:condition>
11     <actm:action>..</actm:action>
12   </actm:rule>
```

---

For extensional aspects it is necessary to define so-called *instance transformation function* $\tau_0$: $I \rightarrow XML$. It defines how a proprietary instance is transformed into an XML document. Moreover an inverse function $\tau_0^{-1}$ must exist to transform an XML document back into a proprietary instance. Note that $\tau_0$ is defined at M2, i.e., independent of a particular schema. Therefore it can be reused across applications.

⊙ *Example 67.* The tailored metaschema below defines the import of an event class by element **actm:importedEventClass**. The exemplary proprietary schema imports event class **announce** from a remote document represented by proxy **u:jobSite**. Finally, the imported event class and events contained therein materialize at the instance level.

---

*(M2) Tailored metaschema with target namespace actm:*
```
01  <xs:element name="importedEventClass" ..>
02    <xs:complexType>
03      <xs:attribute name="name" type="xs:QName" ../>
04      <xs:attribute name="hasMemberType" type="xs:QName" ../> ..
05      <xs:attribute name="proxy" type="xs:QName" ../>
06      <xs:attribute name="remoteEvtCs" type="xs:QName" ../>
07    </xs:complexType>
08  </xs:element>
```

---

*(M1) Exemplary proprietary schema:*
```
09  <actm:importedEventClass name="jobSite.announce"
10    proxy="u:jobSite" remoteEvtCs="j:announce" .. />
```

---

*(M0) Exemplary instance:*
```
11  <jobSite.announce>
12    <actf:event id="e19" ..> ..
13      <actf:deliveryTime pt="2003-03-01T12:14.00.07" ../>
14      <actf:remoteEvent id="r47" ..> ..
15        <actf:occurrenceTime pt="2003-03-01T12:13:14.15" ../>
16      </actf:remoteEvent>
17    </actf:event>
18    <actf:event id="e20" ..> .. </event> ..
19  </jobSite.announce>
```

---

*Semantic expressiveness* is high because using a tailored metaschema makes all semantics explicit at the schema level. However, since these schemas are expressed in a proprietary format, *schema interoperability* is low because standard XML software cannot interpret the proprietary schema. This also affects implementation aspects, since standard XML software cannot be reused to validate instance documents against the proprietary schema. *Locality of change* is high, because the tailored metaschema does not impose redundant or non-atomic schema components.

### 6.3.2   Side by Side Approach

This approach is similar to the Proprietary Schema approach in that it uses an explicit tailored metaschema to define schemas. However, an XML schema is provided in addition, which does not replace the proprietary one but stands side by side to it. Likewise, instance transformation function $\tau_0$ is still used to serialize instances as XML.

The transformation of a proprietary schema into an XML schema is defined at M2 by so-called *schema transformation function* $\tau_1$: $S \to XSD$ and applied at M1 as depicted on the right of Figure 6.2. Function $\tau_1$ can be derived from $\tau_0$, since the latter defines the structure of XML documents implicitly. While $\tau_0$ is used at runtime, i.e., when documents are processed, $\tau_1$ is used at design time, i.e., when schemas are created. Because only extensional aspects of the proprietary schema are transformed to an XML schema, $\tau_1$ is *partial*.

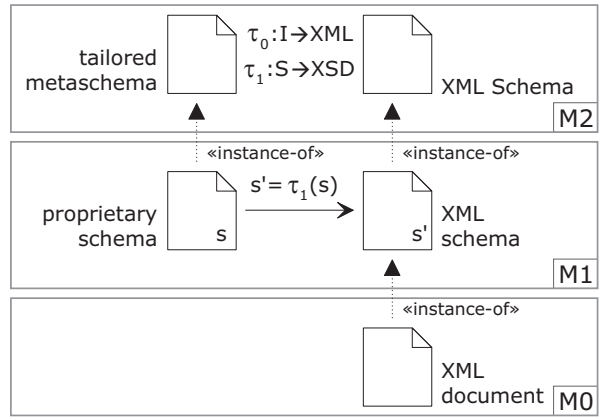

Figure 6.2: Side by Side Approach

⊙ *Example 68.*    The result of transforming extensional aspects of proprietary schema **s** (cf. Example 67) to XML schema **s'** by $\tau_1$ is shown below. It defines the imported event class as element **u:jobSite.announce**, which contains a sequence of **u:event** elements, each representing an imported event and in turn containing the wrapped remote event as element **u:remoteEvent**.

---

*(M1) Exemplary XML schema with target namespace u:*

```
01   <xs:element name="jobSite.announce" actm:eventClass="jobSite.announce">
02     <xs:complexType><xs:sequence>
03       <xs:element name="event" ..>
04         <xs:complexType><xs:sequence>
05           <xs:element name="remoteEvent" ..> .. </xs:element>
06         </xs:sequence></xs:complexType>
07         <xs:attribute name="id" type="xs:ID" ../>
08         <xs:attribute name="deliveryTime" ../>
09         <xs:attribute name="publicationTime" ../>
10         <xs:attribute name="deliveryTime" ../>
11       </xs:element>
12     </xs:sequence></xs:complexType>
13   </xs:element>
```

---

Still using a tailored metaschema to model schemas results in high *semantic expressiveness*. But in contrast to the Proprietary Schema Approach, this approach provides an XML schema for extensional aspects, resulting in higher *schema interoperability*. Thus standard XML software can be used to validate instance documents at the cost of implementing $\tau_1$ to transform schemas. Implementation of applications is supported by providing explicit links from components in s' to components in s (cf. attribute actm:eventClass in Example 68). Having two schemas expressing the same circumstances redundantly by components in terms of different metaschemas makes it necessary to keep them synchronized. Thus *locality of change* is low.

### 6.3.3   Framework Approach

This approach uses only an XML schema that expresses all circumstances formerly modelled by the proprietary schema as shown on the left of Figure 6.3. Thus it eliminates the need for proprietary schema s, transformation $\tau_1$, and synchronization of s with s'.

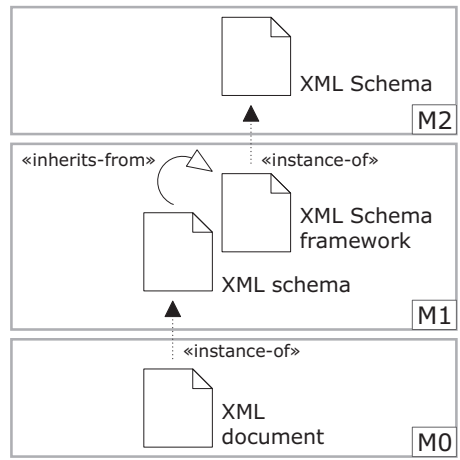

Figure 6.3: Framework Approach

Since intensional aspects are orthogonal to XML Schema, they can be expressed easily using XML Schema extension mechanisms (annotations and foreign attributes). Expressing extensional aspects is more complicated as they must be expressed solely with concepts provided by XML Schema.

The framework concept as known from object-oriented programming [90] can help in this situation. A framework is a means to provide a base schema common to all applications, along with conventions for its adaption and usage in the design of particular schemas. XML Schema provides a set of concepts that can be employed in framework design, such as abstract types, type derivation, abstract elements, and substitution groups (see [106] for a brief overview). Therefore an XML Schema framework comprises a set of reusable and/or specializable elements and types, which form the base schema, and a set of informal conventions describing their reuse and specialization.

⊙ *Example 69.*   The Active XML Schema framework below defines the structure of event classes which are represented by actf:eventClass ele-ments that are of abstract type actf:TEventClass. Moreover, abstract type
actf:TEventType describes events, which comprise an identifier (attribute
id), occurrence time (element occurrenceTime), and status (attribute sta-
tus). Type actf:TEventType is directly or indirectly extended by specialized
event types which are provided by the framework for all kinds of events (e.g.,
actf:TOperationEvtTp for operation events and actf:TImportedEvtTp for im-
ported events). Finally, in addition to event types, reusable event classes
are provided by the framework (e.g., actf:TEvtCs_ImportedEvtTp is a special
event class having event elements of type actf:TImportedEvtTp, a specialized
actf:TEventType).

---

*(M1) XML Schema framework with target namespace actf:*

```
01   <!-- Abstract base event type and class -->
02   <xs:element name="eventClass" type="actf:TEventClass" />
03   <xs:complexType name="TEventClass" abstract="true" />
04   <xs:complexType name="TEventType" abstract="true">
05     <xs:sequence>
06       <xs:element name="occurrenceTime" type="actf:TTimestamp" />
07     </xs:sequence>
08     <xs:attribute name="id" type="xs:ID" ../>
09     <xs:attribute name="status" type="actf:TEvtStatus" ..>
10   </xs:complexType>
11   <!-- Event type and class for operation events -->
12   <xs:complexType name="TOperationEvtTp" abstract="true"> ..
13     <xs:extension base="actf:TEventType"><xs:sequence>
14       <xs:element name="return" nillable="true"> .. </xs:element>
15       <xs:element name="diff" nillable="true" minOccurs="0"> .. </xs:element>
16     </xs:sequence></xs:extension> ..
17   </xs:complexType>
18   <xs:complexType name="TEvtCs_OperationEvtTp"> ..
19     <xs:extension base="actf:TEventClass"><xs:sequence>
20       <xs:element name="event" type="actf:TOperationEvtTp" ../>
21     </xs:sequence></xs:extension> ..
22   </xs:complexType>
23   <!-- Event type and class for imported events -->
24   <xs:complexType name="TImportedEvtTp" abstract="true"> ..
25     <xs:extension base="actf:TEventType"><xs:sequence>
26       <xs:element name="remoteEvent" type="actf:TEventType" />
27     </xs:sequence> .. </xs:extension> ..
28   </xs:complexType>
29   <xs:complexType name="TEvtCs_ImportedEvtTp"> ..
30     <xs:extension base="actf:TEventClass"><xs:sequence>
31       <xs:element name="event" type="actf:TImportedEvtTp" ../>
32     </xs:sequence></xs:extension> ..
33   </xs:complexType>
```

---

To some extent, conventions defining the reuse and specialization of
schema components provided by an XML Schema framework can be en-
forced by mechanisms of XML Schema. For example, an abstract type must be specialized before it can be used, or an abstract substitution group's head
has to be substituted by an element of an appropriate type. Unfortunately,
in many cases these mechanisms are not sufficient to enforce a correct us-
age of the framework. Therefore, schema designers must know and follow
informal conventions regarding the use of framework components.

⊙ *Example 70.*    Modelling the exemplary schema based on the XML
framework requires the definition of the following (as shown below). First,
event type **j:TExecAnnounceEvtTp_Imported** is defined for imported events,
which is done by extending event type **actf:TImportedEvtTp**. Since element
**actf:remoteEvent** cannot be refined by the exemplary schema because it is not
in the framework's namespace, an annotation is provided that indicates that
elements representing remote events shall be of type **j:TExecAnnounceEvtTp**
in instance documents.    Second, a corresponding event class (element
**j:jobSite.announce**) is defined by making it part of the substitution group
headed by **actf:eventClass**. The type of the event class's **actf:event** elements
is defined in an annotation as **j:TExecAnnounceEvtTp_Imported**, due to the
same reasons as with the event type for imported events. Proxy **u:jobSite**,
which is an intensional aspect, is defined within an annotation. Note that
the use of the annotation as well as the definition of parallel type hierarchies
comprising event types and event classes are informal conventions, i.e., not
enforceable by the XML Schema framework.

---

*(M1) Exemplary XML schema with target namespace u:*

```
01  <!-- Imported event type -->
02  <xs:complexType name="TExecAnnounceEvtTp_Imported"> ..
03    <xs:extension base="actf:TImportedEvtTp">
04      <xs:annotation><xs:appinfo>
05        <actm:remoteEvtTp="j:TExecAnnounceEvtTp"/> ..
06      < /xs:annotation></xs:appinfo>
07    </xs:extension> ..
08  </xs:complexType>
09  <!-- Imported event class -->
10  <xs:element name="jobSite.announce"
          type="actf:TEvtCs_ImportedEvtTp" substitutionGroup="actf:eventClass">
11    <xs:annotation><xs:appinfo>
12      <actm:proxy name="jobSite" forDocType="j:JobAnnounce" type="single" / >
13      <actm:hasMemberType="u:TExecAnnounceEvtTp_Imported" /> ..
14    </xs:appinfo></xs:annotation> ..
15  </xs:element>
```

---

Regarding the characteristics of the framework approach, most notably
is the lack of *semantic expressiveness* of extensional aspects. This is exem-
plified by comparing the import of an event class by the proprietary schema
shown in Example 67 with the above schema. Furthermore, informal con-
ventions that must be followed when using a framework severely impact semantic expressiveness. On the positive side, *schema interoperability* is high since schemas (and frameworks) are expressed solely in XML Schema. *Locality of change* is medium since the framework may impose modifications of multiple schema components in order to achieve the modification of a single circumstance.

### 6.3.4 Specialized XML Schema Approach

This approach extends XML Schema with new concepts of the tailored metaschema as shown in Figure 6.4 opposed to the framework approach, which expresses new concepts of the tailored metaschema by XML Schema concepts. To relate new concepts to XML Schema concepts the mechanisms provided by XML Schema itself are used, because XML Schema at M2 is defined by an XML schema, (cf. [157]), which in turn assumes XML Schema at M3 (as one can see, XML Schema is meta-circularly defined [105]). Thus plenty of possibilities exist to relate concepts, such as element composition, type composition, or type derivation. Also redefinition as shown in [126] is an option. Note, however, while [126] focusses on restricting XML Schema, this approach focusses on extending it.

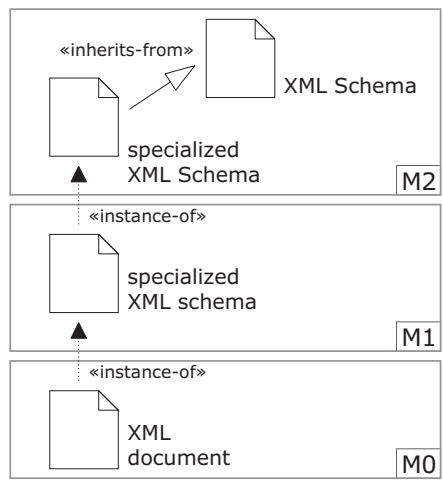

Figure 6.4: Specialized XML Schema Approach

Depending on whether XML Schema is to be extended by intensional or extensional aspects, different procedures are followed. An extension with intensional aspects is simply a matter of adding new concepts to XML Schema without the need to relate them to existing concepts. On the contrary, extensional aspects *must* be defined as specializations of existing concepts such as elements and attributes, in order to inherit the extensional semantics of those concepts. Intensional aspects thus have no standard meaning, i.e., they can be safely ignored by standard XML Schema validators.

This approach has the advantage that an XML Schema validator can interpret specialized XML schemas, because it is possible to derive the basic meaning of a schema component of a specialized concept from the XML Schema concept it is based on. Or in different terms, it is possible to perform a *downcast* according to the principle of type substitutability. Unfortunately, standard XML Schema validators currently do not provide for a plug-able XML Schema necessary for a downcast.

⊙ *Example 71.* In XML Schema, group xs:schemaTop defines the content of element xs:schema, the document element of every XML schema. The group defines a choice of elements xs:element, xs:attribute, and others. It is redefined[2] by the specialized XML Schema to include elements actm:rule and actm:importEventClass. Because a rule has only intensional aspects, element actm:rule can be declared by referencing the respective element declaration of the tailored metaschema depicted in Example 66. Because an imported event class has extensional aspects, element actm:importedEventClass is indirectly derived from xs:topLevelElement (via xs:actm.EventSequence), the type of a global element declaration in XML Schema. The two schema documents with different namespaces shown below form the specialized XML schema.

---

*(M2) Specialized XML Schema with targetNamespace xs:*

```
01   <xs:redefine schemaLocation="XMLSchema.xsd">
02      <xs:group name="schemaTop"><xs:choice>
03         <xs:group ref="xs:schemaTop"/>
04         <xs:element ref="actm:rule"/>
05         <xs:element ref="actm:importedEventClass"/> ..
06      </xs:choice></xs:group>
07   </xs:redefine>
08   <xs:complexType name="actm.EventSequence"> ..
09      <xs:restriction base="xs:topLevelElement"> .. </xs:restriction>
10   </xs:complexType>
```

---

*(M2) Specialized XML Schema with targetNamespace actm:*

```
01   <xs:element name="importedEventClass"> ..
02      <xs:extension base="xs:actm.EventSequence'> ..
03         <xs:attribute name="proxy" type="xs:QName" use="required"/>
04         <xs:attribute name="remoteEvtCsName" type="xs:QName" use="required"/>
05         <xs:attribute name="exported" type="xs:boolean" use="required"/>
06      </xs:extension>
07   </xs:element>
```

---

On the negative side, the power of a downcast is very limited compared to an explicitly defined schema transformation $\tau_1$. In particular, a specialized concept can not arbitrarily modify extensional semantics of its base concept. For instance, the extensional semantics of a specialized element is always limited to that of exactly one element. Therefore it is not possible to define a particular composition of elements by means of one specialized element.

---

[2]The redefinition of xs:schemaTop is a group redefinition that contains a reference to itself. Thus, it is semantically equivalent to a derivation by extension, being applied to an element group instead of a complex type.

Overall *semantic expressiveness* is medium, whereby the semantic expressiveness of intensional and extensional aspects differ. It is high for intensional aspects because they are expressed in terms of their unconstrained metaschema. It is medium for extensional aspects, because their metaschema is constrained by the concepts of XML Schema. If standard XML Schema validators provide for a plug-able XML Schema, *schema interoperability* will be high since they can interpret specialized XML Schemas. Unfortunately, in practice this is not yet the case causing low interoperability. Concerning *locality of change* it is advantageous that only one metaschema is employed. However, one concept of the tailored metaschema is possibly expressed by several concepts of XML Schema, producing several dependent schema components. Therefore locality of change is medium.

## 6.4 Comparison

When comparing the approaches' characteristics summarized in Table 6.1, it gets evident that there is a *tradeoff* between semantic expressiveness and schema interoperability. The Proprietary Schema approach defines new concepts at M2 not defined by XML Schema and thus imposes proprietary schemas at M1, resulting in high expressiveness but low interoperability. The Side by Side Approach tries to overcome this by defining a transformation from new concepts to XML Schema concepts at M2 and applying it to proprietary schemas at M1. This increases interoperability, however, at the cost of locality of change. The Framework approach goes one step further by expressing new concepts by XML Schema concepts at M1, having positive effects on interoperability and locality of change, but negative effects on expressiveness. Finally, the Specialized XML Schema approach directly extends XML Schema with the new concepts at M2. It suffers from the lack of support by existing XML Schema validators and the constraints of the underlying XML Schema.

Since there is no single best approach, one has to choose the most appropriate one based on given requirements. In case schemas will change often, locality of change is the primary criterion with the Proprietary Schema approach being favorable. In case instance documents have to be shared with other partners, schema interoperability is the primary criterion with the Proprietary Schema Approach falling behind. In practice it may be beneficial to use a mixed approach (for an example see Section 6.6).

Table 6.1: Characteristics of the Presented Approaches

| Criteria | Proprietary Schema | Side by Side | Framework | Specialized XML Schema |
|---|---|---|---|---|
| Semantic expressiveness | high | high | low | med. |
| Schema interoperability | low | med. | high | high[†] |
| Locality of change | high | low | med. | med. |

[†]Assuming standard XML validators provide for a plug-able XML Schema

## 6.5  Related Work

Examples for approaches employing the *Proprietary Schema approach* are RDF and JavaBeans Persistence [138]. The RDF standard defines RDF's tailored metaschema and a syntax of RDF instances as XML by an EBNF grammar. This grammar can be seen as a declarative specification of $\tau_0$. In addition, the RDF Schema (RDFS) standard defines a proprietary schema language for RDF. Going beyond XML, JavaBeans Persistence provides for serialization of JavaBean objects as XML documents. It realizes $\tau_0$ by a dedicated Java class. Here, the Java language constitutes the tailored metaschema.

Among approaches following the *Side by Side approach* are [95, 112, 120, 128]. [95] describes transforming OIL [78] ontologies to XML schemas by textually describing $\tau_1$ that transforms OIL concepts to XML Schema concepts. Independent of XML, Microsoft's ADO.NET DataSet [112] implements among others a generic mapping between the relational model (constituting the tailored metaschema) and XML. It allows to read relational data and to write XML data with its XML schema and vice versa, thus implementing $\tau_0$, $\tau_1$, and their inverse. The Side by Side approach has been also extensively explored in [128] recently, yielding an abstract algebra for model mapping [28] (the notion of "model" in [28] corresponds to "schema" in our approach). OMG's XMI [120] is distantly related, because it does not exactly fit the side-by-side approach's structure. However, it has in common that XMI defines a schema transformation, but of metaschemas (i.e., instance of MOF [119]) into XML Schemas. Thus the transformation takes place at M2 instead of M1 and the transformation function, which could be named $\tau_2$, is defined at M3. The intention of XMI is for example to transform the UML metamodel into an XML Schema, and UML models into XML documents. Aligning XMI's intention with the side-by-side approach, the resulting XML document of the former could be seen as the proprietary schema of the latter.

The *Framework approach* has not been employed in the XML field yet. We suspect that a major reason for not employing it is that frameworks usually evolve from simple XML schemas instead of being created from scratch by implementing a tailored metaschema. Going beyond XML, an example of fostering the Framework approach is UML with its extension mechanisms [130]. Thus, instead of extending the UML metaschema at M2 (called metamodel in UML), the extension mechanisms provide a means to customize UML at M1. Extension mechanisms are the main concepts to build reusable frameworks, called profiles in UML. Another example of providing a new semantic concept at M1 is the role pattern [16, 93]. For example, it has been implemented in Smalltalk in terms of a predefined framework [72].

The *Specialized XML Schema approach*, as the Framework approach, has not been employed in the XML field yet. A major reason could be not wanting to lose interoperability. However, tailoring metaschemas is well known in the non-XML literature. So-called open data models have been proposed in the past (e.g., [94, 107]), which consist of a few built-in concepts but which can be extended by additional modeling concepts at M2 for specific application needs.

## 6.6 Employing a Mixed Approach for AXS

When implementing Active XML Schema we decided to follow a mixed approach, mixing Side by Side and Framework approach. We employed the Side by Side approach to provide maximum semantic expressiveness for human modelers. Extensional aspects of proprietary schemas are transformed to XML schemas adhering to a dedicated framework by an XSLT stylesheet representing $\tau_1$. Thus we are fully interoperable and able to reuse standard XML software. Employing both approaches in combination minimizes the required transformation functionality that has to be provided by $\tau_1$. The mixed approach is shown in Figure 6.5. Since Active XML schemas are assumed not to change often, low locality of change is not considered a problem.

In the following the different components of AXS's implementation and its use in the context of the chapter's running example are shown. AXS's implementation comprises its tailored metaschema, the *Active XML Metaschema* m, and its XML Schema framework, the *Active XML Framework* f. The namespace prefixes that are used for them are actm[3] and actf[4] respectively. The exemplary use of m and f for representing a rule and the import of an event class are shown by a proprietary schema s which is an

---

[3]denoting `http://big.tuwien.ac.at/axs/metaschema/1.0`
[4]denoting `http://big.tuwien.ac.at/axs/framework/1.0`

instance of m, and an XML schema s' resulting from a transformation of s
to an instance of f.

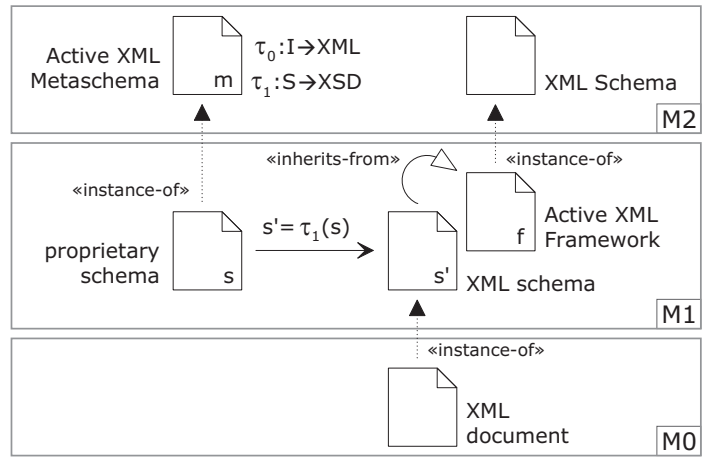

Figure 6.5: Realizing Active XML Schema using a Mixed Approach

First, the Active XML Metaschema for rules is shown below, forming a
part of metaschema m. Here, the metaschema for rules, which is depicted in
Figure 4.7 as an UML diagram, is expressed as an XML schema. The tailored
metaschema has already been sketched in Example 66. A rule comprises a
condition and an action, is described by a name and a priority, and is defined
on an event class.

```
(M2) Active XML Metaschema for rules (with target namespace actm):
<xs:element name="rule">
    <xs:complexType>
        <xs:sequence>
            <xs:element name="condition" type="actm:NativeCode" minOccurs="0" />
            <xs:element name="action" type="actm:NativeCode" />
        </xs:sequence>
        <xs:attribute name="name" type="xs:QName" use="required" />
        <xs:attribute name="priority" type="xs:integer" use="optional" />
        <xs:attribute name="definedOn" type="xs:QName" use="required" />
    </xs:complexType>
</xs:element>
```

Second, the exemplary use of the above metaschema for defining a rule
is shown below, forming a part of proprietary schema s. Again, the schema
has already been sketched in Example 66. Here, it shows the complete rule
introduced in Example 38 and shown in Figure 4.5. All unprefixed elements
are in the namespace of the Active XML Metaschema, the condition and
action are expressed using XSLT. Imported event class u:jobSite.announce on which the rule is defined on, is described later in this section. The **actm:invokeOperation** element is an XSLT extension element that is used to invoke an operation defined by AXS. Bindings **$evt** represents the event that triggered the rule, **$cond** represents the value resulting from the condition's evaluation, and **$staticDoc** represents the contents of the "static" XML document.

---

*(M1) Proprietary schema defining rule announceJobRule (with target namespace u):*

```
<rule name="announceJobRule" definedOn="u:jobSite.announce"
 xmlns="http://big.tuwien.ac.at/axs/metaschema/1.0">
  <condition lang="http://www.w3.org/1999/XSL/Transform">
     <xsl:value-of select="$evt//j:job[j:field=$staticDoc/u:fieldOfInterest]"/>
  </condition>
  <action lang="http://www.w3.org/1999/XSL/Transform">
     <invokeOperation name="postJob">
       <xsl:value-of select="$cond"/>
       </invokeOperation>
  </action>
</rule>
```

---

Third, the Active XML Metaschema for event types is shown below, forming a part of metaschema **m**. Here, the metaschema for event types, which is depicted in Figure 4.4 and Figure 4.6 as UML diagrams, is expressed as an XML schema. This part of the tailored metaschema has not been sketched in this chapter yet.

---

*(M2) Active XML Metaschema for event types (with target namespace actm):*

```
<!-- Base event type -->
<xs:element name="eventType" type="actm:EventType"/>
<xs:complexType name="EventType" abstract="true">
   <xs:attribute name="name" type="xs:QName" use="required"/>
</xs:complexType>
<!-- Operation event type -->
<xs:complexType name="OperationEvtTp">
   <xs:complexContent>
      <xs:extension base="actm:EventType">
        <xs:sequence>
           <xs:element ref="actm:operationRef"/>
        </xs:sequence>
        <xs:attribute name="timeSpec" type="actm:TmSpec" use="required"/>
      </xs:extension>
   </xs:complexContent>
</xs:complexType>
<!-- Imported event type -->
<xs:complexType name="ImportedEvtTp">
```

```
<xs:complexContent>
  <xs:extension base="actm:EventType">
    <xs:attribute name="remoteEvtTp" type="xs:QName" use="required" />
  </xs:extension>
</xs:complexContent>
</xs:complexType>
```

Fourth, the exemplary use of the above metaschema for defining an operation event type is shown below, forming a part of proprietary schemas. The schema has not been sketched in this chapter yet. It defines operation event type j:TExecAnnounceEvtTp for events that are signalled after the execution of operation announce(j:Job) of interface j:JobAnnounce.

*(M1) Proprietary schema defining event type TExecAnnounceEvtTp (with target namespace j):*
```
<eventType name="TExecAnnounceEvtTp" timeSpec="after"
  xsi:type="actm:OperationEvtTp"
  xmlns="http://big.tuwien.ac.at/axs/metaschema/1.0">
  <operationRef interfaceNm="j:JobAnnounce" operationNm="announce">
    </operationRef>
</eventType>
```

Fifth, the Active XML Metaschema for event classes is shown below, forming a part of metaschema m. Here, the metaschema for event classes, which is depicted in Figure 4.4 and Figure 4.6 as UML diagrams, is expressed as an XML schema. The tailored metaschema has already been sketched in Example 67.

*(M2) Active XML Metaschema for event classes (with target namespace actm):*
```
<!-- Base event class -->
<xs:element name="eventClass" type="actm:EventClass" />
<xs:complexType name="EventClass">
  <xs:attribute name="name" type="xs:QName" use="required" />
  <xs:attribute name="exported" type="xs:boolean" use="required" />
  <xs:attribute name="hasMemberType" type="xs:QName" use="required" />
</xs:complexType>
<!-- Proxy -->
<xs:element name="proxy">
  <xs:complexType>
    <xs:attribute name="name" type="xs:QName" use="required" />
    <xs:attribute name="forDocType" type="xs:Name" use="required" />
    <xs:attribute name="type" use="required" type="actm:ProxyTp" />
  </xs:complexType>
</xs:element>
<!-- Imported event class -->
<xs:element name="importedEventClass" type="actm:ImportedEvtCs"
```

```
    substitutionGroup="actm:eventClass" />
<xs:complexType name="ImportedEvtCs">
  <xs:complexContent>
    <xs:extension base="actm:EventClass">
      <xs:sequence>
        <xs:element name="filterExpr" type="actm:NativeCode"
          minOccurs="0" maxOccurs="1" />
      </xs:sequence>
      <xs:attribute name="proxy" type="xs:QName" use="required" />
      <xs:attribute name="remoteEvtCs" type="xs:QName" use="required" />
    </xs:extension>
  </xs:complexContent>
</xs:complexType>
```

Sixth, the exemplary use of the above metaschema for defining an imported event type and an imported event class is shown below, forming a part of proprietary schema s. The schema has already been sketched in Example 67. It defines remote event type j:TExecAnnounceEvtTp_Imported for wrapping events of operation event j:TExecAnnounceEvtTp. Proxy u:jobSite is used in importing remote event class j:announce as imported event class u:jobSite.announce.

*(M1) Proprietary schema defining an imported event type and event class (with target namespace u):*
```
<actm:eventType name="TExecAnnounceEvtTp_Imported"
  remoteEvtTp="j:TExecAnnounceEvtTp" xsi:type="actm:ImportedEvtTp" />
<actm:proxy name="jobSite" forDocType="j:JobAnnounce" type="single" />
<actm:importedEventClass name="jobSite.announce" exported="false"
  proxy="u:jobSite" remoteEvtCs="j:announce"
  hasMemberType="u:TExecAnnounceEvtTp_Imported" />
```

Seventh, the Active XML Framework for event types is shown below, forming a part of framework f. The framework has already been sketched in Example 69. Here, the complete definitions of the abstract base event type actf:TEventType and the abstract operation event type actf:TOperationEvtTp are shown, where the latter is derived from the former. These types are part of the event type hierarchy at the schema layer (M1) mentioned in Section 4.1.2.

*(M1) Active XML Framework for event types (with target namespace actf):*
```
<xs:element name="event" type="actf:TEventType" />
<xs:complexType name="TEventType" abstract="true">
  <xs:sequence>
    <xs:element name="occurrenceTime" type="actf:TTimestamp" />
  </xs:sequence>
  <xs:attribute name="id" type="xs:NMTOKEN" use="required" />
```

```
<xs:attribute name="status" type="actf:TEvtStatus" use="optional" />
</xs:complexType>
<xs:complexType name="TOperationEvtTp" abstract="true">
  <xs:complexContent>
    <xs:extension base="actf:TEventType">
      <xs:sequence>
        <xs:element name="return" nillable="true">
          <xs:complexType mixed="true">
            <xs:sequence>
              <xs:any namespace="##any" processContents="lax" />
            </xs:sequence>
          </xs:complexType>
        </xs:element>
        <xs:element name="diff" nillable="true" minOccurs="0">
          <xs:complexType mixed="true">
            <xs:sequence>
              <xs:any namespace="##any" processContents="lax" />
            </xs:sequence>
          </xs:complexType>
        </xs:element>
      </xs:sequence>
    </xs:extension>
  </xs:complexContent>
</xs:complexType>
```

Eighth, XML schema **s'** as the result of transforming the part of proprietary schema **s** that defines an operation event type (cf. fourth), is shown below. It defines how the extensional aspects of events are represented at the instance layer (M0).

*(M1) XML schema for operation event type TExecAnnounceEvtTp (with target namespace j):*

```
<xs:complexType name="TExecAnnounceEvtTp">
  <xs:complexContent>
    <xs:extension base="actf:TOperationEvtTp">
      <xs:sequence>
        <xs:element name="job" type="j:Job" />
      </xs:sequence>
    </xs:extension>
  </xs:complexContent>
</xs:complexType>
```

Ninth, the Active XML Framework for imported event types and imported event classes is shown below, forming a part of framework **f**. The framework has already been sketched in Example 69. Here, the complete definitions of the abstract imported event type **actf:TImportedEvtTp** and abstract imported event class **actf:TEvtCs_ImportedEvtTp** are shown. As one can see, the event class stores events of the imported event type, i.e., its member type is the imported event type.

```
(M1) Active XML Framework for imported event types and classes (with target
       namespace actf):
<!-- Imported event type -->
<xs:complexType name="TImportedEvtTp" abstract="true">
  <xs:complexContent>
    <xs:extension base="actf:TEventType">
      <xs:sequence>
        <xs:element name="publicationTime" type="actf:TTimestamp" />
        <xs:element name="deliveryTime" type="actf:TTimestamp" />
        <xs:element name="remoteEvent" type="actf:TEventType" />
      </xs:sequence>
    </xs:extension>
  </xs:complexContent>
</xs:complexType>
<!-- Imported event class -->
<xs:element name="eventClass" type="actf:TEventClass" abstract="true" />
<xs:complexType name="TEvtCs_ImportedEvtTp">
  <xs:complexContent>
    <xs:extension base="actf:TEventClass">
      <xs:sequence>
        <xs:element name="event" type="actf:TImportedEvtTp" minOccurs="0"
               maxOccurs="unbounded" />
      </xs:sequence>
    </xs:extension>
  </xs:complexContent>
</xs:complexType>
```

Tenth, XML schema **s'** as the result of transforming the part of proprietary schema **s** that defines an imported event type and imported event class (cf. sixth), is shown below. It defines how the extensional aspects of imported events and imported event classes are represented at the instance layer (M0).

```
(M1) XML schema for an imported event type and class (with target namespace u):
<xs:complexType name="TExecAnnounceEvtTp_Imported">
  <xs:complexContent>
    <xs:extension base="actf:TImportedEvtTp" />
  </xs:complexContent>
</xs:complexType>
<xs:element name="jobSite.announce" type="actf:TEvtCs_ImportedEvtTp"
  substitutionGroup="actf:eventClass" />
```

# Part III

# Consistency in Document Flows on the Web

# Chapter 7

# Traceable Document Flows (TDF)

## Contents

This chapter describes the first layer of the two-layered extension to AXS. It provides for traceable document flows, abbreviated *TDFs*, and is also referred to as *layer IM-1*.

To exemplify the layer's, i.e., TDFs' usefulness, consider the following problems with which on is regularly confronted in today's personal ad-hoc data exchange, e.g., via email. First, it is hardly possible to determine where a document in one's file system originated. Usually answers to this question are found after human inquiries using one's email system. Second, it is even more difficult to determine who read or edited a document before. Usually this involves lengthy inquiries and requires interviewing other people. Third, metadata about the document, if not stored proprietarily as part of the document, e.g., as it is the case with documents in Sun StarOffice or Microsoft Office format, are lost during data exchange between individuals.

Again, harvesting metadata is a human task carried out by determining and interviewing people who read or edited the document before.

Section 7.2 continues to exemplify TDFs' usefulness by describing four application scenarios where TDFs can be beneficial. The first scenario in Subsection 7.2.1 shows how TDFs can be used to trace documents exchanged via email, e.g., for the use by a program chair who sends papers and review forms to program committee members. The second scenario in Subsection 7.2.2 describes the use of TDFs for collaborative authoring, e.g., when editing or writing a book. The third scenario in Subsection 7.2.3 shows how TDFs can be used as a light-weight infrastructure to support workflows, either to discover existing workflows (bottom-up) or to model workflows and constrain document flows (top-down). The last scenario in Subsection 7.2.4 shows how TDFs can be utilized to share papers in a scientific community, e.g., to share metadata about papers and determine domain experts.

The chapter is organized as follows. Section 7.1 presents the infrastructure model for traceable document flows. It comprises documents, versions, locations, and interactions thereon. In parallel, an ontological framework is presented with which artifacts such as a concrete document or a concrete version can be described. Section 7.2 presents the above application scenarios, Section 7.3 sketches a prototypical implementation and finally Section 7.4 discusses related work.

## 7.1 Infrastructure Model and Ontological Framework

This section presents the infrastructure model for TDFs, which employs versioning as known from configuration management [52]. To make instantiations of the model's concepts interoperable and thus available for other applications, we employ Semantic Web technology to describe them, i.e., OWL [165]. We define an ontological framework for document flows comprising these concepts, which are used when describing instantiations. Being a framework, any other ontology can be used in addition, e.g., for annotating documents. The use of additional ontologies is influenced by the given application as is exemplified in Section 7.2.

### 7.1.1 Documents and Versions

A document is identified by a *document identifier* (DID) and keeps its identity across modifications. Since the model is intended to be implemented on the Web, a DID is a URI [80]. Between two subsequent modifications a document is represented by a static concrete occurrence having certain contents. In the presented model a document exists both as an abstract concept represented by its DID (referred to as *document*) and as set of concrete occurrences (referred to as *versions*).

A version is identified by a *version identifier* (VID) which is a URI. A modification to a version in the presented model does not necessarily result in a new version, i.e., a version's content may change over time. All versions of a document together represent its history. A version's VID is globally unique, i.e., unique across documents so that it suffices to identify and retrieve the version.

⊙ *Example 72.*   A document representing a purchase order and containing items to be ordered is identified by its DID, e.g., $d_{order}$. Each time an order item is added, modified, or deleted, a version of the document is modified. Selected versions may be made persistent and have an assigned VID, e.g., $v_1$ and $v_2$.

The underlying version model is organized as a two-level *acyclic version graph* (cf. [52]) that is composed of branches, each consisting of a sequence of versions. This model is employed in several version management systems such as WebDAV [81, 82] and CVS [1]. While versions constitute the nodes in the graph, three relationships between versions are distinguished, namely successor, offspring (starting a new branch), and predecessor. Each version has at most one successor and possibly several offsprings in different branches. Successor and offspring versions together are also referred to as following versions. Because a version can be merged into a version of another branch, a version can have several predecessors.

Additionally to the used concepts known from version models, we introduce the concepts of a current version and a frozen version. A *current version* is a version that has no succeeding version. Multiple current versions may exist per document in different branches. To provide for a consistent version graph, versions that are non-current can be *frozen* to prevent them from further being modified, avoiding that they become inconsistent with succeeding versions. Whether non-current versions are frozen is determined by configuration for each document separately (see below). By not freezing a version a very unconstrained exchange of versions can take place, because a succeeding version does not prevent the version from being modified.

⊙ *Example 73.*   An exemplary version graph of $d_{order}$ is depicted on the left of Figure 7.1, showing five versions and three branches. Version numbers used throughout the thesis have an odd amount of numbers (digits) separated by periods, whereas numbers at odd digits count versions (e.g., $v_2$ for the second version) and numbers at even digits count branches (e.g., $v_{1.2.3}$ refers to the third version in branch $v_{1.2}$, which is the second branch following version $v_1$). In Figure 7.1 versions $v_1$, $v_{1.1.1}$, $v_{1.2.1}$, and $v_2$ are frozen, while $v_3$ is a current version.

To talk about the topology of a version graph, *path $p = (v_i, ..., v_j)$* is defined as a sequence of versions starting with $v_i$ and ending with $v_j$, where version $v_k$ $(i \leq k < j)$ is followed by $v_{k+1}$ in the version graph. Since version $v_k$ can have one successor and multiple offspring versions, several paths between $v_i$ and $v_j$ may exist.

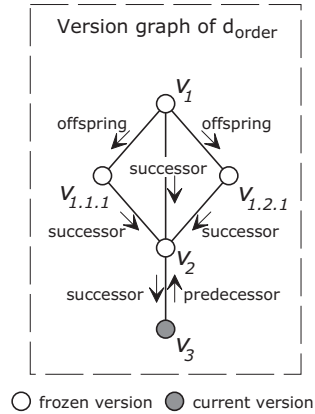

Figure 7.1: Exemplary Version Graph

⊙ *Example 74.* Concerning the version graph of document $d_{order}$ shown in Figure 7.1, paths $p_1 = (v_1, v_{1.1.1}, v_2, v_3)$, $p_2 = (v_1, v_{1.2.1}, v_2, v_3)$, and $p_3 = (v_1, v_2, v_3)$ exist between $v_1$ and $v_3$.

Documents and versions are modelled as classes tdf:Document[1] and tdf:Version and are described by several properties. At minimum a document as wells as a version must be described by properties tdf:title and tdf:description. Being URIs, a document's DID as well as a version's VID are used directly as identifiers in semantic descriptions. Each document must be additionally described by property tdf:freezeNonCurrentVersions, which defines whether its non-current versions are frozen. Its value is specified when the document is created. Each version must be additionally described by property tdf:isVersionOf, which relates it to the document it is a version of. Moreover, a version may have successor versions (property tdf:hasSuccessorVersion), offspring versions (property tdf:hasOffspringVersion), and preceding versions (property tdf:hasPrecedingVersion). A version may be stored at a peer (property tdf:isStoredAt). Of course, additional ontologies can be used to further describe documents and versions, i.e., to describe a user's annotations.

Classes and properties describing documents and versions are closely linked to elements and qualifiers of the Dublin Core [84], thereby grounding them on commonly understood semantics. The relationships are as follows:

---

[1]prefix "tdf" denotes namespace http://www.big.tuwien.ac.at/research/documentflows/1.0/

1. tdf:name $\sqsubseteq$ dc:title[2],
   tdf:title $\sqsubseteq$ dc:title, and
   tdf:description $\sqsubseteq$ dc:description;

2. tdf:isVersionOf $\sqsubseteq$ dcterms:isVersionOf[3];

3. tdf:hasSuccessorVersion $\sqsubseteq$ tdf:hasFollowingVersion,
   tdf:hasOffspringVersion $\sqsubseteq$ tdf:hasFollowingVersion, and
   tdf:hasFollowingVersion $\sqsubseteq$ dcterms:isReplacedBy;

4. tdf:hasPrecedingVersion $\sqsubseteq$ dcterms:replaces, where
   tdf:hasPrecedingVersion $\equiv$ tdf:hasFollowingVersion$^{-}$, and
   dcterms:replaces $\equiv$ dcterms:isReplacedBy$^{-}$.

### 7.1.2   Locations

Since a document as an abstract concept cannot be allocated itself, versions are allocated at *locations*. Each version is allocated at a single location, while one location can host multiple versions of the same document. Note that the concept of location is new to version models and thus makes the presented version model more expressive. Locations form the basis for modelling the distribution aspect of documents.

The meaning that is associated with a location can be very diverse. It is assumed, however, that a location is in a close relationship to a natural or juristic person such as a version's creator and owner. This relationship may also be "weaker", e.g., to persons who process a version or are simply interested in its contents. Usually, locations have autonomy in how to handle allocated versions, e.g., in how to store, to version, and to control access. Thus a modified relationship to a person, such as modified ownership, is likely to go in parallel with a modified location.

A location is represented by a document peer, short *peer*, that serves as a repository for versions, communicates with other peers, and performs interactions requested by users on versions. Peers may differ in the interactions they support, however, they must minimally support the interactions described in Subsection 7.1.3. A peer is identified by a *peer identifier* (PID) which is a URI.

A peer is modelled by class tdf:Peer and described by properties tdf:name and tdf:description. Moreover, property tdf:serves has a peer as its domain and class tdf:ServedEntity as its range. The latter is the superclass of a set of predefined classes wdn:Person[4], wdn:Social_group (in turn superclass of, e.g., wdn:Organization), and wdn:Role. The set of predefined subclasses

---

[2]prefix "dc" denotes namespace `http://purl.org/dc/elements/1.1/`, see [83]

[3]prefix "dcterms" denotes namespace `http://purl.org/dc/terms/`, see [83]

[4]prefix "wdn" denotes namespace `http://xmlns.com/wordnet/1.6/` and identifies concepts defined by WordNet [98]

can be extended (by adding sibling classes) as well as further refined (by adding subclasses). The predefined classification reflects addressees to whom document delivery in real life is targeted: (1) to persons; (2) to social groups such as organizations, departments of organizations, or clubs; and (3) to roles that are fulfilled by agents as in workflows. If a peer does not have property `tdf:serves`, the entity the peer serves is unknown as it is the case with post office boxes. By using classes defined by WordNet [98] the description is based on commonly understood semantics.

To maintain a document's version graph, the dependencies between multiple versions, which are likely stored at different locations, have to be maintained. Following a P2P model, the version graph is maintained distributed. Because of the requirement of low coupling in a Web context, only references to directly preceding and following versions are maintained with each version stored at a peer. This decision is also influenced by the assumption that navigation and communication from a version to its directly preceding and following versions is more likely to be possible (e.g., by a firewall configuration) and will occur more often than therefrom to other versions, e.g., the initial one.

Having introduced locations, the rationale behind employing versioning can be refined which is different from the rationale in version management. While in the latter capturing modifications and identifying configurations is of primary concern, i.e., the *evolution aspect* of documents, the presented model employs versions to distinguish between and keep track of versions stored at different locations in a network, i.e., the *distribution aspect* of documents. Nevertheless, because the presented model is more expressive than known version models, it captures the evolution aspect as well.

A peer stores semantic descriptions of individuals of the concepts presented so far – documents, versions, and peers – and of interactions (see Subsection 7.1.3). It can be queried about locally stored individuals of these concepts, e.g., remote versions can be discovered by querying a local version to determine its preceding versions, offspring versions, and successor version. Aside of querying for other versions, semantic descriptions with application semantics of local versions can be queried as well.

The presented approach can well be combined with related approaches to provide enhanced functionality. First, for the replication of semantic descriptions which are currently only available locally, e.g., Edutella's replication service [116] can be used. Second, for answering queries that search for an arbitrary version (i.e., not by navigating, starting from a known version), existing discovery mechanisms, e.g., provided by Gnutella[5] or JXTA [113] can be used. Third, for the resolution of a DID to a VID, e.g., of the initial version or current versions, existing mechanisms for resolving location-independent identifiers, e.g., URNs, can be used (cf. [169] for an overview). Fourth, for querying semantic descriptions of a set of peers, related approaches such as Edutella [116] can be used.

---

[5]`http://rfc-gnutella.sourceforge.net`

### 7.1.3   Interactions on Documents and Versions

Because the presented model for TDFs is based on a version model, the interactions that are provided by peers mostly have counterparts in version management systems such as WebDAV and CVS, however, there are major differences. First, the presented model is richer by dealing with locations, and second, there is no central control. Thus the semantics of interactions with counterparts differ to cope with locations and decentralization. Moreover, interactions without counterparts are introduced.

Depending on a version's environment it is in a certain state. When being allocated at a peer, the version is *checked-in*. Using appropriate interactions, a following version can be retrieved from a peer. The retrieved version has status *checked-out*. A checked-in version is *online* when the peer it is allocated at can communicate with other peers. Otherwise, or when the version is checked-out, it is *offline*. A checked-out version can be checked in using interaction `checkin`.

On checked-in version $v_i$ basically three basic interactions can be performed. First, interaction `read` retrieves the version's content (or in Web terms a "representation" of the version). Second, interaction `checkout` retrieves successor version $v_{i+1}$ or offspring version $v_{i.j.1}$ when used with option `successor` or `offspring`, respectively. Third, checked-in version $v_k$ that resides in another branch than $v_i$ can be merged into $v_i$ using interaction `merge`. Thereby, $v_i$ becomes the successor of $v_k$. The merged versions may be allocated at different peers. Merging the versions' contents can be done manually or automatically by approaches such as [61].

When a version is checked out two files are retrieved from the peer its predecessor is allocated at. The first file contains the contents of the version and is thus called the *data file*. The second file comprises semantic descriptions concerning the version and the version graph (i.e., at least a single statement that uses the `tdf:precedingVersion` property). The file is thus called the *metadata file*. Semantic descriptions that are available for the checked-out version's predecessor are taken over to the metadata file of the checked-out version. Which data is taken over depends on (a) what the user performing the checkout is allowed to read from the data available for the checked-out version's preceding version, and (b) which data the user performing the checkout is interested in. For a checkin at a later date, both the data and the metadata file are necessary. The descriptions that were taken over are marked as such using appropriate statements.

Note that different to WebDAV and CVS the `checkout` interaction in the presented model retrieves a following version not a representation, thus creating a following version at checkout time instead of the time when the modified representation is checked-in at a later date. It is essential to check out a version, which has its own VID assigned to it, so that it can be recognized as a version by humans and machines, making it possible, e.g., to send it to other people via email and most important to make statements about it using an appropriate language, e.g., RDF or OWL. Moreover, if non-current versions are frozen, a checked-out succeeding version cannot become inconsistent with its predecessor. Whereby consistency means that the succeeding version cannot lack modifications that have been performed on its predecessor.

Aside of the interactions presented so far, which allow to construct version graphs, two interactions allow to modify version graphs. First, interaction `delete` removes a version from the version graph. Second, interaction `reallocate` stores a version at another peer which may involve modifying its VID if the VID is a location dependent identifier. A user who requests to perform an interaction on a version which has been deleted or reallocated is notified of the modification. Both interactions have to be used with caution, since external references to the version become "broken".

The effects of a `delete` interaction are more complex than at first sight. When version $v_i$ is deleted, the event graph has to be modified accordingly by connecting versions from the set of preceding versions $V_p$ with versions of the set of following versions $V_f$. If $v_i$ is an offspring version of preceding version $v_p \in V_p$, then $v_p$ is connected with each $v_f \in V_f$ by an offspring relationship. If $v_i$ is the successor version of preceding version $v_p$, and a following version $v_f$ is the successor version of $v_i$, $v_p$ and $v_f$ are connected by a successor relationship, otherwise by an offspring relationship. This is shown in Figure 7.2, where the upper part depicts version graphs before performing interaction `delete` $v_2$ and the lower part after the deletion. Furthermore, on the left side 1:1 connections are shown, while on the right side other combinations (n:1, 1:n, and m:n) are shown which can be inferred from from 1:1 connections. Note that, while the descriptions of preceding and following versions are updated automatically, replications of descriptions of the affected part of the version graph maintained by other applications may become inconsistent.

For convenience three interactions are provided that are composed of previously mentioned interactions and which are therefore called *composite interactions*. Like the previously mentioned interactions, to which we refer to as simple interactions henceforth, composite interactions are atomic, i.e., they are performed as a whole or not at all. Composite interactions determine the flow of a document, which is caused by allocating following versions at possibly different peers, more directly than simple interactions. Namely, interaction `proceed` proceeds a version with a succeeding version, `branch` proceeds a version with an offspring version, and `copy` copies a version to a new document. As one can easily see, each of `proceed` and `branch` composes a `checkout` and `checkin` interaction, while `copy` composes a `read` and

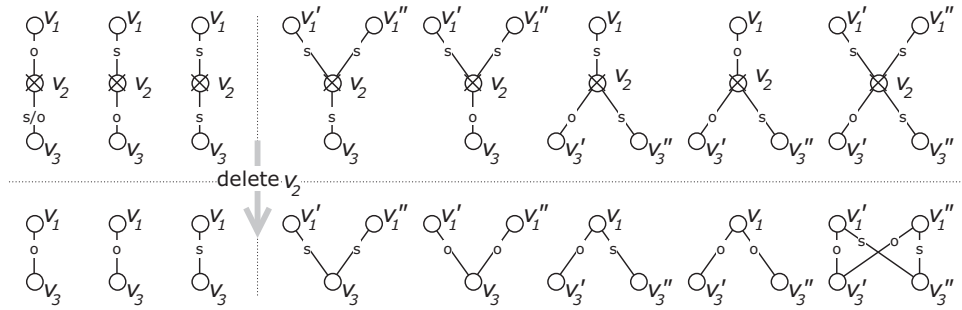

Figure 7.2: Effects of `delete` Interaction on Version Graphs

`checkin` interaction. The composite interactions correspond to how paper forms and documents are handled in real life, i.e., they can be handed on or copied and distributed. For a more precise characterization of composite interactions see the Appendix.

A peer does not only store semantic descriptions of documents and versions it hosts but also of semantic descriptions of interactions it has performed. An interaction is represented by an individual of class tdf:Interaction, which is described by the user or peer who issued the interaction (property tdf:isIssuedBy), by the peer that performed the interaction (property tdf:isPerformedBy), by the date and time it was performed (property tdf:isPerformedAt), and by the version the interaction is targeted at (property tdf:hasTargetVersion). Particular interactions are represented by pairwise disjunct subclasses, e.g., tdf:Read ⊑ tdf:Interaction and tdf:CheckIn ⊑ tdf:Interaction, where tdf:Read ⊓ tdf:CheckIn = ∅. Interactions tdf:CheckOut, tdf:Reallocate, and tdf:Merge are additionally described by property tdf:hasSourceVersion, which identifies (a) the preceding version of the target version being checked out, (b) the version being merged into the target version, or (c) the version being reallocated to the target version in case the VID changes due to reallocation (i.e., if the VID is a location dependent identifier). Composite interactions are represented by individuals of class tdf:CompositeInteraction and are composed of two simple interactions (property tdf:isComposedOf). Three pairwise disjunct subclasses are distinguished for each particular composite interaction, e.g., tdf:Proceed ⊑ tdf:CompositeInteraction.

Like with documents and versions, additional semantic descriptions can be stored along each interaction. They can can be either application-independent, such as a comment in natural text (e.g., using property dc:description), or application-specific, e.g., if the interaction corresponds to some state change in a workflow description.

Semantic descriptions of interactions can be useful in several situations. This strongly applies for descriptions of `read`s, which can be used, e.g., for recommendations (see Subsection 7.2.4), as well as for descriptions of `deletes` and `reallocates` can be used to inform users of deleted versions or to navigate to reallocated ones. Descriptions of `checkins`, `checkouts`, and `merges` are less useful (which, however, depends on the application domain), because they also manifest in the version graph, which is not entirely the case for `reads`, `deletes`, and `reallocates`. Therefor, their semantic descriptions primarily enable one to query versions' histories describing performed interactions.

## 7.2 Application Scenarios

This section present four exemplary application scenarios for TDFs. Namely, these are traceable email, collaborative authoring, workflows, and information exchange. In the second and the third scenario, it can safely be assumed that the contents of versions are not private and that people are willing to let others know where their documents originated, i.e., it is agreed that user specific data is available for others. In the first and the fourth scenario, however, privacy concerns may arise. There are several possibilities which can be used alone or in combination to resolve them: First, user data can be made anonymous while still exploiting its value. This is possible by aggregating the metadata and providing statistical data interpretation, e.g., for features like "users who read this document also read...". Second, one can implement authentification and authorization to specify and reveal what users are allowed to read. It is not dealt with details of privacy issues, which is assumed to be outside the focus of this work.

### 7.2.1 Traceable Email

Usually when sending and receiving files via email knowledge is lost. This knowledge comprises, among others, who originally created the file, who else received and read or edited the file before or after myself, and what other people know about this file in the form of semantic annotations. The loss is caused by (1) missing support of file systems, e.g., for globally unique identifiers; (2) by missing support of email clients, e.g., for functionality like querying where a file originated or querying remotely stored annotations; and (3) by missing integration of the two, e.g., for determining whether I have sent a file to someone else or for determining that a received file is a successor version of a locally stored file.

By providing a component that transparently enables people to use TDFs for files they have sent and received via email we provide them with *traceable email*. The component we have developed is described in more detail in Section 7.3. Basically it filters outgoing email messages and in case it detects a version's datafile attached to an email, a succeeding version is checked out from the sender's peer and sent to the recipient. There, an agent filters incoming email messages and in case a version is attached to an email it automatically checks it in at the recipient's peer.

When using traceable email for tracing files that are exchanged via email otherwise lost knowledge is preserved and new knowledge is unveiled. Using traceable email, users can easily determine not only from whom a file was received, but also where it originated and who else was in possession of the file before. Users can determine to which persons a file they have sent is re-distributed further on. They can also read the contents of any of these versions of the file. Moreover, traceable email provides a non-proprietary possibility for storing and querying also remotely stored annotations about files that were received or sent as email attachments, thereby unveiling annotations of other people.

An exemplary use of traceable email is the management of a paper review process. After submission, e.g., via a web application, which created a document and an initial version, the program chair checks out a following version of each paper and sends it (i.e., the data file) and the review form (i.e., the metadata file) to program committee members. Note that the review form is an annotation to the version, expressed according to an ontology for reviews. Committee members possibly forward received papers to additional (sub-) reviewers. After each reviewer has filled out his/her review form, he/she sends it back to the super-reviewer or to the chair where it is checked. This approach is very flexible and allows the chair to query for the status of each review at any time. Also an aggregated report summarizing all reviews of a paper can be easily created by querying the annotations.

Comparing traceable email with related work [57], the latter focusses on defining machine-processable email contents, while we focus on revealing a document's flow that is established via email.

## 7.2.2   Collaborative Authoring

TDFs can be employed for collaboratively editing documents like version systems in general can be used for this purpose. Different from version systems, however, TDFs do not imply the need for a central shared document repository, which may be not available. Moreover, the annotation mechanism of TDFs is superior compared to the log facility of version systems such as CVS. While in both cases annotations marked up using an ontology can be made, where the ontologies used can be freely chosen, only with TDFs the annotations are hooked into the provided ontological framework. Thereby, semantic descriptions of version graphs, not available in CVS, are provided along and aligned with annotations.

For example, consider the complex collaborative task of writing an edited book. For each book chapter, there are multiple authors working together. Hence it is necessary to create a document for each chapter. For all of these documents, **df:FreezeNonCurrentVersions** is set to true. Thus only the author in possession of a current version owns the "edit token". She/he can edit the draft of the chapter and hand it on to a co-author afterwards. It is also possible that authors work in parallel on branched versions and merge their changes afterwards. At any moment, the editor can determine the status of each chapter. In addition, the authors can have a look at the current versions of the other chapters, e.g. to align their terminology or their references.

### 7.2.3    Workflows

In environments where a workflow management system (WfMS) is not available and/or where it is impossible to deploy one (e.g., when a shared, central WfMS cannot be negotiated) or where it is unreasonable to deploy a WfMS (e.g. too costly or time-consuming), TDFs are a light-weight alternative infrastructure for workflow management. Such environments are faced more and more often with the increase of workflows crossing organizational boundaries. And, starting from the observation that documents are the basis of collaborative work applications, it is natural to think about documents as basic coordination entities for workflows.

Employed *top-down*, TDFs enables one for the definition of workflows in the form of document flows. In particular, a workflow type at the schema level is defined by describing a document flow type to which certain document flows must adhere to. This assumes that the workflow can be modelled in term of a document flow, however, if a document flow from the application domain cannot be used, an artificial "control" document flow can be introduced. As an example (assuming every user has its own peer), user A can instruct user B to proofread a document by handing it over to him, followed by user C who has to add a reference into an internal library catalog, and user D who has to give her permission for its publication (e.g., stored as signed annotation), where user C and D can work in parallel. With TDFs, users can query for subsequent steps at any time and peers can restrict the document flow to a valid one, i.e., to one that accords to its type.

Obviously, an extension to TDFs is needed to model and execute document flow types. First, for modelling document flow types, an appropriate ontology has to be defined and used. One could start from the workflow patterns defined in [143] and map the control flows defined there to document flows. This may involve defining the concept of composite document or record, which contains multiple documents and might control the document flow, e.g., by using reactive rules as described in [92] and the functionality provided by the active extension to TDFs described in Section 8.1. Second, for executing document flow types, peers must be extended to either interpret descriptions of document flow types directly or indirectly. A possibility for the latter would be, as mentioned before, to map document flow types to reactive rules and to interpret the rules as provided by the active extension to TDFS described in Section 8.1.

TDFs can also be used in a *bottom-up* way for workflows. Since the meta-data describing document flows logs how versions were distributed across a network, it can be analyzed ex post to discover document flows that took place regularly. They can be possibly abstracted to document flow types further on. This way, workflow analysis can be carried out by investigating existing document flows.

### 7.2.4   Information Exchange

This scenario is comparable to well-known P2P filesharing applications like Gnutella[6]. The focus is on exchanging *static* documents, meaning that all versions of a document have the same content. Thus annotations are not longer specific for a single version of a document but for all of them. Here, the revelation of document flows, the possibility to annotate versions, and the possibility to query this metadata distinguishes TDFs from other file-sharing applications. Of course users are not forced to annotate their documents, but decide by themselves how much time they spend for annotation.

An example of this application scenario is that of scientists working on the same subject. Typically, they all use a folder in their local file system to store and manage their collection of scientific papers related to their research field. If they use TDFs to share their folder and annotations with others, every user can benefit from observing which documents other users have read, e.g., for answering questions like "users who read this document also read..." or for determining researchers working in the same field. Furthermore, using TDFs it is possible to find out all existing versions of a given document or all documents having at least a single version that share the same datafile. The total number of these can then be used to rank documents, and the peers where these are stored may form a list of potential contact addresses for comments and questions. Summarized, most value of employing TDFs for information exchange can be expected to come from sharing annotations and semantic descriptions of stored files, not from revealing particular document flows.

## 7.3   Implementation

We have implemented a proof-of-concept prototype using JXTA as our P2P platform. Regarding identifiers, PIDs are URNs in the jxta namespace, DIDs and VIDs are UUIDs expressed as URNs. Thereby the prototype is independent of physical network addresses, i.e., peers and versions can be physically moved on a network without affecting their identifiers. The resolution of URNs to network addresses is provided by JXTA. Using the prototype, the user can navigate to preceding and following versions starting from any peer and any version, can retrieve more information including annotations about the version, and can read the version's contents. Because, among others, peers and versions are published using so called JXTA advertisements, they can be dynamically discovered using JXTA's Peer Discovery Protocol as well.

---

[6]`http://rfc-gnutella.sourceforge.net`

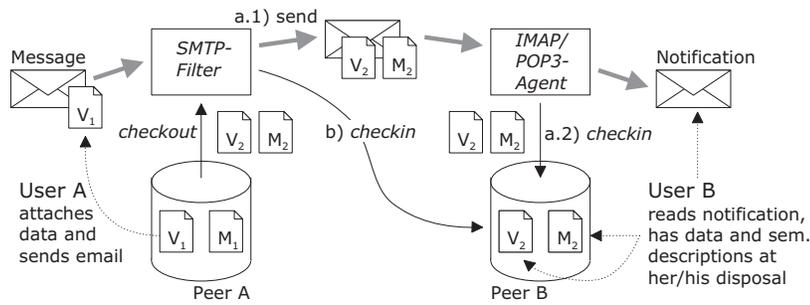

Figure 7.3: Sending and Receiving Traceable and Queryable Email

For a different user interface to peers and a seamless integration with standard email clients, we have implemented a dedicated component. It comprises an SMTP-Filter and an IMAP/POP3-Agent, as shown in Figure 7.3. They are transparent to users and operate in a non-intrusive way. When a user wants to re-distribute a version to another user, she simply creates an email and attaches the version's data file to it. The SMTP-Filter checks out a following version on behalf of the user, thereby receiving the data- and metadata file from the peer. Subsequently it either sends the version to the recipient by email (step a.1 in the figure) or checks it in at the destination peer (alternative step b). In the former case, the recipient's IMAP/POP3-Agent detects the received email and checks the version in on behalf of the recipient (step a.2). In either case the recipient is notified by an email which also contains the original message sent by user A (not comprising the attached file). Other user interface to peers, such as via HTTP and Web Services, are reasonable, but have not been implemented yet.

## 7.4 Related Work

Regarding traceable document flows many distantly related approaches exist which provide the generic functionality of exchanging documents. Among them are e-Mail systems, ebXML, BPEL, and SOAP. When using any of these, however, identities of exchanged documents are not preserved and thus the flow is not traceable. Approaches that are closer related are those from office information systems in the 1980s dealing with form management, e.g., [141] presents among others forms that can flow through an organization. It differs from traceable document flows with respect to flowing arti-facts (only proprietary forms can flow), the data model (versioning is not employed), and the architecture (using central nodes).

Traceable document flows are new to P2P applications, which provide for distributed computing, file sharing, and online collaboration [115]. Among them, the closest related application is instant messaging, but again, identities of exchanged documents are not preserved. Nevertheless, techniques of existing approaches can be employed, e.g., for version discovery using a central index as in Napster[7], flooded requests as in Gnutella[8], or distributed index structures using super-peers as in FastTrack[9]. Approaches for directed routing using distributed hash tables (DHTs), which assign an identifier to a version based on a hash of its content and name and store it at peers with similar identifier (cf. [14, 115] for an overview), cannot be used for discovery because versions in our model cannot be allocated freely.

Concerning distributed concurrent versioning systems, the most prominent existing systems are CVSup [4] and DCVS [3]. However, both of them are based on the client/server-paradigm and provide repository replication. They can thus only provide for virtual document flows and can not be compared to TDFs which are based a P2P model and support physical document flows.

There is a model for encoding semantic information in P2P networks, namely the SWAP metadata model [34] that takes a similar approach of annotating information with metadata about its origin. Unlike our approach which does not change the format of documents, all information is additionally converted to RDF representations. The SWAP model assumes that data is not physically replicated between peers but rather queries are used for information exchange. Since document distribution is an important enabler for TDFs, this a a major difference. An interesting approach is the query routing algorithm REMINDIN' [140] designed for the SWAP platform. It uses observation of other peers' queries and answers to determine their domain knowledge and to decide who is the right peer to answer a certain query. An adaption of this algorithm for TDFs would be useful for the information exchange scenario presented in Subsection 7.2.4.

---

[7] http://napster.com
[8] http://rfc-gnutella.sourceforge.net
[9] http://www.fasttrack.nu

# Chapter 8

# Enriching TDFs with Active Behavior (ATDF)

## Contents

The chapter describes the second layer of the two-layered extension to AXS. It provides for traceable document flows that are enriched with active behavior, abbreviated *ATDF*, and is also referred to as *layer IM-2*.

The chapter is organized as follows. First, Section 8.1 describes the enrichment of TDFs with active behavior by employing AXS. Therefore, additional interactions on documents and version are introduced (see Subsection 8.1.2) and a basic event routing algorithm is presented (see Subsection 8.1.3). Second, Section 8.2 presents event routing algorithms that use local indices of version graphs to optimize event routing. Finally, Section 8.3 discusses related work.

## 8.1 Enriching the Infrastructure Model

This section presents layer IM-2 on top of the layer for document flows (IM-1) to enrich them by active behavior, thus providing for active traceable document flows. Basically, it is the result of mapping the publish/subscribe protocol to a P2P model. By implementing this layer, peers become active and are henceforth called *active peers*. To define active behavior, the layer employs Active XML Schema (AXS).

The running example is a document flow from the workflow domain as mentioned in Subsection 7.2.3.

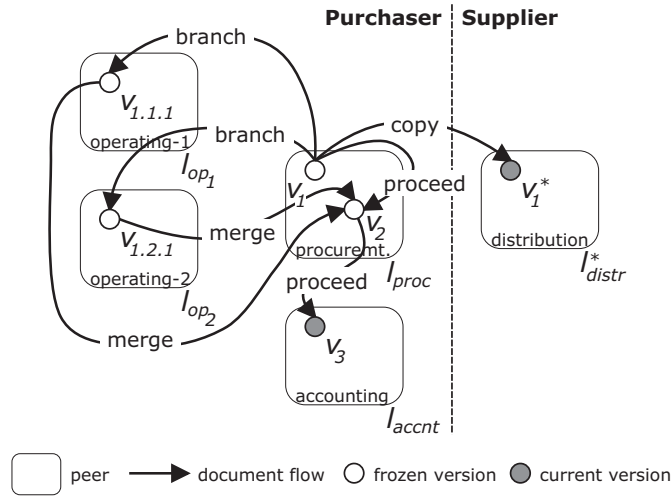

Figure 8.1: Exemplary Document Flow

⊙ *Example 75.* The procurement department quarterly determines the need for higher priced office supplies of operating departments. Therefore, it creates order document $d_{order}$ and initial version $v_1$. Furthermore, offspring versions $v_{1.1.1}$, $v_{1.2.1}$, etc. are created, one for each operating department and sent to them as shown in Figure 8.1. After a month, the offspring versions are merged into version $v_2$ which succeeds $v_1$, consolidating the overall need (they cannot be merged into $v_1$ since this would cause cycles in the version graph). Different locations represent different departments, i.e., $l_{proc}$ denotes the procurement department and $l_{op_1}$, $l_{op_2}$, etc. denote operating departments. Versions are persistent and can be used at a later date, e.g., operating departments can use $v_{1.1.1}$, $v_{1.2.1}$, etc. later on to look up office supplies ordered in that quarter.

To perform this document flow, the procurement department performs multiple interactions "`checkout offspring of` $v_1$" of document $d_{order}$ to retrieve offspring version $v_{1.i.1}$ for each operating department $l_{op_i}$. These versions are checked in at $l_{op_i}$ using interaction `checkin`. A successor version $v_2$ of $v_1$ is retrieved by performing "`checkout successor of` $v_1$" followed by a `checkin` at the same location. After one month branched versions $v_{1.i.1}$ are merged into $v_2$ by performing "`merge` $v_{1.i.1}$ `into` $v_2$" for each branched version, resulting in a single version $v_2$ at the procurement department com-prising all items to be ordered. Instead of using simple interactions as above, composite interactions `proceed`, `branch`, and `copy` can be used to perform the same document flow as done in Figure 8.1.

After determining the overall need in office supplies the procurement department sends the order to the distribution department of a supplier company at $l^*_{distr}$ (again, cf. Figure 8.1). To assure that the supplier company can handle the order independently of the purchasing company, the latter performs interaction "`copy` $v_2$ `to` $v^*_1$ `at` $l^*_{distr}$", which copies the version to the distribution department, assigning it to a new document. Throughout the rest of the chapter, locations, documents, and versions of the supplier company are marked by an superscripted asterisk.

### 8.1.1 Employing Active XML Schema

Active XML Schema (AXS) is employed on top of layer IM-1. Since AXS provides active behavior to XML documents only, the format of data files participating in document flows for which active behavior can be provided is restricted to XML.

*Briefly recalling AXS*, it allows one to specify active behavior by ECA rules and passive behavior by operations with XML schemas, thereby specifying behavior of XML documents that are valid instances of the schema. When an operation is executed on an XML document, an according event occurs, being of an event type (the event's intensional aspect) and stored in an event class (the event's extensional aspect). The event class in turn is stored as part of the XML document. When an event occurs, all rules defined upon the event class it is stored in are triggered, i.e., executed.

⊙ *Example 76.* AXS is used to define operations with the document type of $d_{order}$. Among others, operation `addItem(my:OrderItemType)` adds the item given via the parameter to an order and operation `removeItem(xs:integer)` removes the item with the given part-number from an order. When one of these operations is invoked on an XML document, an event occurs therein which is stored in an event class of the same name.

In addition to the event types already provided by AXS new *interaction event types* are specified. Where the event types already provided by AXS comprise operation events, mutation (primitive and composite) events, and calendar events (see Subsection 4.1.2). The new interaction event types are defined as specialization of the abstract event type. When an interaction is performed, being either one on TDFs (see previous Subsection 7.1.3) or on ATDFs (see following Subsection 8.1.2), an according event occurs which is stored in an event class.

In AXS an event class of an XML document can be subscribed by remote XML documents which replicate the event class. Events that subsequently occur in the subscribed document are delivered to the subscribing documents. This provides for distributed active behavior, allowing a document to react to events that occurred in other documents. By using the publish/subscribe protocol documents are loosely coupled.

Summarized, an active peer is capable of detecting events occurred in a version, of receiving events from versions of subscribed documents, of distributing events to versions of subscribing documents, of executing operations, and of executing ECA rules. In the following, Subsection 8.1.2 introduces new interactions on active documents and versions, and Subsection 8.1.3 presents the realization of the publish/subscribe protocol on top of the P2P model of layer IM-1 and describes the event routing algorithm.

### 8.1.2   Interactions on Active Documents and Versions

Three additional interactions are provided by active peers to deal with active behavior defined by AXS. First, interaction `invoke` executes an operation on a checked-in current version. Second, interactions `subscribe` and `unsubscribe` are used to establish and release subscriptions between documents. The latter two are special in that subscriptions have effects on possibly all versions of a document (depending on the employed event routing model, see Subsection 8.1.3 and Subsection 8.2.2), i.e., when document $d_s$ subscribes document $d_p$, possibly all versions of $d_s$ are notified of events that occur in some version of $d_p$. A document may subscribe itself.

Interactions `subscribe` and `unsubscribe` can only be issued against a particular version of the subscribing document, called the event recipient. Thereupon this version notifies a particular version of the publishing document, called the event distributor (see Subsection 8.1.3 for these two particular versions). This ensures that data can be exchanged freely between those two versions. Moreover, restricting the versions these interactions can be issued against facilitates control over authentification, authorization, etc. by centralization, which accords to having effects on all current versions.

A peer stores semantic descriptions of interactions on active documents and versions as it stores descriptions of interactions on (passive) documents and versions (see Subsection 7.1.3). Each of the above interactions are represented by an according class, namely tdf:Invoke, tdf:Subscribe, and tdf:Unsubscribe which are all subclasses of tdf:ActiveInteracion $\sqsubseteq$ tdf:Interaction. Thus they are described by the properties describing tdf:Interactions (see Subsection 7.1.3). Moreover, the invocation of an operation is described by the name of the invoked operation (property tdf:ofOperation), by the passed parameters (property tdf:withParamValue), and by the operation's return value (property tdf:hasReturnValue). Interactions subscribe and unsubscribe are described by property tdf:ofPublishingDocument.

When an interaction is performed an according event occurs. Every AXS schema defines per default an event type for each interaction from Subsection 7.1.3, be it simple or composite, except for operation invocations, and for interactions `subscribe` and `unsubscribe`. Event classes storing events of those types can be materialized on demand, they must be named after the invoked interactions their events represent, e.g., `CheckIn` or `Merge`. For operation invocations an event type is defined for every single operation specified by the AXS schema (and not for the `invoke` interaction). An event representing an interaction is stored in an event class of the version it was targeted at. Interested documents can subscribe these event classes to be informed of interactions.

⊙ *Example 77.*    Because the procurement department wants to be alerted when a department has orders above average or when the overall budget is used up, it creates document $d_{alert}$ and an initial version. This document subscribes event classes of document $d_{order}$ that reflect executions of operations `addItem` and `removeItem`. Thus, if any current version of $d_{order}$ is modified, the initial version of $d_{alert}$ is notified of the occurred event. By aggregating data of such events alerting situations can be detected.

It is possible to embed interactions on documents and versions in operations and rules defined by AXS. Thus interactions cannot only be performed by users but also by versions, enabling them to automatically perform an interaction in reaction to occurred events.

⊙ *Example 78.*    Upon copying $v_2$ to $l^*_{distr}$, which creates $v^*_1$ of newly created document $d^*_{order}$ at $l^*_{distr}$, the purchaser's document $d_{order}$ subscribes the supplier's document $d^*_{order}$ to be informed of the order's acceptance, shipment, and billing, which are represented by invocations of operations `accept`, `reject`, `ship`, and `bill`.

### 8.1.3   Basic Event Routing (Layer IM-2.1B)

As shown in Figure 8.2, layer IM-2.1B provides for event routing by adding a layer on top of the base of layer of IM-2.

A document plays multiple publisher and subscriber roles during its lifetime, producing and consuming events. Since a document's current versions may be allocated at changing locations, events have to be routed from publishers to subscribers dynamically.

A naive event routing algorithm would be to send an event from the version of a document in publisher role where the event occurs to all current versions of all subscribing documents. This would involve determining all target versions (a) either ad-hoc, i.e., by querying the P2P network using some discovery mechanism, (b) to store an index of all target versions with each current version of a document in publisher role, or (c) a mixture of both. While alternative (a) would result in high network traffic, alternative

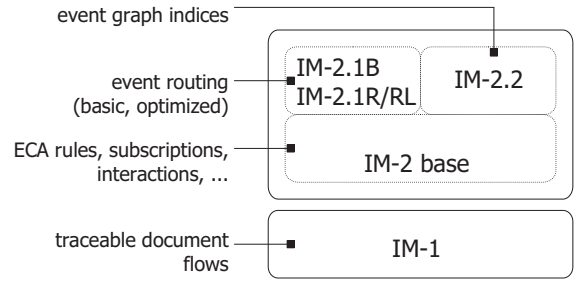

Figure 8.2: Architectural Layers of ATDFs

(b) would yield high complexity in maintaining the indices. The deficiency that all alternatives share is that a version is not notified of an event because it is off-line, not reachable due to the P2P overlay network topology or a firewall, or because of an out-dated index structure.

Instead, the presented event routing algorithm tries to find a balance between network traffic and complexity while guaranteeing event delivery to all subscribers. This is achieved by employing a form of centralization as in hybrid P2P systems such as FastTrack[1] and [38]. Therefor, two central nodes are introduced for each document, namely two distinguished versions, the event distributor and event recipient (see later in this section). It is based on the following two assumptions. First, there exists a network path between at least two distinguished versions of different documents so that data can be exchanged between the two documents via these versions. A network path may be defined explicitly, e.g., by a firewall configuration. Second, data can be exchanged between two directly following versions of the same document.

The initial version of a document in publisher role is called the document's *event distributor*. It is responsible of distributing occurred events to event recipients of subscribed documents and thus has to maintain an index on them. If the event distributor is followed by other versions and an event occurs therein, the event is forwarded to the event distributor. This is done by forwarding the event to the respective preceding version until the event distributor is reached. If a preceding version was offline when an event was to be forwarded to it, the peer that tried to forward the event buffered it until the version was back online. In case of a merged version, which has $n$ preceding versions ($n > 1$), a peer decides autonomously on the subset of them to which an event is forwarded (the number of the subset's members is defined by the peer's configuration parameter $\rho$). If $\rho > 1$ events are duplicated and have to be filtered out when the first common version is reached (which performed the branch). When an event is forwarded from the version where it occurred towards the event distributor it is said to be in the *sending phase*.

---
[1] http://www.fasttrack.nu

⊙ *Example 79.*   After acceptance of the order by the supplier's distribution department, $v_1^*$ is proceeded at the storage department by $v_2^*$ as shown in Figure 8.3. Upon the order's shipment by the storage department, operation `ship` is invoked, causing an operation event to occur in $v_2^*$ at $l_{store}^*$, which is forwarded to the event distributor $v_1^*$ at $l_{distr}^*$. Thereafter, $v_2^*$ is proceeded at the accounting department $l_{accnt}^*$ by $v_3^*$ where operation `bill` is invoked later on. Again, an operation event occurs which is forwarded via $v_2^*$ at $l_{store}^*$ to the event distributor $v_1^*$ at $l_{distr}^*$. Both events are distributed to subscribing documents by $v_1^*$ at $l_{distr}^*$.

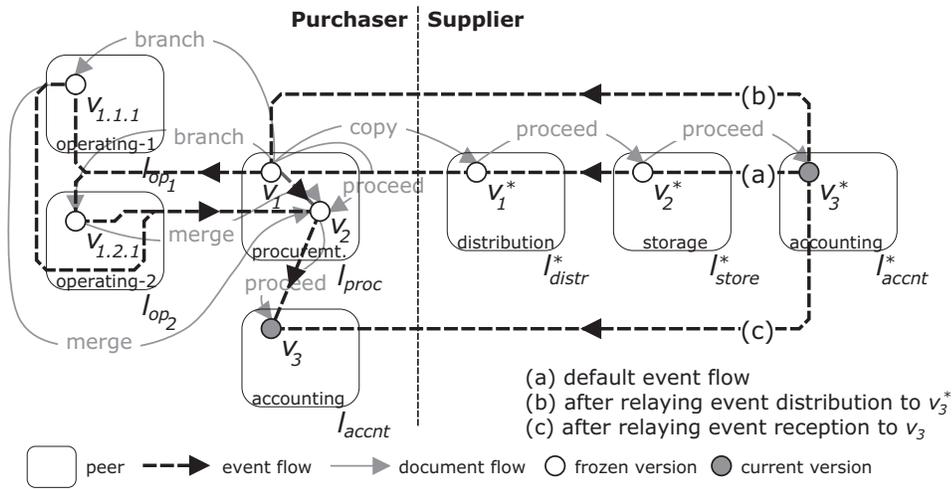

Figure 8.3: Exemplary Event Flows

The initial version of a document in subscriber role is called the document's *event recipient*. It receives events from event distributors of subscribed documents and is responsible of forwarding them to all current versions. If the event recipient is followed by other versions, received events are forwarded to its successor version and offspring versions. This is recursively applied until all current versions are reached. By forwarding events to the successor version as well as to offspring versions, events are duplicated and have to be filtered out when versions are merged. Duplication by forwarding events to *all* following versions is necessary because a version cannot determine whether all its offspring versions are (indirectly) merged into a single version later on. If a directly following version was offline when an event was to be forwarded to it, the peer that tried to forward the event buffered it until the version was back online. When an event is forwarded from the event recipient towards current versions it is said to be in the *receiving phase*.

⊙ *Example 80.*    After shipment the procurement department's current version $v_2$ is proceeded by $v_3$ at the accounting department $l_{accnt}$, which awaits the supplier's billing before booking and paying. The event representing the invocation of operation `bill` at $l^*_{accnt}$ in Example 79 is delivered to event recipient $v_1$ at $l_{proc}$ which forwards it to $v_{1.1.1}$, $v_{1.2.1}$, and $v_2$. The latter version filters out the same event it subsequently receives from $v_{1.1.1}$ and $v_{1.2.1}$. Finally, version $v_2$ forwards the event to $v_3$ at $l_{accnt}$. The event routing is shown in (a) of Figure 8.3.

Network traffic caused by event routing as described before can be reduced by *manually optimizing event flows*. This is supported by interaction `relay` (see the Appendix for the interaction's definition). Because `relay` can modify a document's event distributor and event recipient, active peers provide for querying a version of whether it is the event distributor or event recipient.

First, network traffic can be reduced by *relaying event distribution* from $v_i$ to $v_j$, where the latter becomes the new event distributor. This makes forwarding events that occur in versions (indirectly) following $v_j$ from $v_j$ to $v_i$ obsolete. To relay event distribution from $v_i$ to $v_j$, every path (cf. Subsection 7.1.1) between every current version and $v_i$ must contain $v_j$. This ensures that events can be routed from current versions where they occur to the event distributor after relaying.

⊙ *Example 81.*    Relaying event distribution from version $v^*_1$ to $v^*_3$ saves forwarding events occurring at $l^*_{accnt}$ via $l^*_{store}$ to $l^*_{distr}$. Events originating at $l^*_{accnt}$ thus flow directly via $l_{proc}$, $l_{op_1}$, and $l_{op_2}$ to $l_{accnt}$ as shown in (b) of Figure 8.3.

Second, network traffic can be reduced by *relaying event reception* from $v_i$ to $v_j$, where the latter becomes the new event recipient. This makes forwarding events that occur in subscribed documents from $v_i$ to $v_j$ obsolete, because events are delivered from event distributors directly to $v_j$ instead of $v_i$. To relay event reception from $v_i$ to $v_j$, a path must exist from $v_j$ to every current version. This ensures that all current versions can be notified of newly occurred events after relaying.

⊙ *Example 82.*    Relaying event reception from version $v_1$ to $v_3$ of document $d_{order}$ makes event distributor $v^*_3$ send occurred events directly to $v_3$, which saves forwarding events from $v_1$ via $v_2$, $v_{1.1.1}$, and $v_{1.2.1}$ to $v_3$ as shown in (c) of Figure 8.3.

## 8.2 Optimization

The model for document flows in Section 7.1 as well as the model for event flows in Subsection 8.1.3 assume high availability of peers, however, this assumption restricts peers in their autonomy. Moreover, low coupling of peers is gained at the cost of network traffic, since events can only be forwarded to directly preceding or following versions, resulting in unnecessary event forwarding. This section presents an optimized model that overcomes these two properties. First, it provides for higher autonomy of peers by waiving the expectation of high availability, and second, it provides for event flows that are in need of less network traffic by avoiding unnecessary event forwarding.

Both improvements of the optimized model are gained by having *version graph indices* available locally at each peer. Thereby each peer does not only know a version's direct predecessors and successors but also indirect ones. Version graph indices are maintained by employing active behavior provided by layer IM-1 and the basis of IM-2 so that the infrastructure model presented so far does not have to be adapted. How version graph indices are maintained is presented in detail in Subsection 8.2.1.

The improvement of *higher autonomy* is gained by by-passing offline peers. By by-passing them in the sending and receiving phase of event forwarding a peer can be offline without blocking event flows. By by-passing offline peers during manual version discovery, a peer can be offline without preventing people from determining an online version's (indirectly) following versions, e.g., when one wants to determine current versions by querying a frozen version she knows of.

The improvement of *reduced network* traffic is gained by by-passing peers that have no responsibilities regarding event distribution of a certain document. Having a version graph available locally at each peer, it is possible to forward an event not only to a directly preceding or following version as employed in the basic model for event flows, but to any version. Thereby unnecessary event forwarding can be avoided and network traffic reduced.

How the active extension is employed to realize the optimized model is described in detail in the following two subsections.

### 8.2.1 Maintaining Version-Graph Indices (Layer IM-2.2)

This layer, which is situated side-to-side with the basic event routing model in layer IM-2.1 (see Figure 8.2), maintains a *version graph index* at each peer for every version the peer stores, which is a local view on the version graph the version is part of. To be able to construct the version graph index, denoted as $G_{v_i}$, for every version $v_i$ of document $d$ the peer stores, the document subscribes event classes that reflect document flows of itself, i.e., $d$ subscribes $d$. Thereby each of the document's versions receives an event upon a modification of the version graph. Subscription of the set of event classes $E^G := \{\mathsf{CheckIn}, \mathsf{CheckOut}, \mathsf{Merge}, \mathsf{Delete}, \mathsf{Reallocate}\}$ is sufficient to construct $G_{v_i}$. Note that due to outstanding event deliveries caused by offline peers, version graph $G_{v_i}$ at one peer may differ from version graph $G_{v_j}$ of the same document at another peer.

A version graph index is defined as a view over the set of event classes $E^G$. Simplified, the view defining query selects the latest event for each version from the $\mathsf{CheckOut}$ and the $\mathsf{Merge}$ event class to determine dependencies between versions, the latest from $\mathsf{CheckIn}$ and $\mathsf{Reallocate}$ to determine the version's location, and uses $\mathsf{Delete}$ to filter out deleted versions. As views in databases, a version graph index can be either virtual or materialized. If it is materialized it may be maintained incrementally, depending on the peer's capabilities, meaning that with a newly imported event $e \in E^G$ the view is updated according to $e$ without access to any other event in $E^G$.

⊙ *Example 83.* Consider the version graph depicted in Figure 8.1 when comprising $v_1$, $v_{1.1.1}$, and $v_2$. Further assume that the document subscribes the set of event classes $E^G$ of itself that reflect document flows. Version $v_{1.1.1}$ then comprises a checkin event reflecting interaction "`checkin` $v_1$ `of` $d$", and a checkin and checkout event reflecting each of the two interactions "`branch` $v_1$ `to` $v_{1.1.1}$" and "`proceed` $v_1$ `by` $v_2$". Upon performing interaction "`branch` $v_1$ `to` $v_{1.2.1}$", two according events are imported into and stored in $v_{1.1.1}$. From the events in event classes $\mathsf{CheckIn}$ and $\mathsf{CheckOut}$ version graph index $G_{v_{1.1.1}}$ can be constructed.

If a peer is not capable of maintaining version graph indices incrementally, frozen versions need special attention. By definition, these versions are immutable and thus cannot store any event that occurred after the time they became frozen. Such events, however, are necessary to construct $G_{v_i}$ for frozen version $v_i$ if the index is not maintained incrementally. To resolve this issue, a peer stores such events in so called *auxiliary event classes* in addition to the event classes of a frozen version. By providing an integrated view over these event classes, a version graph index can be transparently specified over them as if all event classes were stored within the version.

⊙ *Example 84.* Continuing Example 83, when version $v_{1.1.1}$ is merged into $v_2$ next, an according merge event occurs, which is first forwarded to the document's event distributor $v_1$ which is also the event recipient. From there it is delivered to current version $v_2$ and $v_{1.2.1}$. The event also passes frozen versions $v_1$ and $v_{1.1.1}$ where the event is stored as well (possibly in auxiliary event classes).

Summarized, the optimized model for document flows provides higher autonomy for peers with respect to their availability. The model offers the advantage of having a version graph index available at each peer where a version of the document is allocated at. This allows to query a version arbitrarily about its version graph an not only about directly preceding and directly following versions. The model, however, inherently provides version graph indices only for versions that (indirectly) follow the event recipient. Thus one can only benefit from this layer when looking for related versions that (indirectly) follow the event recipient. Note that this layer assumes that the event routing layer IM-2.1 guarantees that an imported event (caused by self-subscription) passes every version following the event recipient either during the event distribution or event reception phase.

### 8.2.2   Optimized Event Routing (Layers IM-2.1R/RL)

This section presents two optimized event routing algorithms provided by layers IM-2.1R and IM-2.1RL which both reduces network traffic and deliver events more robustly (i.e., unaffected by offline peers) compared to basic event routing IM-2.1B presented in Subsection 8.1.3. For short we refer to the event routing algorithms as {B|R|RL} event routing. The layers are intended to supplement IM-2.1B, nevertheless, for a single document a dedicated event routing algorithm has to be chosen. Optimization builds upon version graph indices and can thus only supplement IM-2.1B if IM-2.2 is deployed (also see Figure 8.2).

R event routing provides _Robust event delivery_ by forwarding imported events from the event recipient directly to all versions (indirectly) following the event recipient using its version graph index. Thus event delivery can no longer be blocked by offline peers. Network traffic in R routing is reduced by forwarding an event occurring in some version immediately to the event distributor, i.e., without using intermediate versions.

RL event routing is equally robust as R routing and moreover results in _Lowest network traffic._ In principal, its event routing is a specialized form of R routing, in that imported events are only delivered to current versions, resulting in the lowest network traffic possible.

The optimized event routing models need to know which version in a version graph index is the event distributor and which one is the event recipient. Thus in addition to the event classes mentioned in Section 8.2.1, a document subscribes event class **Relay** of itself. Then, due to self-subscription, this data is available at every version (indirectly) following the event recipient (and of course the event recipient itself). If the event distributor precedes the event recipient, it has to observe the events it distributes to keep track of possible reallocations or relays affecting the event recipient.

A choice between the event routing models is likely to depend on an application's requirements because they differ in whereto events are delivered. Remember, R event routing delivers events to all versions following the event recipient, while RL only delivers them to current versions to minimize network traffic. If non-current versions are frozen, which is defined by property **tdf:freezeNonCurrentVerions** (see Subsection 7.1.1), it is likely that RL event routing will be employed, since frozen versions are immutable and can neither store events nor trigger rules. Vice versa, if non-current versions are not frozen, it is likely that R event routing will be employed, since events may be of interest at every version, e.g., to keep the versions consistent.

*Optimizing the sending phase* in R and RL event routing, events are forwarded directly from the version where they occur to the event distributor instead of using other version as intermediaries (as shown in Figure 8.4). This works well as long as the event distributor is online, however, it may change its status to offline and vice versa. Thus the following two cases have to be dealt with (remember that peers hosting event distributors and recipients should be online most of the time):

1. Event distributor $v_d$ may be offline. If an event is to be forwarded to $v_d$ from current version $v_c$, the version's peer which tries to forward events buffers them until $v_d$ is back online, except when the event is a `checkin` event. In the latter case, the event is forwarded to the $v_c$'s directly preceding version, so that some other version "knows" about it and buffers it. This version is then responsible of forwarding events during the receiving phase to $v_c$ as long as $v_d$ is offline. If the directly preceding version is offline as well, the same action is applied to some other current version. If neither $v_d$ nor any current version is online, the peer of the checked-in version periodically tests whether one of them went online. As soon as this happens, an appropriate action is performed so that the checked-in version becomes part of the version graph and is notified of events. If an event is to be forwarded from a non-current version to (offline) $v_d$, it is forwarded to a current version's peer and is buffered there so that it is forwarded to the event distributor later on (see below).

2. If event distributor $v_d$ goes online after being offline, it performs the following steps: (1) it notifies all peers that host a current version according to the version's (possibly out-dated) $G_{v_d}$ to receive buffered events therefrom, (2) if such received events reflect changes to the version graph, e.g., reveal new versions, step 1, i.e., the notification and reception is recursively applied to those versions as well. Thereby $G_{v_d}$ becomes up-to-date and buffered events are harvested from the versions where they occurred. And (3) harvested events are sent to event recipients of subscribing documents.

*Optimizing the receiving phase*, one has to distinguish between R and RL event routing. In R event routing, all events are forwarded from event recipient $v_r$ directly to all (indirectly) following versions using $G_{v_r}$ instead of using other versions as intermediaries. In RL event routing, events that do not reflect modifications to $v_r$'s version graph are forwarded from $v_r$ directly to current versions (as shown in Figure 8.4). If a version is offline when events are to be forwarded to it, event recipient $v_r$ buffers them until the version is back online. Again, an event recipient's peer may go offline (although it should not or only rarely), thus the following two cases have to be dealt with.

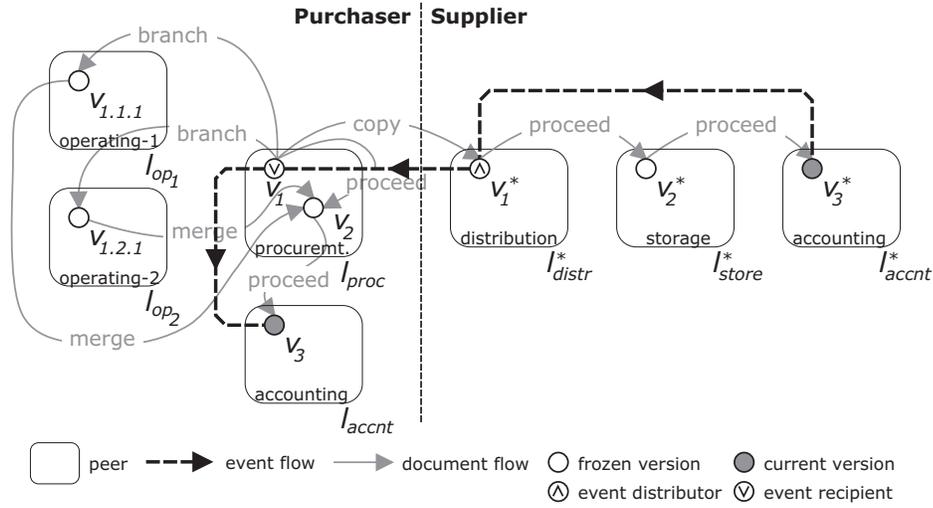

Figure 8.4: Optimized Event Flow using RL Event Routing

1. Event recipient $v_r$ may be offline. If an event is to be forwarded to it from an event distributor, the latter buffers the event until $v_r$ is back online.

2. If event recipient $v_r$ goes online after being offline, it performs the following steps: (1) it notifies the peers of subscribed documents' event distributors to receive buffered events, (2) it updates $G_{v_r}$ on the basis of events that reflect document flows which have been received during step 1 due to self-subscription (if $v_r \neq v_d$), and (3) it forwards events directly to (indirectly) following versions (in R routing) or directly to current versions only (in RL routing). In RL routing, events that reflect modifications to $v_r$'s version graph are delivered to all (indirectly) following versions using B or R routing (configured with the document, see below).

If event distributor $v_d$ is offline while event recipient $v_r$ is online and both belong to the same document, $G_{v_r}$ may be outdated and thus events that are received by $v_r$ from subscribed documents are possibly sent to (a) versions that are no longer a current version, in which case the receiving version forwards it to its directly following versions, i.e., falls back to B event routing; (b) versions that have been reallocated, in which case the receiving peer is responsible of forwarding the event to the reallocated version, or (c) versions that have been deleted, in which case $v_r$ is notified which sends the event to the (online) preceding version that is closest to the deleted version instead; that version in turn forwards the event to its directly following versions, i.e., falls back to B event routing.

For layer IM-2.2 to maintain version graph indices, events that reflect modifications to the version graph have to be delivered to every version (indirectly following the event recipient) of the version graph. While this naturally happens when using B or R event routing, this is not the case with RL event routing. Hence B event routing has to be employed for forwarding events in the sending phase and/or B or R event routing Choosing B/R event routing in both phases makes event delivery more robust to offline event distributors and offline event recipients by keeping the version graph indices up to date irrespective of the two versions' online status. Thus a fall back to B event routing in the receiving phase is less likely to occur. We expect B/R routing employed in both phases to result in less network traffic if the peers are more likely to be offline, however, this remains to be proved by experiments. Meanwhile it can be stated that it suffices to employ B/R event routing in either the sending or receiving phase.

The event routing chosen to deliver events is specified by configuration with the document. There, property **tdf:useEventRouting** with a value out of {B|R|RL} defines the event routing employed. If it is RL, the two additional properties **tdf:useInSendingPhase** and **tdf:useInReceivingPhase** with a possible values of {B|R} define the according routing algorithms to use in the sending and receiving phase, respectively, of which one must be specified.

Summarized, the optimized event routing model for event flows automatically makes event delivery more robust and reduce network traffic, while still being manually controllable by `relay` interactions. The optimized algorithms, however, assume that a document's event recipient and event distributor can exchange data with any (current) version (not an assumption in B event routing). A mixture of the two models, where events are forwarded to the most distant version of those with which data can be exchanged, would not be based on this assumption and may be the most appropriate one in a given setting. Note that such a mixed model does not have to be developed necessarily, but can be partly achieved manually by employing the optimized model and setting event distributors and event recipients accordingly.

## 8.3    Related Work

The ATDF approach employs Active XML Schema (AXS) as an approach for active XML on top of the TDF approach. It was chosen because it provides the most comprehensive set of feature that are useful in a distributed environment. For brief disussions of these approaches see Subsection 1.3.2 and Subsection 1.4.5.

Regarding distributed active behavior related approaches focus on different aspects of distribution. First, [47, 97] independently propose systems that support distributed event sources which are integrated by a central server. The server is responsible of event composition, condition evaluation, and rule execution. Second, the idea first presented by [69] to unbundle databases' active behavior into software components is applied to distributed environments by [50, 51]. They propose among others distributed services for the tasks that [47, 97] assign to a central server. Different from all of these approaches we propose a P2P architecture and focus on issues when the data for which active behavior is defined for is distributed, i.e., dynamically allocated in a network.

The presented event routing algorithm is tailored to the application, i.e., it uses version graphs to route events between versions. It cannot be compared to other algorithms, e.g., used by Gnutella and FastTrack, because they do not employ the publish/subscribe protocol and thus cannot ensure that messages are received by all recipients. Comparing it to P2P approaches that use the publish/subscribe protocol, it is different from [171], which uses a single centralized event server to handle subscriptions. However, while [38]'s underlying model uses DHTs and thus cannot be directly compared, it is similar in that it is a hybrid P2P model by using multiple central nodes (there, one for each topic).

# Chapter 9

# Outlook

The thesis presented a combined approach for maintaining consistency of data on the Web. While the SMWP approach is considered mature, there are still some research issues that seem worth addressing in the area of Active XML Schema (AXS) in general and in the area of composite mutation events and active document flows (ATDF) in particular. These are enlisted in the following.

- In the area of AXS, triggers are a procedural way of manually defining the logic for incremental view maintenance. Naturally, a more declarative way is desirable. From a declarative view definition, triggers could be derived that realize view maintenance, as it has been done in [44] by deriving triggers for view maintenance from SQL queries. For the declarative definition of views, a query language such as XQuery [167], consistency constraints such as in [176] and view correspondence assertions such as in [144] could be used.

- An extension to AXS to provide for class-based modelling would ease the definition of data replication and maintenance. Again, procedural triggers could be derived from declarative class-based models and view definitions to maintain consistency. As we were not able to identify general replication patterns in XML documents for designing class-based models, it is likely that they will be domain dependent. A good starting point may be [33], which describes a fragmentation technique for XML documents. Employing this technique, one could analogously to SMWP define an approach for parameterized fragments and map them to parameterized page classes, e.g., to define a Web site on top of a native XML database.

- A promising extension to the proposed approach for composite mutation events is the automatic derivation of composite event type definitions from XML schemas. With these event type definitions, an application engineer can start working with. We assume that if schemas use

XML schema concepts such as type definitions, type hierarchies and model groups in a meaningful way, automatically derived composite event type definitions will be meaningful as well.

- There is still work to be done in fine-tuning the approach for composite events. The model and implementation can be optimized, e.g., often it is sufficient when an operator node does not compare whole path instances but only its last step. Moreover, one can think of various extensions to the model, such as how to handle wildcards in path types and path instances, or how to deal with phantom events, i.e., composite events that are raised although a constituent event reflects a modification that has been undone by a subsequent event.

- To better understand requirements for composite mutation events, and thus being better able to judge on the approach's possible extensions, a case study is currently undertaken. It uses composite events to deserialize RDF/XML documents as specified in [166] into RDF graphs. The specification uses events to define the deserialization (see Section 6 and 7 therein) and thus neatly fits the functionality provided by composite mutation events.

- A work worth exploring in the area of composite mutation events is the combination of the hierarchical context with contexts from Snoop. This is possible since the presented approach is fully compatible with Snoop and thus provides for combined expressiveness. First, the hierarchical context presented herein is orthogonal to Snoop's contexts and can thus be arbitrarily combined with the latter, providing for *simultaneous* event combination by hierarchical position and time. Second, composite XML event types can be combined by operators from Snoop, providing for *subsequent* event detection based on time (e.g., NOT detects non-occurrences of events in time intervals).

- To explore the usage of traceable document flows in personal ad-hoc data exchange, we are currently undertaking a case study by employing the approach to support communication within student teams who are collaboratively working on software engineering projects. We hope to detect frequent patterns in document exchange and get useful feedback. Moreover, we are experimentally evaluating the prototype, comparing the presented basic event routing algorithm to an approach that solely uses the JXTA [113] infrastructure. The latter supports flooded requests only, meaning that one peer in order to communicate with another one sends a message to all peers he knows, which in turn forward the message to all peers they know so that finally the message reaches the peer it was targeted at. By the experimental evaluation we hope to gain insight into the algorithm's implications on network traffic.

- To explore the use of ATDF in a workflow setting, mapping workflow patterns that define control flows as, e.g., proposed in [143], to document flows. Thereby existing flow constructs such as sequence, exclusive choice, and synchronization would be leveraged to document flows. The idea is to describe document flows top down using an appropriate ontology, i.e., the patterns, and to derive necessary events and rules on distributed documents to enforce the document flow.

# List of Figures

# List of Tables

# Appendix

## A  SMWP

### Schema Definition Language

This section contains a complete specification of the schema definition language as introduced in Section 2.2. Terms "/*" and "*/" are used to delimit comments.

*SMWPStmt* := (CreateStmt | DropStmt | ShowStmt | AlterStmt) ";".

*/* Create Statement */*
*CreateStmt* :=
    "CREATE" (FCCreateStmt | PCCreateStmt | ParamCreateStmt).
*FCCreateStmt* : = PFCCreateStmt | DFCCreateStmt.
*PFCCreateStmt* :=
    "PRIMARY FRAGMENT CLASS" FCSignature
    "FRAGMENTATION BASE CLASS" FCSignature
    ["TUPLE SELECTION PREDICATE" SQLExpr]
    ["FRAGMENT SELECTION PREDICATE" SQLExpr].
*DFCCreateStmt* :=
    "DERIVED FRAGMENT CLASS" FCSignature
    "FRAGMENTATION BASE CLASS" FCSignature ["AS" Ident]
    "DERIVATION BASE CLASS" FCSignature ["AS" Ident]
    "JOIN BY" SQLExpr.

*PCCreateStmt* :=
    "PAGE CLASS" PCSignature
    ["FILENAME" FileName]
    "FOUNDATION FRAGMENT CLASS" PCFCMapping
    {"FRAGMENT CLASS" PCFCMapping}.
*PCFCMapping* := FCSignature ["AS" Ident]
    [ParameterMap] {InternalPageRef | ExternalPageRef}.
*ParameterMap* :=
    "PARAMETER MAP" Ident "AS" Ident {"," Ident "AS" Ident}.

*InternalPageRef* :=
 "INTERNAL LINK TO FRAGMENT CLASS" FCSignature
 ["AS" Ident] "JOIN BY" SQLExpr.
*ExternalPageRef* :=
 "EXTERNAL LINK TO PAGE CLASS" PCSignature ["AS" Ident]
 "CONTAINING FRAGMENT CLASS" FCSignature ["AS" Ident]
 "JOIN BY" SQLExpr.

*ParamCreateStmt* := VBPCreateStmt | PBPCreateStmt.
*VBPCreateStmt* :=
 "VALUE BASED PARAMETER" Ident "ON" FCSignature
 (UseReferenceRelationStmt | CreateReferenceRelationStmt).
*UseReferenceRelationStmt* :=
 "USE REFERENCE RELATION" Ident "(" Ident ")".
*CreateReferenceRelationStmt* := "CREATE REFERENCE RELATION".
*PBPCreateStmt* :=
 "PREDICATE BASED PARAMETER" Ident "ON" FCSignature
 "PREDICATES {" "("" Ident ""," SQLExpr ")"
 {"," "("" Ident ""," SQLExpr ")"} "}".

*SQLExpr* := "{" {ANY} "}".

*/* Drop Statement */*
*DropStmt* := "DROP" (FCDropStmt | PCDropStmt | ParamDropStmt).
*FCDropStmt* := "FRAGMENT CLASS" FCSignature.
*PCDropStmt* := "PAGE CLASS" PCSignature.
*ParamDropStmt* := "PARAMETER" Ident "DEFINED UPON" FCSignature.

*/* Show Statement */*
*ShowStmt* := "SHOW" (FCShowStmt | PCShowStmt | ParamShowStmt).
*FCShowStmt* := "FRAGMENT CLASS" (FCSignature | "*").
*PCShowStmt* := "PAGE CLASS" (PCSignature | "*").
*ParamShowStmt* := "PARAMETER" (Ident | "*")
 ["DEFINED UPON" FCSignature].

*/* Alter Statement */*
*AlterStmt* := "ALTER" PCAlterStmt.
*PCAlterStmt* :=
 "PAGE CLASS" PCSignature (PCAlterAddStmt | PCAlterDropStmt).
*PCAlterAddStmt* :=
 [PCAlterAliasStmt] "ADD FRAGMENT CLASS" PCFCMapping.
*PCAlterAliasStmt* :=
 "REFERRING TO FRAGMENT CLASS"
 FCSignature "AS" Ident {"," FCSignature "AS" Ident}.

Table A.1: Prefixes used in Naming SMWP Artifacts

| Prefix | Description |
|--------|-------------|
| FC | *Content relation* storing a fragment class' data, e.g., FC_PremWines_region stores data of fragment class PremWines<region>. |
| FR | *Fragmentation relation* storing fragmentation, e.g., FR_PremWines_region stores fragmentation of fragment class PremWines<region>. |
| RV | *Reference relation* storing a parameter's domain, e.g., RV_Wineries_region stores the domain of parameter region defined on root fragment class Wineries<>. |
| AT | *Auxiliary trigger* used in propagating modifications of a derivation base class' content relation to a derived fragment class' content relation. |
| CT | *Content trigger* propagating modifications between content relations. |
| FT | *Fragmentation trigger* propagating modifications between fragmentation relations and from reference relations to fragmentation relations. |
| ST | *Serialization trigger* propagating modifications of data and fragmentation to pages. |

*PCAlterDropStmt* := "DROP FRAGMENT CLASS" FCSignature.

*/* Miscellaneous */*
*/* Signature of a fragment class */*
*FCSignature* := FCName "<" [ParamName {"," ParamName}] ">".
*/* Signature of a page class */*
*PCSignature* := PCName "<" [ParamName {"," ParamName}] ">".
*ParamName* := Ident. */* Name of a parameter */*
*FCName* := Ident. */* Name of a fragment class */*
*PCName* := Ident. */* Name of a page class */*
*FileName* := Ident "." Ident. */* Filename-Template for pages */*
*Ident* := letter {letter | digit | "_" | "-"}.
*letter* := "A|B|..|Z|a|b|..|z".
*digit* := "0|1|..|9".

## Naming Conventions

When referring to artifacts of the SMWP approach such as fragment classes and content relations, placeholders and variables are written in italics (e.g., *F<L>* or *FC_F_L*) while concrete artifacts are denoted in sans-serif (e.g., PremWines<region> or FC_PremWines_region).

Artifacts of the realization model (i.e., relations and triggers) are named using prefixes to explicitly distinguish their kind. Table A.1 describes the prefixes used.

# B   ATDF

This section shows the syntax of interactions to control active document flows by EBNF productions. Comments are used for annotation and are enclosed between "/\*" and "\*/".

After opening a connection to a peer by using interaction `open`, arbitrary interactions can be issued by a user affecting local versions, which are stored at the peer the user is connected to, and remote versions, which are stored at other, remote peers. While every interaction affects a local version, e.g., by modifying its content or deleting it, interactions that may affect remote versions are limited to `read`, `checkin`, `reallocate`, `merge`, and `relay`.

The implemented prototype takes advantage of the interactions' structure and implements a basic security mechanism by authenticating the user upon opening a connection which authorizes her/him to manipulate local versions arbitrarily. When an interaction is issued that affects a remote version, the remote peer tests whether the user is authorized to perform it.

Stmt := (ConnectStmt | DocFlowStmt | ActiveDocFlowStmt) ';'.
*/\* Connect to a peer \*/*
ConnectStmt := ('`open`' PID | '`close`').
*/\* Basic Artefacts \*/*
DID := `xs:anyURI`.
VID := `xs:anyURI`.
RemoteVID := VID ('`at`' | '`@`') PID.
RemoteDID := DID ('`at`' | '`@`') PID.
PID := `xs:anyURI`.
Variable := '`$`' Ident.
Ident := ('`A`'..'`Z`'|'`a`'..'`z`') {'`A`'..'`Z`'|'`a`'..'`z`'|'`0`'..'`9`'|'`-`'|'`_`'|'`.`'}.
Filename := '`file://`' {'`A`'..'`Z`'|'`a`'..'`z`'|'`0`'..'`9`'|'`-`'|'`_`'|'`.`'|'`/`'|'`:`'}.
*/\* Document flows \*/*
DocFlowStmt :=
    CheckinStmt | CheckoutStmt | ReadStmt |
    MergeStmt | DeleteStmt | ReallocateStmt | BindStmt |
    ProceedStmt | BranchStmt | CopyStmt.

## Simple Interactions

CheckinStmt := '`checkin`' (VID | RemoteVID) ['`of`' DID] '`from`' Variable.
CheckoutStmt :=
    '`checkout`' ('`successor`' | '`offspring`') '`of`' VID ['`into`' Variable].
ReadStmt := '`read`' (VID | RemoteVID) ['`into`' Variable].
MergeStmt := '`merge`' (VID) '`into`' (VID | RemoteVID).
DeleteStmt := '`delete`' VID.
ReallocateStmt := '`reallocate`' VID '`to`' (VID | RemoteVID).
BindStmt := '`bind`' Filename '`to`' Variable.

**Composite Interactions**

ProceedStmt := 'proceed' VID 'by' (VID | RemoteVID).
BranchStmt := 'branch' VID 'to' (VID | RemoteVID).
CopyStmt := 'copy' VID 'to' (VID | RemoteVID) ['of' DID].
/* *Composite interactions are formed from basic ones as follows:*

  *proceed* $v_1$ *by* $v_2$ $\equiv$
      *checkout successor of* $v_1$ *into* $f$; *checkin* $v_2$ *from* $f$;
  *branch* $v_1$ *to* $v_{1.i.1}$ $\equiv$
      *checkout offspring of* $v_1$ *into* $f$; *checkin* $v_{1.i.1}$ *from* $f$;
  *copy* $v_1$ *to* $v_2$ *of* $d_2$ $\equiv$
      *read* $v_1$ *into* $f$; *checkin* $v_2$ *of* $d_2$ *from* $f$;
*/

**Interactions on Active Documents and Versions**

ActiveDocFlowStmt := InvokeStmt | SubscribeStmt | RelayStmt.
InvokeStmt := 'invoke' OperationInvoc 'on' VID ['into' Variable].
OperationInvoc := Ident (LiteralValue {',' LiteralValue}).
LiteralValue := ' " ' String ' " ' | Number | .. | XML | Variable.
SubscribeStmt :=
    ('subscribe' | 'unsubscribe')
    EventClass 'of' (DID | RemoteDID) 'by' DID.
EventClass := Ident.
RelayStmt :=
    'relay' ('distribution' | 'reception')
    'from' VID 'to' (VID | RemoteVID).

# Acknowledgements

*Es war einmal vor langer Zeit und das war eine sehr gute Zeit da war eine Muhkuh die kam die Straße herunter gegangen und diese Muhkuh die da die Straße heruntergegangen kam die traf einen schönen tleinen Tnaben und der hieß Tucktuck-Baby...*

Indem ich also den Einstieg, den ersten Satz, mit obigem Zitat antrete, das übrigens seinerseits der erste Satz eines Buches von James Joyce ist, habe ich den für mich schwierigsten Teil der Danksagung bewältigt. Dass durch die Auswahl des einen eine Unzahl anderer nicht berücksichtig werden kann, muss verziehen werden. Dabei ist unter Umständen zu bedenken, dass die Wissenschaft einer persönlichen Eigenschaft förderlich zu sein scheint, die Flaubert in seinem Wörterbuch der Gemeinplätze den Advokaten zuschreibt: "Vor lauter Für und Wider ist ihre Urteilskraft verdorben".

Wem mein Dank gilt, ist ungleich einfacher zu bestimmen, denn alleine hätte ich diese Arbeit nicht zu Stande gebracht:

Ich danke meinen Betreuern, Gerti Kappel und Michael Schrefl, die einen wesentlichen Beitrag zu dieser Arbeit, direkt und indirekt, geleistet haben. Danke für euren Einsatz, eure uneingeschränkte Hilfsbereitschaft, für das produktive Umfeld, in dem es ausreichend Freiheit für die Erfüllung abwechslungsreicher und ambitionierter Aufgaben gab, und für das Vorleben eines wissenschaftlichen Enthusiasmus und eines Ideenreichtums, der ständiger Motivator war.

Danke meinen Kolleginnen und Kollegen, für die interessante Zusammenarbeit in Forschung und Lehre. Für die fruchtbaren Diskussion, die ausgetauschten Ideen, für die sinnvollen als auch die sinnlosen, die Anregungen und die Unterstützung. Namentlich bedanke ich mich bei Margit Brandl, Sabine Graf, Stefan Lechner, Elke Michlmayr, Günther Preuner, Michael Schadler, Sonja Willinger und insbesondere meinen jahrelangen Bürokollegen Gerhard Kramler, der mir durch seinen Blick hinter die Dinge sehr viel gezeigt hat, und Thomas Thalhammer, der mir mit seiner typischen Tatkraft beim Einstieg immer hilfsbereit zur Seite stand.

Danke auch an die Studierenden, die durch die prototypische Realisierung der Ansätze und ihr Feedback ihren Anteil an der Arbeit haben, namentlich Werner Enser, Jan Wenger, Florian Sonntag und Christian Sokop.

# Curriculum Vitae

| | |
|---|---|
| **Personal Record** | Martin Bernauer<br>born on April 28, 1975<br>in Salzburg, Austria.<br>`mailto:bernauer@big.tuwien.ac.at` |
| **Education** | Feb. 2001 - Jan. 2005<br>Ph.D. studies in Business Informatics<br>at the Johannes Kepler University Linz and<br>at the Vienna University of Technology.<br><br>Oct. 1994 - Oct. 1999<br>M.S. studies in Business Informatics<br>at the Johannes Kepler University Linz. |
| **Job Experience** | Jul. 2002- Jan. 2005<br>Faculty member at the Business Informatics Group,<br>Institute for Software Technology and Interactive Systems,<br>at the Vienna University of Technology.<br><br>Feb. 2001 - Jun. 2002<br>Faculty member at the Data and Knowledge Engineering Group,<br>Institute for Business Informatics,<br>at the Johannes Kepler University Linz.<br><br>Oct. 1999 - Jan. 2000<br>Consultant for Web-Technologies at<br>Porsche Informatik Austria, Salzburg. |
| **Publications** | see `http://www.big.tuwien.ac.at/research/publications` |